\documentclass{CSML}

\def\dOi{12(2:6)2016}
\lmcsheading%
{\dOi}
{1--28}
{}
{}
{Jul.~\phantom03, 2014}
{Jun.~13, 2016}
{}

\ACMCCS{[{\bf Theory of computation}]: Logic---Automated reasoning;
  Formalisms---Rewrite systems}

\pdfoutput=1

\usepackage{amsmath}
\usepackage{url}
\usepackage[utf8]{inputenc}
\usepackage{latexsym}
\usepackage{stmaryrd}
\usepackage{bussproofs}
\EnableBpAbbreviations
\usepackage{tabularx}
\usepackage[sectionbib]{natbib}
\usepackage{multirow}
\usepackage{hyperref}

\newcommand{\comment}[1]{}

\newtheorem{theorem}{Theorem}
\numberwithin{theorem}{section}

\newtheorem{definition}[theorem]{Definition}

\newcommand{\done}[2][?]{}

\DeclareUnicodeCharacter{00A0}{~} 
\DeclareUnicodeCharacter{00A7}{\S} 
\DeclareUnicodeCharacter{00AC}{\ensuremath{\neg}} 
\DeclareUnicodeCharacter{00B0}{^{\circ}} 
\DeclareUnicodeCharacter{00B1}{^1}
\DeclareUnicodeCharacter{00B2}{^2} 
\DeclareUnicodeCharacter{00B7}{\ensuremath{\cdot}} 
\DeclareUnicodeCharacter{00B9}{\textsuperscript{l}} 
\DeclareUnicodeCharacter{00D7}{\ensuremath{\times}} 
\DeclareUnicodeCharacter{00F7}{\ensuremath{\div}} 
\DeclareUnicodeCharacter{02B3}{\ensuremath{{^r}}} 
\DeclareUnicodeCharacter{02E1}{\ensuremath{{^l}}} 
\DeclareUnicodeCharacter{0393}{\ensuremath{\Gamma}} 
\DeclareUnicodeCharacter{0394}{\ensuremath{\Delta}} 
\DeclareUnicodeCharacter{0397}{\ensuremath{\textrm{H}}} 
\DeclareUnicodeCharacter{0398}{\ensuremath{\Theta}} 
\DeclareUnicodeCharacter{039B}{\ensuremath{\Lambda}} 
\DeclareUnicodeCharacter{039E}{\ensuremath{\Xi}} 
\DeclareUnicodeCharacter{03A3}{\ensuremath{\Sigma}} 
\DeclareUnicodeCharacter{03A6}{\ensuremath{\Phi}} 
\DeclareUnicodeCharacter{03A8}{\ensuremath{\Psi}} 
\DeclareUnicodeCharacter{03A9}{\ensuremath{\Omega}} 
\DeclareUnicodeCharacter{03B1}{\ensuremath{\mathnormal{\alpha}}} 
\DeclareUnicodeCharacter{03B2}{\ensuremath{\beta}} 
\DeclareUnicodeCharacter{03B3}{\ensuremath{\mathnormal{\gamma}}} 
\DeclareUnicodeCharacter{03B4}{\ensuremath{\mathnormal{\delta}}} 
\DeclareUnicodeCharacter{03B5}{\ensuremath{\mathnormal{\varepsilon}}} 
\DeclareUnicodeCharacter{03B6}{\ensuremath{\mathnormal{\zeta}}} 
\DeclareUnicodeCharacter{03B7}{\ensuremath{\mathnormal{\eta}}} 
\DeclareUnicodeCharacter{03B8}{\ensuremath{\mathnormal{\theta}}} 
\DeclareUnicodeCharacter{03B9}{\ensuremath{\mathnormal{\iota}}} 
\DeclareUnicodeCharacter{03BA}{\ensuremath{\mathnormal{\kappa}}} 
\DeclareUnicodeCharacter{03BB}{\ensuremath{\mathnormal{\lambda}}} 
\DeclareUnicodeCharacter{03BC}{\ensuremath{\mathnormal{\mu}}} 
\DeclareUnicodeCharacter{03BD}{\ensuremath{\mathnormal{\mu}}} 
\DeclareUnicodeCharacter{03BE}{\ensuremath{\mathnormal{\xi}}} 
\DeclareUnicodeCharacter{03C0}{\ensuremath{\mathnormal{\pi}}} 
\DeclareUnicodeCharacter{03C1}{\ensuremath{\mathnormal{\rho}}} 
\DeclareUnicodeCharacter{03C3}{\ensuremath{\mathnormal{\sigma}}} 
\DeclareUnicodeCharacter{03C4}{\ensuremath{\mathnormal{\tau}}} 
\DeclareUnicodeCharacter{03C6}{\ensuremath{\mathnormal{\varphi}}} 
\DeclareUnicodeCharacter{03C7}{\ensuremath{\mathnormal{\chi}}} 
\DeclareUnicodeCharacter{03C8}{\ensuremath{\mathnormal{\psi}}} 
\DeclareUnicodeCharacter{03C9}{\ensuremath{\mathnormal{\omega}}} 
\DeclareUnicodeCharacter{03D5}{\ensuremath{\mathnormal{\phi}}} 
\DeclareUnicodeCharacter{03F5}{\ensuremath{\mathnormal{\epsilon}}} 
\DeclareUnicodeCharacter{10627}{\ensuremath{\lbana}}
\DeclareUnicodeCharacter{10628}{\ensuremath{\rbana}}
\DeclareUnicodeCharacter{1D62}{_i} 
\DeclareUnicodeCharacter{2020}{\ensuremath{\dagger}} 
\DeclareUnicodeCharacter{2022}{\ensuremath{\text{\textbullet}}} 
\DeclareUnicodeCharacter{2026}{\ensuremath{\ldots}}
\DeclareUnicodeCharacter{202F}{{\,}}
\DeclareUnicodeCharacter{2032}{\ensuremath{^\prime}}  
\DeclareUnicodeCharacter{2033}{\ensuremath{^\second}} 
\DeclareUnicodeCharacter{2034}{\ensuremath{^\third}}  
\DeclareUnicodeCharacter{2070}{\ensuremath{^0}} 
\DeclareUnicodeCharacter{2071}{\ensuremath{^1}}
\DeclareUnicodeCharacter{2072}{\ensuremath{^2}}
\DeclareUnicodeCharacter{2073}{\ensuremath{^3}}
\DeclareUnicodeCharacter{2074}{\ensuremath{^4}}
\DeclareUnicodeCharacter{2075}{\ensuremath{^5}}
\DeclareUnicodeCharacter{2076}{\ensuremath{^6}}
\DeclareUnicodeCharacter{2077}{\ensuremath{^7}}
\DeclareUnicodeCharacter{2078}{\ensuremath{^8}}
\DeclareUnicodeCharacter{2079}{\ensuremath{^9}}
\DeclareUnicodeCharacter{207A}{\ensuremath{^+}} 
\DeclareUnicodeCharacter{207B}{\ensuremath{^-}} 
\DeclareUnicodeCharacter{2080}{\ensuremath{{}_0}} 
\DeclareUnicodeCharacter{2081}{\ensuremath{{}_1}}
\DeclareUnicodeCharacter{2082}{\ensuremath{{}_2}}
\DeclareUnicodeCharacter{2083}{\ensuremath{{}_3}}
\DeclareUnicodeCharacter{2084}{\ensuremath{{}_4}}
\DeclareUnicodeCharacter{2085}{\ensuremath{{}_5}}
\DeclareUnicodeCharacter{2086}{\ensuremath{{}_6}}
\DeclareUnicodeCharacter{2087}{\ensuremath{{}_7}}
\DeclareUnicodeCharacter{2088}{\ensuremath{{}_8}}
\DeclareUnicodeCharacter{2089}{\ensuremath{{}_9}}
\DeclareUnicodeCharacter{2096}{\ensuremath{{}_k}} 
\DeclareUnicodeCharacter{208A}{\ensuremath{_+}} 
\DeclareUnicodeCharacter{208B}{\ensuremath{_-}} 
\DeclareUnicodeCharacter{2099}{\ensuremath{_n}} 
\DeclareUnicodeCharacter{2102}{\ensuremath{\mathbb{C}}} 
\DeclareUnicodeCharacter{2115}{\ensuremath{\mathbb{N}}} %
\DeclareUnicodeCharacter{211D}{\ensuremath{\mathbb{R}}} 
\DeclareUnicodeCharacter{2124}{\ensuremath{\mathbb{Z}}} 
\DeclareUnicodeCharacter{214B}{\ensuremath{\parr}} 
\DeclareUnicodeCharacter{2190}{\ensuremath{\leftarrow}} 
\DeclareUnicodeCharacter{2191}{\ensuremath{\uparrow}} 
\DeclareUnicodeCharacter{2192}{\ensuremath{\rightarrow}} 
\DeclareUnicodeCharacter{2193}{\ensuremath{\downarrow}} 
\DeclareUnicodeCharacter{2194}{\ensuremath{\leftrightarrow}} 
\DeclareUnicodeCharacter{2196}{\nwarrow} 
\DeclareUnicodeCharacter{2197}{\nearrow} 
\DeclareUnicodeCharacter{219D}{\ensuremath{\leadsto}} 
\DeclareUnicodeCharacter{21A6}{\ensuremath{\mapsto}} 
\DeclareUnicodeCharacter{21C6}{\ensuremath{\leftrightarrows}} 
\DeclareUnicodeCharacter{21D0}{\ensuremath{\Leftarrow}} 
\DeclareUnicodeCharacter{21D2}{\ensuremath{\Rightarrow}} 
\DeclareUnicodeCharacter{21D4}{\ensuremath{\Leftrightarrow}} 
\DeclareUnicodeCharacter{2200}{\ensuremath{\forall}} 
\DeclareUnicodeCharacter{2203}{\ensuremath{\exists}} 
\DeclareUnicodeCharacter{2205}{\ensuremath{\varnothing}} 
\DeclareUnicodeCharacter{2208}{\ensuremath{\in}} 
\DeclareUnicodeCharacter{2209}{\ensuremath{\not\in}} 
\DeclareUnicodeCharacter{220B}{\ensuremath{\ni}}
\DeclareUnicodeCharacter{220E}{\ensuremath{\qed}} 
\DeclareUnicodeCharacter{2211}{\ensuremath{\sum}}
\DeclareUnicodeCharacter{2215}{\mathbb{N}} 
\DeclareUnicodeCharacter{2217}{\ensuremath{\ast}} 
\DeclareUnicodeCharacter{2218}{\ensuremath{\circ}} 
\DeclareUnicodeCharacter{2219}{\ensuremath{\bullet}} 
\DeclareUnicodeCharacter{221E}{\ensuremath{\infty}} 
\DeclareUnicodeCharacter{2223}{\ensuremath{\mid}} 
\DeclareUnicodeCharacter{2227}{\wedge}
\DeclareUnicodeCharacter{2228}{\vee}
\DeclareUnicodeCharacter{2229}{\ensuremath{\cap}} 
\DeclareUnicodeCharacter{222A}{\ensuremath{\cup}} 
\DeclareUnicodeCharacter{2237}{::} 
\DeclareUnicodeCharacter{223C}{\ensuremath{\sim}} 
\DeclareUnicodeCharacter{2243}{\ensuremath{\simeq}} 
\DeclareUnicodeCharacter{2245}{\ensuremath{\cong}} 
\DeclareUnicodeCharacter{2248}{\ensuremath{\approx}} 
\DeclareUnicodeCharacter{225C}{\ensuremath{\stackrel{\scriptscriptstyle {\triangle}}{=}}} 
\DeclareUnicodeCharacter{225D}{\ensuremath{\stackrel{\scriptscriptstyle {def}}{=}}} 
\DeclareUnicodeCharacter{225F}{\ensuremath{\stackrel{\scriptscriptstyle ?}{=}}} 
\DeclareUnicodeCharacter{2260}{\ensuremath{\neq}}
\DeclareUnicodeCharacter{2261}{\ensuremath{\equiv}}
\DeclareUnicodeCharacter{2264}{\ensuremath{\le}} 
\DeclareUnicodeCharacter{2265}{\ensuremath{\ge}} 
\DeclareUnicodeCharacter{226B}{\ensuremath{\gg}} 
\DeclareUnicodeCharacter{227A}{\ensuremath{\prec}} 
\DeclareUnicodeCharacter{2282}{\ensuremath{\subset}} 
\DeclareUnicodeCharacter{2283}{\ensuremath{\supset}} 
\DeclareUnicodeCharacter{2286}{\ensuremath{\subseteq}} 
\DeclareUnicodeCharacter{2287}{\ensuremath{\supseteq}} 
\DeclareUnicodeCharacter{2293}{\ensuremath{\sqcup}} 
\DeclareUnicodeCharacter{2293}{\sqcap} 
\DeclareUnicodeCharacter{2294}{\sqcup} 
\DeclareUnicodeCharacter{2295}{\ensuremath{\oplus}} 
\DeclareUnicodeCharacter{2297}{\ensuremath{\otimes}} 
\DeclareUnicodeCharacter{22A1}{\ensuremath{\boxdot}} 
\DeclareUnicodeCharacter{22A2}{\ensuremath{\vdash}}
\DeclareUnicodeCharacter{22A4}{\ensuremath{\top}} 
\DeclareUnicodeCharacter{22A5}{\ensuremath{\bot}} 
\DeclareUnicodeCharacter{22A7}{\models} 
\DeclareUnicodeCharacter{22A8}{\models} 
\DeclareUnicodeCharacter{22A9}{\Vdash} 
\DeclareUnicodeCharacter{22B8}{\ensuremath{\multimap}} 
\DeclareUnicodeCharacter{22C3}{\ensuremath{\bigcup}} 
\DeclareUnicodeCharacter{22C4}{\ensuremath{\diamond}} 
\DeclareUnicodeCharacter{22C6}{\ensuremath{\star}}
\DeclareUnicodeCharacter{22EE}{\ensuremath{\vdots}} 
\DeclareUnicodeCharacter{22EF}{\ensuremath{\cdots}} 
\DeclareUnicodeCharacter{2308}{\ensuremath{\lceil}}
\DeclareUnicodeCharacter{2309}{\ensuremath{\rceil}}
\DeclareUnicodeCharacter{230A}{\ensuremath{\lfloor}}
\DeclareUnicodeCharacter{230B}{\ensuremath{\rfloor}}
\DeclareUnicodeCharacter{25A1}{\ensuremath{\square}} 
\DeclareUnicodeCharacter{25AF}{\mathop{\talloblong}} 
\DeclareUnicodeCharacter{25B7}{\ensuremath{\triangleright}} 
\DeclareUnicodeCharacter{25B9}{\ensuremath{\rhd}} 
\DeclareUnicodeCharacter{25C7}{\ensuremath{\diamond}} 
\DeclareUnicodeCharacter{2605}{\ensuremath{\star}}   
\DeclareUnicodeCharacter{2713}{\ensuremath{\checkmark}} 
\DeclareUnicodeCharacter{27C2}{\ensuremath{^\bot}} 
\DeclareUnicodeCharacter{27E6}{\ensuremath{\llbracket}} 
\DeclareUnicodeCharacter{27E7}{\ensuremath{\rrbracket}} 
\DeclareUnicodeCharacter{27E8}{\ensuremath{\langle}} 
\DeclareUnicodeCharacter{27E9}{\ensuremath{\rangle}} 
\DeclareUnicodeCharacter{27F6}{{\longrightarrow}} 
\DeclareUnicodeCharacter{27F7}{{\longleftrightarrow}} 
\DeclareUnicodeCharacter{27F9}{\ensuremath{\Longrightarrow}} 
\DeclareUnicodeCharacter{29F8}{\ensuremath{\textfractionsolidus}} 
\DeclareUnicodeCharacter{2A02}{\ensuremath{\bigotimes}} 
\DeclareUnicodeCharacter{2A04}{\mathop{\dot{\cup}}} 
\DeclareUnicodeCharacter{2AEB}{\ensuremath{\perp\!\!\!\!\perp}} 
\DeclareUnicodeCharacter{2AFE}{\mathop{\talloblong}} 
\DeclareUnicodeCharacter{1D7D9}{\ensuremath{\mathds{1}}} 
\DeclareUnicodeCharacter{1D538}{\ensuremath{\mathds{A}}} 
\DeclareUnicodeCharacter{1D539}{\ensuremath{\mathds{B}}} 
\DeclareUnicodeCharacter{1D49F}{\ensuremath{\mathcal{D}}} 
\DeclareUnicodeCharacter{1D4A6}{\ensuremath{\mathcal{K}}} 

\DeclareUnicodeCharacter{220E}{\ensuremath{\blacksquare}} 

\DeclareMathAlphabet{\mathbfsf}{\encodingdefault}{\sfdefault}{bx}{n}

\newcommand{\kw}[1]{\ensuremath{\mathbf{#1}}}
\makeatletter
\@ifundefined{lhs2tex.lhs2tex.sty.read}%
  {\@namedef{lhs2tex.lhs2tex.sty.read}{}%
   \newcommand\SkipToFmtEnd{}%
   \newcommand\EndFmtInput{}%
   \long\def\SkipToFmtEnd#1\EndFmtInput{}%
  }\SkipToFmtEnd

\newcommand\ReadOnlyOnce[1]{\@ifundefined{#1}{\@namedef{#1}{}}\SkipToFmtEnd}
\usepackage{amstext}
\usepackage{amssymb}
\usepackage{stmaryrd}
\DeclareFontFamily{OT1}{cmtex}{}
\DeclareFontShape{OT1}{cmtex}{m}{n}
  {<5><6><7><8>cmtex8
   <9>cmtex9
   <10><10.95><12><14.4><17.28><20.74><24.88>cmtex10}{}
\DeclareFontShape{OT1}{cmtex}{m}{it}
  {<-> ssub * cmtt/m/it}{}

\DeclareFontShape{OT1}{cmtt}{bx}{n}
  {<5><6><7><8>cmtt8
   <9>cmbtt9
   <10><10.95><12><14.4><17.28><20.74><24.88>cmbtt10}{}
\DeclareFontShape{OT1}{cmtex}{bx}{n}
  {<-> ssub * cmtt/bx/n}{}

\newcommand{\Conid}[1]{\mathit{#1}}
\newcommand{\Varid}[1]{\mathit{#1}}
\newcommand{\anonymous}{\kern0.06em \vbox{\hrule\@width.5em}}

\usepackage{polytable}

\@ifundefined{mathindent}%
  {\newdimen\mathindent\mathindent\leftmargini}%
  {}%

\def\resethooks{%
  \global\let\SaveRestoreHook\empty
  \global\let\ColumnHook\empty}
\newcommand*{\savecolumns}[1][default]%
  {\g@addto@macro\SaveRestoreHook{\savecolumns[#1]}}
\newcommand*{\restorecolumns}[1][default]%
  {\g@addto@macro\SaveRestoreHook{\restorecolumns[#1]}}
\newcommand*{\aligncolumn}[2]%
  {\g@addto@macro\ColumnHook{\column{#1}{#2}}}

\resethooks

\newcommand{\onelinecommentchars}{\quad-{}- }
\newcommand{\commentbeginchars}{\enskip\{-}
\newcommand{\commentendchars}{-\}\enskip}

\newcommand{\visiblecomments}{%
  \let\onelinecomment=\onelinecommentchars
  \let\commentbegin=\commentbeginchars
  \let\commentend=\commentendchars}

\newcommand{\invisiblecomments}{%
  \let\onelinecomment=\empty
  \let\commentbegin=\empty
  \let\commentend=\empty}

\visiblecomments

\newlength{\blanklineskip}
\newcommand{\hsindent}[1]{\quad}
\let\hspre\empty
\let\hspost\empty

\EndFmtInput
\makeatother
\ReadOnlyOnce{polycode.fmt}%
\makeatletter

\newcommand{\hsnewpar}[1]%
  {{\parskip=0pt\parindent=0pt\par\vskip #1\noindent}}

\newcommand{\hscodestyle}{}

\newcommand{\sethscode}[1]%
  {\expandafter\let\expandafter\hscode\csname #1\endcsname
   \expandafter\let\expandafter\endhscode\csname end#1\endcsname}

  {\par\noindent
   \advance\leftskip\mathindent
   \hscodestyle
   \let\\=\@normalcr
   \let\hspre\(\let\hspost\)%
   \pboxed}%
  {\endpboxed\)%
   \par\noindent
   \ignorespacesafterend}

  {\hsnewpar\abovedisplayskip
   \advance\leftskip\mathindent
   \hscodestyle
   \let\hspre\(\let\hspost\)%
   \pboxed}%
  {\endpboxed%
   \hsnewpar\belowdisplayskip
   \ignorespacesafterend}

  {\hsnewpar\abovedisplayskip
   \advance\leftskip\mathindent
   \hscodestyle
   \let\\=\@normalcr
   \(\pboxed}%
  {\endpboxed\)%
   \hsnewpar\belowdisplayskip
   \ignorespacesafterend}

\newcommand{\plainhs}{\sethscode{plainhscode}}

\plainhs

  {\hsnewpar\abovedisplayskip
   \advance\leftskip\mathindent
   \hscodestyle
   \let\\=\@normalcr
   \(\parray}%
  {\endparray\)%
   \hsnewpar\belowdisplayskip
   \ignorespacesafterend}

  {\parray}{\endparray}

  {\(\parray}{\endparray\)}

\def\codeframewidth{\arrayrulewidth}
\RequirePackage{calc}

  {\parskip=\abovedisplayskip\par\noindent
   \hscodestyle
   \arrayrulewidth=\codeframewidth
   \tabular{@{}|p{\linewidth-2\arraycolsep-2\arrayrulewidth-2pt}|@{}}%
   \hline\framedhslinecorrect\\{-1.5ex}%
   \let\endoflinesave=\\
   \let\\=\@normalcr
   \(\pboxed}%
  {\endpboxed\)%
   \framedhslinecorrect\endoflinesave{.5ex}\hline
   \endtabular
   \parskip=\belowdisplayskip\par\noindent
   \ignorespacesafterend}

\newcommand{\framedhslinecorrect}[2]%
  {#1[#2]}

  {\(\def\column##1##2{}%
   \let\>\undefined\let\<\undefined\let\\\undefined
   \newcommand\>[1][]{}\newcommand\<[1][]{}\newcommand\\[1][]{}%
   \def\fromto##1##2##3{##3}%
   }{\) }%

  {\let\orighscode=\hscode
   \let\origendhscode=\endhscode
   \def\endhscode{\def\hscode{\endgroup\def\@currenvir{hscode}\\}\begingroup}
   \orighscode\def\hscode{\endgroup\def\@currenvir{hscode}}}%
  {\origendhscode
   \global\let\hscode=\orighscode
   \global\let\endhscode=\origendhscode}%

\makeatother
\EndFmtInput
\ReadOnlyOnce{jfpcompat.fmt}%
\makeatletter
\def\@authortable{%
  \leavevmode \hbox \bgroup $\col@sep\tabcolsep
  \let\d@llarbegin\begingroup
  \let\d@llarend\endgroup
  \let\\\author@tabcrone
  \ignorespaces
  \@tabarray}

\makeatother
\EndFmtInput
\newcommand{\un}{\mbox{\texttt{\_}}}   
\newcommand{\unopun}[1]{\ensuremath{\,\un{#1}\un\,}}

\usepackage[plain]{fancyref}
\usepackage{tikz}
\usetikzlibrary{external}
\usetikzlibrary{matrix,snakes,patterns}

\tikzstyle{inChart}=[anchor=north east]
\tikzstyle{grid}=[loosely dotted] 

\newcommand{\subt}[3] {
  \draw[grid] (#1,#1) -- (#1,#2) node[inChart] {#3} -- (#2,#2);
  \fill[color=black] (#1,#2) circle (2pt)
 }

\newcommand{\subc}[3] {
  \draw[thin] (#1,#1) -- (#1,#2) node[inChart] {#3} -- (#2,#2);
 }

\renewcommand\Varid[1]{\ensuremath{\mathsf{#1}}}
\renewcommand\Conid[1]{\ensuremath{\mathsf{#1}}}
\newcommand{\closure}[1]{{#1}^+}
\newcommand{\initial}[1]{I(#1)}
\newcommand{\sing}[1]{σ_{#1}}
\newcommand\powerset[1]{\ensuremath{\mathcal P(#1)}}
\newcommand{\grammarRule}[1]{(#1)}
\newcommand{\generates}[1][]{\stackrel{#1}{\longrightarrow}}
\newcommand{\generatestrans}{\generates[*]}

\newcommand{\tritwo}[4][]{
\twobytwo{#2}{#3}{0}{#4}
}

\newcommand{\twobytwo}[4]{
  \begin{bmatrix}
    #1&#2\\#3&#4
  \end{bmatrix}
}
\begin{document}

\title[Certified Context-Free Parsing]{Certified Context-Free Parsing:\\ A formalisation of Valiant's Algorithm in Agda}
\author[J.-P.~Bernardy]{Jean-Philippe Bernardy}
\address{Chalmers University of Technology \& University of Gothenburg, Sweden}
\email{\{bernardy, patrikj\}@chalmers.se}
\author[P.~Jansson]{Patrik Jansson}
\address{\vspace{-18 pt}}


\begin{abstract}
  Valiant (1975) has developed an algorithm for recognition of context
  free languages.
  As of today, it remains the algorithm with the best asymptotic
  complexity for this purpose.
  In this paper, we present an algebraic specification,
  implementation, and proof of correctness of a generalisation of
  Valiant's algorithm.
  The generalisation can be used for recognition, parsing or generic
  calculation of the transitive closure of upper triangular matrices.
  The proof is certified by the Agda proof assistant.
  The certification is representative of state-of-the-art methods for
  specification and proofs in proof assistants based on type-theory.
  As such, this paper can be read as a tutorial for the Agda system.
\end{abstract}

\maketitle

\section*{Introduction}

Context-free grammars \citep{chomsky_syntactic_1957} are the standard
formalism to express and study the syntactic structure of programming
languages.  While numerous algorithms are used for parsing
context-free inputs, the subject of this paper is a generalisation of
the recognition algorithm discovered by \citet{valiant_general_1975}.  (For
simplicity we call the generalisation simply ``Valiant's algorithm''
below.)

Valiant's algorithm has many qualities. First, it is efficient:
it is the parsing algorithm with the best worst-case asymptotic complexity \citep{lee_fast_2002} ($O(n^3)$).
Further, it has recently been identified that its performance in
common average cases is also excellent \citep{bernardy_efficient_2015}:
in the presence of a hierarchical input, and given some care in the representation of grammars and data
structures, it behaves linearly, and can be parallelized.
Second, Valiant's algorithm is abstract and general. While its main
application is parsing, it solves the problem of finding the
transitive closure of a generalised relation $W$ given as an
upper-triangular matrix, where the underlying element operation is not
necessarily associative.
Third, Valiant's algorithm is relatively simple: we will see that it
can be expressed as two mutually defined functions by induction on the
size of the matrix.

The combined qualities of Valiant's algorithm (importance of the
problem solved, efficiency, generality and simplicity) make it, in our
opinion, one of the top ten algorithms that every computer scientist
should learn. As such, it is a good candidate for being given a fully
certified-correct implementation. Such an implementation is
the main contribution of this paper.

The medium chosen for our proof is the Agda proof assistant
\citep{norell_practical_2007}. One motivation for this choice is that
Agda is based on type-theory, and using type-theory as a core means that
our proof can be verified by a
type-checker, which can itself be subject to formal
verification. This chain of certification, relying ultimately on a
small trusted base shared with many other proofs makes us very
confident that our development is correct.  (Even though the core of
Agda is not currently verified, there is ongoing effort in this
direction.)

A secondary goal of this work is to provide an exemplary proof: we
have taken particular care to make the proof approachable. In
particular:
1. the algorithm is derived from its specification (instead of being
first implemented and proved after the fact).
2. the core of formal proof is close to a
semi-formal proof developed earlier \citep{bernardy_efficient_2013}.
Further, because we assume little knowledge of the Agda proof
assistant, our development can be used as a tutorial on algorithm
specification and derivation in Agda.


The rest of the paper is organised as follows.  In \fref{sec:charts},
we review how Valiant's algorithm reduces parsing to the computation
of transitive closure. \Fref{sec:prel} introduces
Agda's syntax and some basic formalisation concepts. Sections
\ref{sec:overview} to \ref{sec:smallest} contain the formal
development. We review related work in \fref{sec:related} and conclude
with possible extensions in \fref{sec:extensions}.

\section{Parsing as Transitive Closure}\label{sec:charts}
In this section we review the basic idea underlying Valiant's
algorithm, namely that context-free parsing can be specified as
computing a transitive closure.
\subsection{Chomsky Normal Form}
The simplest implementation of Valiant's algorithm takes as
input a grammar in Chomsky Normal Form \citep{chomsky_certain_1959}. In
Chomsky Normal From, hereafter abbreviated CNF, the production rules
are restricted to one the following forms.
\begin{align*}
A_0 &::= A_1 A_2 & \text{(binary)} \\
A &::= t       & \text{(unary)}  \\
S &::= \epsilon  & \text{(nullary)}
\end{align*}
Any context-free grammar $\mathcal G = (N, \Sigma, P, S)$ generating a
given language can be converted to a grammar $\mathcal G'$ in CNF
generating the same language.
($N$ is the set of non-terminals, $\Sigma$ is the alphabet, $P$ is the
set of productions, $S$ is the start symbol and we use $A$ to range
over elements of $N$.)
Hence we will assume from now on a grammar provided in CNF.
Moreover, because it is easy to handle the empty string specially, we
conventionally exclude it from the input language and thus exclude the
nullary rule $S ::= \epsilon$ from the set of production rules.
The reader avid of details is directed to \citet{lange_cnf_2009} for a
pedagogical account of the process of reduction to CNF.

Given a grammar specified as above, the problem of parsing is reduced
to finding a binary tree such that each leaf corresponds to a symbol
of the input (and a suitable unary rule) and such that each branch corresponds to
a suitable binary rule. Essentially, parsing is equivalent to considering
all possible bracketings of the input, and finding one (or more) that form a
valid parse.

\subsection{Charts}
Let $w$ be a vector of input symbols.
We define the operations $0,+,\cdot$ and $\sing{}$ as follows.
\begin{definition}[$0,+,\cdot$ on $\powerset N$]
\begin{align*}
  0 & = ∅ &
  x + y & = x \cup y &
  x \cdot y & = \{ A_0 ~|~ A_1 ∈ x, A_2 ∈ y, \grammarRule{A_0 ::= A_1 A_2} ∈ P \} \\
  & & & & \sing i & = \{A ~|~ \grammarRule{A ::= w[i]} ∈ P \}
\end{align*}
\end{definition}
The above operations can be lifted to matrices, in the usual way (and
we do so formally in \fref{sec:matrices}.)
The $(\cdot)$ operation fully characterises the binary production
rules of the grammar, while $\sing{}$ captures the unary ones.
We will then use a matrix of sets of non-terminals to record which
non-terminals can generate a given substring. Such a matrix is called a chart.
If $C$ is a (complete) chart, $A ∈ C_{ij}$ iff $A \generates[*] w[i..j]$.
See \fref{fig:chart} for an illustration.
The operation $\sing{}$ is used to construct an initial chart $\initial w$ such that
\begin{align*}
  \initial w_{i,i+1} & = \sing i \\
  \initial w_{i,j} & = 0 & \text{if $j ≠ i+1$}
\end{align*}
The matrix $W = \initial w$ is a partial chart:
it contains the correct non-terminals for strings of length one,
stored at positions $(i,i+1)$ in the chart.
All other positions are empty (zero).
Computing the transitive closure of $W$ will complete the chart;
so that at position $(0,n)$ one will find the non-terminals generating
the whole input.
\begin{definition}[Transitive closure]
  If it exists, the transitive closure of a matrix $W$, written $W^+$,
  is the smallest matrix $C$ such that
$$ C = W + C · C$$
\end{definition}
\noindent
The equation means that $C$ contains all possible associations of
$W$ multiplied by itself
\footnote{If $(·)$ were associative, then a simpler formula for the
  transitive closure could be given, and a much more efficient
  technique could be used to compute it, but then all bracketings
  would be equivalent and $(·)$ could not capture the binary rules of
  a context-free grammar.}:
\begin{align*}
C & = W + C · C \\
  & = W + (W + C · C) · (W + C · C) \\
  & = W + WW + W·(C·C) + (C·C)·W + (C · C)·(C · C)&%
\!\!\!\!\!\!\text{by distributivity}\\
  & = \ldots \\
  & = W + WW + W·(W·W) + (W·W)·W + (W · W)·(W · W) + \ldots &%
\!\!\!\text{by dist.\ \& comm.}\\
\end{align*}
Therefore, if a parse tree (possible bracketing) exists, the algorithm
will find it.
Furthermore, because $C$ is the least matrix satisfying the equation,
it will not contain any non-terminal which does not generate the
input.

\newcommand{\mrk}[2]{\node[inChart] at (#1,#1) {#2}}
\begin{figure}
\centering
  \begin{tikzpicture}
    \pgftransformrotate{-90}
    \pgftransformscale{0.4}
    \draw (0,0) -- (8,8);
    \subt 1 5 {$A$};
    \subt 4 7 {$B$};
    \mrk 1 {$i$};
    \mrk 5 {$j$};
    \mrk 4 {$k$};
    \mrk 7 {$l$};
  \end{tikzpicture}
  \quad
  \begin{tikzpicture}
    \pgftransformrotate{-90}
    \pgftransformscale{0.4}
    \draw (0,0) -- (8,8);
    \subt 1 3 {$X$};
    \subt 3 7 {$Y$};
    \subt 1 7 {$Z$};
  \end{tikzpicture}
  \caption{Example charts.
    In each chart a point at position $(r,c)$ corresponds to a substring starting at $r$ and ending at $c$.
    The first parameter ($r$ for row) grows downwards and the second
    one ($c$ for column) rightwards.
    The input string $w$ is represented by the diagonal line.
    Dots in the upper-right part represent non-terminals.
    The first chart witnesses $A \generatestrans w[i..j]$ and
    $B \generatestrans w[k..l]$.
    An instance of the rule $Z ::= X Y$ is exemplified on the second chart.
  }
  \label{fig:chart}
\end{figure}

The above procedure specifies a recogniser: by finding the closure of
$I(w)$ one finds if $w$ is parsable, but not the corresponding parse
tree.  However, one can obtain a proper parser by using sets of parse
trees (instead of non-terminals) and extending $(\cdot)$ to combine
parse trees.

\section{Agda preliminaries}\label{sec:prel}

This section introduces the elements of Agda necessary to
understand the upcoming sections.
We present both the language itself, and some definitions which are
part of the standard library.
Throughout the paper, we will use a literate-programming style.
The body of this section is a single module \ensuremath{\Conid{Preliminaries}} which
contains ``specification building blocks'' --- a number of definition
for later use.
\begin{hscode}\SaveRestoreHook
\column{B}{@{}>{\hspre}l<{\hspost}@{}}%
\column{E}{@{}>{\hspre}l<{\hspost}@{}}%
\>[B]{}\kw{module}\;\Conid{Preliminaries}\;\kw{where}{}\<[E]%
\ColumnHook
\end{hscode}\resethooks
In Agda code, scoping is indicated by indentation.
Indentation is rather hard to visualise in a paper which spans over
several pages, so we will instead give scoping hints in the text when
necessary.
The scope of the current module extends to the end of the section.

\subsubsection*{Propositions-as-types}
The philosophy behind Agda is that each proposition is expressed as a
type.
That is, proving that a proposition \ensuremath{\Conid{P}} holds means finding an
inhabitant (an element) of \ensuremath{\Conid{P}} (read as a type).
With this in mind, we define the type of relations over an underlying
type \ensuremath{\Conid{A}} as functions mapping two elements of type \ensuremath{\Conid{A}} to a another
type (a \ensuremath{\Conid{Set}} in Agda parlance).
\begin{hscode}\SaveRestoreHook
\column{B}{@{}>{\hspre}l<{\hspost}@{}}%
\column{8}{@{}>{\hspre}l<{\hspost}@{}}%
\column{11}{@{}>{\hspre}l<{\hspost}@{}}%
\column{E}{@{}>{\hspre}l<{\hspost}@{}}%
\>[B]{}\Conid{Rel}\;\!:\!\;\Conid{Set}\;\to \;\Conid{Set₁}{}\<[E]%
\\
\>[B]{}\Conid{Rel}\;\Conid{A}\;{}\<[8]%
\>[8]{}\mathrel{=}\;{}\<[11]%
\>[11]{}\Conid{A}\;\to \;\Conid{A}\;\to \;\Conid{Set}{}\<[E]%
\ColumnHook
\end{hscode}\resethooks
Hence, an element of type \ensuremath{\Conid{Rel}\;\Conid{A}} is a binary relation on \ensuremath{\Conid{A}}.
For example \ensuremath{\Conid{Rel}\;\Conid{A}} will be inhabited by equivalence relations and orderings.
A consequence of the above definition is that our relations are
constructive:
to show that a pair \ensuremath{(\Varid{x,y})} is in a relation \ensuremath{\Conid{R}} we must provide a
witness of type \ensuremath{\Conid{R}\;\Varid{x}\;\Varid{y}}.
(In passing, the type of relations is a so-called big set (\ensuremath{\Conid{Set₁}}),
because it contains \ensuremath{\Conid{Set}}s itself.
This distinction is necessary for the consistency of the logical
system.)

As another example, we can define the existential quantifier
connective as follows:
\pagebreak 
\begin{hscode}\SaveRestoreHook
\column{B}{@{}>{\hspre}l<{\hspost}@{}}%
\column{3}{@{}>{\hspre}l<{\hspost}@{}}%
\column{5}{@{}>{\hspre}l<{\hspost}@{}}%
\column{E}{@{}>{\hspre}l<{\hspost}@{}}%
\>[B]{}\kw{record}\;\Varid{∃}\;\{\mskip1.5mu \Conid{A}\;\!:\!\;\Conid{Set}\mskip1.5mu\}\;(\Conid{P}\;\!:\!\;\Conid{A}\;\Varid{→}\;\Conid{Set})\;\!:\!\;\Conid{Set₁}\;\kw{where}{}\<[E]%
\\
\>[B]{}\hsindent{3}{}\<[3]%
\>[3]{}\Varid{constructor}\;\un,\!\un{}\<[E]%
\\
\>[B]{}\hsindent{3}{}\<[3]%
\>[3]{}\kw{field}{}\<[E]%
\\
\>[3]{}\hsindent{2}{}\<[5]%
\>[5]{}\Varid{proj₁}\;\!:\!\;\Conid{A}{}\<[E]%
\\
\>[3]{}\hsindent{2}{}\<[5]%
\>[5]{}\Varid{proj₂}\;\!:\!\;\Conid{P}\;\Varid{proj₁}{}\<[E]%
\ColumnHook
\end{hscode}\resethooks
That is, the existence of an element satisfying \ensuremath{\Conid{P}} can be written
\ensuremath{\Varid{∃}\;(\Varid{\char92 x}\;\to \;\Conid{P}\;\Varid{x})} (or, equivalently, \ensuremath{\Varid{∃}\;\Conid{P}}), and proving this
proposition means to find a witness \ensuremath{\Varid{x}\;\!:\!\;\Conid{A}} and an inhabitant \ensuremath{\Varid{p}\;\!:\!\;\Conid{P}\;\Varid{x}}.
The \ensuremath{\Varid{constructor}} keyword here introduces the name \ensuremath{\un,\!\un} as the
two-argument constructor of the record type.
Infix operators are declared using underscores on both sides of the
name so an infix comma (\ensuremath{\Varid{,}}) can now be used as in the following
example: \ensuremath{(\Varid{x}\;\Varid{,}\;\Varid{p})\;\!:\!\;\Varid{∃}\;\Conid{P}}.
Note that only the last argument (\ensuremath{\Conid{P}}) of the \ensuremath{\Varid{∃}} symbol is written, the
first (\ensuremath{\Conid{A}}) is left for Agda to infer: we say that it is \emph{implicit}.
Implicit parameters are marked by placing them in braces at the
declaration site.
An implicit argument can be supplied explicitly at a call site, if the
programmer encloses it with braces.
This syntax can be useful if Agda fails to infer the argument in a
certain context.

\subsubsection*{Entire relations}
A binary relation from a set \ensuremath{\Conid{A}} to a set \ensuremath{\Conid{B}} is called \emph{entire} if
every element of \ensuremath{\Conid{A}} is related to at least one element of \ensuremath{\Conid{B}}, and we
can encode this definition as follows.
\begin{hscode}\SaveRestoreHook
\column{B}{@{}>{\hspre}l<{\hspost}@{}}%
\column{E}{@{}>{\hspre}l<{\hspost}@{}}%
\>[B]{}\Conid{Entire}\;\!:\!\;\{\mskip1.5mu \Conid{A}\;\Conid{B}\;\!:\!\;\Conid{Set}\mskip1.5mu\}\;\to \;(\Varid{\char95 R\char95 }\;\!:\!\;\Conid{A}\;\to \;\Conid{B}\;\to \;\Conid{Set})\;\to \;\Conid{Set}{}\<[E]%
\\
\>[B]{}\Conid{Entire}\;\Varid{\char95 R\char95 }\;\mathrel{=}\;\Varid{∀}\;\Varid{a}\;\to \;\Varid{∃}\;\lambda \;\Varid{b}\;\to \;\Varid{a}\;\Conid{R}\;\Varid{b}{}\<[E]%
\ColumnHook
\end{hscode}\resethooks
Here, again, the use of underscores around \ensuremath{\Conid{R}} makes it an infix
operator (in its scope).
Fixity is just a presentation issue, so an equivalent, but shorter,
definition is \ensuremath{\Conid{Entire}\;\Conid{R}\;\mathrel{=}\;\Varid{∀}\;\Varid{a}\;\to \;\Varid{∃}\;(\Conid{R}\;\Varid{a})} where \ensuremath{\Conid{R}} is prefix and \ensuremath{\Conid{R}\;\Varid{a}} is a partial application.
A consequence of proving that a relation \ensuremath{\Conid{R}} is entire in our constructive
setting is that we get a function contained in \ensuremath{\Conid{R}}.
We can extract the function using the first field of the \ensuremath{\Varid{∃}} record.
\begin{hscode}\SaveRestoreHook
\column{B}{@{}>{\hspre}l<{\hspost}@{}}%
\column{E}{@{}>{\hspre}l<{\hspost}@{}}%
\>[B]{}\Varid{fun}\;\!:\!\;\{\mskip1.5mu \Conid{A}\;\Conid{B}\;\!:\!\;\Conid{Set}\mskip1.5mu\}\;\to \;\{\mskip1.5mu \Varid{\char95 R\char95 }\;\!:\!\;\Conid{A}\;\to \;\Conid{B}\;\to \;\Conid{Set}\mskip1.5mu\}\;\to \;\Conid{Entire}\;\Varid{\char95 R\char95 }\;\to \;\Conid{A}\;\to \;\Conid{B}{}\<[E]%
\\
\>[B]{}\Varid{fun}\;\Varid{ent}\;\Varid{a}\;\mathrel{=}\;\Varid{proj₁}\;(\Varid{ent}\;\Varid{a}){}\<[E]%
\ColumnHook
\end{hscode}\resethooks
The proof that the function is contained in \ensuremath{\Conid{R}} can be obtained from
the second field of \ensuremath{\Varid{∃}}:
\begin{hscode}\SaveRestoreHook
\column{B}{@{}>{\hspre}l<{\hspost}@{}}%
\column{12}{@{}>{\hspre}l<{\hspost}@{}}%
\column{32}{@{}>{\hspre}l<{\hspost}@{}}%
\column{E}{@{}>{\hspre}l<{\hspost}@{}}%
\>[B]{}\Varid{correct}\;\!:\!\;{}\<[12]%
\>[12]{}\{\mskip1.5mu \Conid{A}\;\Conid{B}\;\!:\!\;\Conid{Set}\mskip1.5mu\}\;\to \;\{\mskip1.5mu \Varid{\char95 R\char95 }\;\!:\!\;\Conid{A}\;\to \;\Conid{B}\;\to \;\Conid{Set}\mskip1.5mu\}\;\to \;(\Varid{ent}\;\!:\!\;\Conid{Entire}\;\Varid{\char95 R\char95 })\;\to {}\<[E]%
\\
\>[12]{}\kw{let}\;\Varid{f}\;\mathrel{=}\;\Varid{fun}\;\Varid{ent}\;\kw{in}\;{}\<[32]%
\>[32]{}\Varid{∀}\;\{\mskip1.5mu \Varid{a}\;\!:\!\;\Conid{A}\mskip1.5mu\}\;\to \;\Varid{a}\;\Conid{R}\;(\Varid{f}\;\Varid{a}){}\<[E]%
\\
\>[B]{}\Varid{correct}\;\Varid{ent}\;\{\mskip1.5mu \Varid{a}\mskip1.5mu\}\;\mathrel{=}\;\Varid{proj₂}\;(\Varid{ent}\;\Varid{a}){}\<[E]%
\ColumnHook
\end{hscode}\resethooks
The above pattern generalises to relations of any number of arguments.
In this paper we need the following version:
\begin{hscode}\SaveRestoreHook
\column{B}{@{}>{\hspre}l<{\hspost}@{}}%
\column{E}{@{}>{\hspre}l<{\hspost}@{}}%
\>[B]{}\Conid{Entire3}\;\!:\!\;\{\mskip1.5mu \Conid{A}\;\Conid{B}\;\Conid{C}\;\Conid{D}\;\!:\!\;\Conid{Set}\mskip1.5mu\}\;\to \;(\Conid{R}\;\!:\!\;\Conid{A}\;\to \;\Conid{B}\;\to \;\Conid{C}\;\to \;\Conid{D}\;\to \;\Conid{Set})\;\to \;\Conid{Set}{}\<[E]%
\\
\>[B]{}\Conid{Entire3}\;\Conid{R}\;\mathrel{=}\;\Varid{∀}\;\Varid{x}\;\Varid{y}\;\Varid{z}\;\to \;\Varid{∃}\;(\Conid{R}\;\Varid{x}\;\Varid{y}\;\Varid{z}){}\<[E]%
\ColumnHook
\end{hscode}\resethooks
with corresponding definitions of \ensuremath{\Varid{fun3}} and \ensuremath{\Varid{correct3}}.

\subsubsection*{Uniqueness} An element of \ensuremath{\Conid{UniqueSolution}\;\unopun{\Varid{≃}}\;\Conid{P}} is a proof that the predicate \ensuremath{\Conid{P}}
has (at most) a \emph{unique} solution relative to some underlying
relation \ensuremath{\unopun{\Varid{≃}}}.
%
%
\begin{hscode}\SaveRestoreHook
\column{B}{@{}>{\hspre}l<{\hspost}@{}}%
\column{47}{@{}>{\hspre}l<{\hspost}@{}}%
\column{51}{@{}>{\hspre}l<{\hspost}@{}}%
\column{E}{@{}>{\hspre}l<{\hspost}@{}}%
\>[B]{}\Conid{UniqueSolution}\;\!:\!\;\{\mskip1.5mu \Conid{A}\;\!:\!\;\Conid{Set}\mskip1.5mu\}\;\to \;\Conid{Rel}\;\Conid{A}\;\to \;(\Conid{A}\;\to \;\Conid{Set})\;\to \;\Conid{Set}{}\<[E]%
\\
\>[B]{}\Conid{UniqueSolution}\;\unopun{\Varid{≃}}\;\Conid{P}\;\mathrel{=}\;\Varid{∀}\;\{\mskip1.5mu \Varid{x}\;\Varid{y}\mskip1.5mu\}\;\to \;\Conid{P}\;\Varid{x}\;\to \;\Conid{P}\;\Varid{y}\;{}\<[47]%
\>[47]{}\to \;{}\<[51]%
\>[51]{}\Varid{x}\;\Varid{≃}\;\Varid{y}{}\<[E]%
\ColumnHook
\end{hscode}\resethooks
A proof \ensuremath{\Varid{usP}\;\!:\!\;\Conid{UniqueSolution}\;\unopun{\Varid{≃}}\;\Conid{P}} is thus a function which given
two hidden arguments of type \ensuremath{\Conid{A}} and two proofs that they satisfy \ensuremath{\Conid{P}}
returns a proof that they are related by \ensuremath{\unopun{\Varid{≃}}}.

\subsubsection*{Least solutions} In optimisation problems, one often wants to find the least solution with
respect to some order \ensuremath{\unopun{\mathrel{≤}}}.
We use \ensuremath{\Conid{LowerBound}\;\unopun{\mathrel{≤}}\;\Conid{P}} for the predicate that holds for an \ensuremath{\Varid{a}\;\!:\!\;\Conid{A}}
iff \ensuremath{\Varid{a}} is smaller than all elements satisfying \ensuremath{\Conid{P}}.
If a lower bound is in the set (satisfies the predicate \ensuremath{\Conid{P}}) it is
called \emph{least}.
\begin{hscode}\SaveRestoreHook
\column{B}{@{}>{\hspre}l<{\hspost}@{}}%
\column{9}{@{}>{\hspre}l<{\hspost}@{}}%
\column{16}{@{}>{\hspre}l<{\hspost}@{}}%
\column{19}{@{}>{\hspre}l<{\hspost}@{}}%
\column{21}{@{}>{\hspre}l<{\hspost}@{}}%
\column{26}{@{}>{\hspre}l<{\hspost}@{}}%
\column{29}{@{}>{\hspre}l<{\hspost}@{}}%
\column{E}{@{}>{\hspre}l<{\hspost}@{}}%
\>[B]{}\Conid{LowerBound}\;\!:\!\;\{\mskip1.5mu \Conid{A}\;\!:\!\;\Conid{Set}\mskip1.5mu\}\;\to \;\Conid{Rel}\;\Conid{A}\;\to \;(\Conid{A}\;\to \;\Conid{Set})\;\to \;(\Conid{A}\;\to \;\Conid{Set}){}\<[E]%
\\
\>[B]{}\Conid{LowerBound}\;\unopun{\mathrel{≤}}\;\Conid{P}\;{}\<[19]%
\>[19]{}\Varid{a}\;\mathrel{=}\;\Varid{∀}\;\Varid{z}\;\to \;(\Conid{P}\;\Varid{z}\;\to \;\Varid{a}\;\mathrel{≤}\;\Varid{z}){}\<[E]%
\\[\blanklineskip]%
\>[B]{}\Conid{Least}\;{}\<[9]%
\>[9]{}\!:\!\;\{\mskip1.5mu \Conid{A}\;\!:\!\;\Conid{Set}\mskip1.5mu\}\;\to \;\Conid{Rel}\;\Conid{A}\;\to \;(\Conid{A}\;\to \;\Conid{Set})\;\to \;(\Conid{A}\;\to \;\Conid{Set}){}\<[E]%
\\
\>[B]{}\Conid{Least}\;{}\<[9]%
\>[9]{}\unopun{\mathrel{≤}}\;\Conid{P}\;{}\<[16]%
\>[16]{}\Varid{a}\;\mathrel{=}\;{}\<[21]%
\>[21]{}\Conid{P}\;\Varid{a}\;{}\<[26]%
\>[26]{}\mathbin{\!\times\!}\;{}\<[29]%
\>[29]{}\Conid{LowerBound}\;\unopun{\mathrel{≤}}\;\Conid{P}\;\Varid{a}{}\<[E]%
\ColumnHook
\end{hscode}\resethooks
Note that a proof \ensuremath{\Varid{alP}\;\!:\!\;\Conid{Least}\;\unopun{\mathrel{≤}}\;\Conid{P}\;\Varid{a}} is a pair of a proof that \ensuremath{\Varid{a}}
is in \ensuremath{\Conid{P}} and a function \ensuremath{\Varid{albP}\;\!:\!\;\Conid{LowerBound}\;\unopun{\mathrel{≤}}\;\Conid{P}\;\Varid{a}}.
And, in turn, \ensuremath{\Varid{albP}} is a function that takes any \ensuremath{\Varid{z}\;\!:\!\;\Conid{A}} (with a proof that
\ensuremath{\Varid{z}} is in \ensuremath{\Conid{P}}) to a proof that \ensuremath{\Varid{a}\;\mathrel{≤}\;\Varid{z}}.


\subsubsection*{Records and modules}
The upcoming proof makes extensive use of
\emph{records}, which we review now in detail.
Agda record types contain fields and helper definitions.
Fields refer to data which is stored in the record, while helper
definitions provide values which can be computed from such
data.
Because Agda treats proofs (of propositions) as data, one can require
the fields to satisfy some laws, just by adding (proofs of) those laws as fields.
Our first record type example, \ensuremath{\Varid{∃}\;\Conid{P}}, has two fields; one element
\ensuremath{\Varid{proj₁}} and a proof that it satisfies the property \ensuremath{\Conid{P}}.
As a more complex record type example we use the following
(simplified) version of \ensuremath{\Conid{IsCommutativeMonoid}} from
\ensuremath{\Conid{Algebra.Structures}} in the standard library.
The record type is parametrised over a carrier set, a relation, a
binary operation and its identity element:
\savecolumns
%
\begin{hscode}\SaveRestoreHook
\column{B}{@{}>{\hspre}l<{\hspost}@{}}%
\column{3}{@{}>{\hspre}l<{\hspost}@{}}%
\column{5}{@{}>{\hspre}l<{\hspost}@{}}%
\column{7}{@{}>{\hspre}l<{\hspost}@{}}%
\column{20}{@{}>{\hspre}l<{\hspost}@{}}%
\column{31}{@{}>{\hspre}l<{\hspost}@{}}%
\column{36}{@{}>{\hspre}l<{\hspost}@{}}%
\column{41}{@{}>{\hspre}l<{\hspost}@{}}%
\column{44}{@{}>{\hspre}l<{\hspost}@{}}%
\column{E}{@{}>{\hspre}l<{\hspost}@{}}%
\>[3]{}\kw{record}\;\Conid{IsCommutativeMonoid}\;{}\<[31]%
\>[31]{}\{\mskip1.5mu \Conid{A}\;\!:\!\;\Conid{Set}\mskip1.5mu\}\;(\unopun{\mathrel{≈}}\;\!:\!\;\Conid{Rel}\;\Conid{A})\;{}\<[E]%
\\
\>[31]{}(\unopun{\mathbin{∙}}\;\!:\!\;\Conid{A}\;\to \;\Conid{A}\;\to \;\Conid{A})\;(\Varid{ε}\;\!:\!\;\Conid{A})\;\!:\!\;\Conid{Set₁}\;\kw{where}{}\<[E]%
\\
\>[3]{}\hsindent{2}{}\<[5]%
\>[5]{}\kw{field}{}\<[E]%
\\
\>[5]{}\hsindent{2}{}\<[7]%
\>[7]{}\Varid{isSemigroup}\;{}\<[20]%
\>[20]{}\!:\!\;\Conid{IsSemigroup}\;{}\<[36]%
\>[36]{}\unopun{\mathrel{≈}}\;{}\<[44]%
\>[44]{}\unopun{\mathbin{∙}}{}\<[E]%
\\
\>[5]{}\hsindent{2}{}\<[7]%
\>[7]{}\Varid{identityˡ}\;{}\<[20]%
\>[20]{}\!:\!\;\Conid{LeftIdentity}\;{}\<[36]%
\>[36]{}\unopun{\mathrel{≈}}\;{}\<[41]%
\>[41]{}\Varid{ε}\;{}\<[44]%
\>[44]{}\unopun{\mathbin{∙}}{}\<[E]%
\\
\>[5]{}\hsindent{2}{}\<[7]%
\>[7]{}\Varid{comm}\;{}\<[20]%
\>[20]{}\!:\!\;\Conid{Commutative}\;{}\<[36]%
\>[36]{}\unopun{\mathrel{≈}}\;{}\<[44]%
\>[44]{}\unopun{\mathbin{∙}}{}\<[E]%
\ColumnHook
\end{hscode}\resethooks
The fields capture the requirements of being a commutative monoid, in
terms of three other properties.
Here, the first field (\ensuremath{\Varid{isSemigroup}}) is also of record type; it is in
fact common in Agda to define deeply nested record structures.

In Agda every record type also doubles as a module parametrised over
a value of that type.
%
For example, within the scope of the above record, given a value \ensuremath{\Varid{isNP}\;\!:\!\;\Conid{IsSemigroup}\;\{\mskip1.5mu \Conid{ℕ}\mskip1.5mu\}\;\unopun{\mathrel{≈}}\;\unopun{\Varid{+}}}, the phrase \ensuremath{\Conid{IsSemigroup}\;\Varid{isNP}} is
meaningful in a context where Agda expects a module.
It denotes a module containing a declaration for each field and helper
definition of the \ensuremath{\Conid{IsSemigroup}} record type.
Hence, within the scope of the above record, one can access the
(nested) fields of \ensuremath{\Varid{isSemigroup}} in the module \ensuremath{\Conid{IsSemigroup}\;\Varid{isSemigroup}}.
In fact, it is very common for a record type to re-export all the
definitions of inner records. This can be done with the following
declaration (still inside the \ensuremath{\kw{record}}):
\restorecolumns
\begin{hscode}\SaveRestoreHook
\column{B}{@{}>{\hspre}l<{\hspost}@{}}%
\column{5}{@{}>{\hspre}l<{\hspost}@{}}%
\column{E}{@{}>{\hspre}l<{\hspost}@{}}%
\>[5]{}\kw{open}\;\Conid{IsSemigroup}\;\Varid{isSemigroup}\;\kw{public}{}\<[E]%
\ColumnHook
\end{hscode}\resethooks
\ensuremath{\kw{open}} means that the new names are brought into scope for later
definitions inside the record type (module) and
\ensuremath{\kw{public}} means the new names are also exported (publicly visible).
This means that the user can ignore the nesting when fetching nested fields (but not when constructing them).

\subsubsection*{Equality proofs}
We finish this section with an example of a helper definition that also serves as an introduction to \emph{equality-proof notation}.
A helper definition of \ensuremath{\Conid{IsCommutativeMonoid}} is the right identity
proof \ensuremath{\Varid{identityʳ}}, which is derivable from commutativity and left
identity.
We can define it as follows (still inside the \ensuremath{\kw{record}\;\Conid{IsCommutativeMonoid}}):
%
\restorecolumns
\begin{hscode}\SaveRestoreHook
\column{B}{@{}>{\hspre}l<{\hspost}@{}}%
\column{5}{@{}>{\hspre}l<{\hspost}@{}}%
\column{7}{@{}>{\hspre}l<{\hspost}@{}}%
\column{9}{@{}>{\hspre}l<{\hspost}@{}}%
\column{25}{@{}>{\hspre}l<{\hspost}@{}}%
\column{E}{@{}>{\hspre}l<{\hspost}@{}}%
\>[5]{}\Varid{identityʳ}\;\!:\!\;\Varid{∀}\;\Varid{x}\;\to \;{}\<[25]%
\>[25]{}(\Varid{x}\;\mathbin{∙}\;\Varid{ε})\;\mathrel{≈}\;\Varid{x}{}\<[E]%
\\
\>[5]{}\Varid{identityʳ}\;\Varid{x}\;\mathrel{=}\;{}\<[E]%
\\
\>[5]{}\hsindent{2}{}\<[7]%
\>[7]{}\Varid{begin}\;{}\<[E]%
\\
\>[7]{}\hsindent{2}{}\<[9]%
\>[9]{}\Varid{x}\;\mathbin{∙}\;\Varid{ε}\;{}\<[E]%
\\
\>[5]{}\hsindent{2}{}\<[7]%
\>[7]{}≈\!\!⟨\;\Varid{comm}\;\Varid{x}\;\Varid{ε}\;\Varid{⟩}\;{}\<[E]%
\\
\>[7]{}\hsindent{2}{}\<[9]%
\>[9]{}\Varid{ε}\;\mathbin{∙}\;\Varid{x}\;{}\<[E]%
\\
\>[5]{}\hsindent{2}{}\<[7]%
\>[7]{}≈\!\!⟨\;\Varid{identityˡ}\;\Varid{x}\;\Varid{⟩}\;{}\<[E]%
\\
\>[7]{}\hsindent{2}{}\<[9]%
\>[9]{}\Varid{x}\;{}\<[E]%
\\
\>[5]{}\hsindent{2}{}\<[7]%
\>[7]{}\Varid{∎}{}\<[E]%
\ColumnHook
\end{hscode}\resethooks
The above is merely the composition by transitivity of \ensuremath{(\Varid{comm}\;\Varid{x}\;\Varid{ε})} and
\ensuremath{(\Varid{identityˡ}\;\Varid{x})}, but by using special purpose operators the user can keep
track of the intermediate steps in the proof in a style close to pen-and-paper proofs.
%


\section{Formal Development Overview}
\label{sec:overview}

In the following sections we expose our formalisation of the
specification of the transitive closure algorithm, its implementation
(Valiant's algorithm) and the proof of correctness.
The algorithm falls out from a calculational refinement of the specification
rather than being exposed \textit{ex nihilo} and proved separately.
The development is presented in a number of stages:
\begin{itemize}
\item We define the ring-like algebraic structure where we set our
  development (\fref{sec:algebra}),

\item We give the specification of the transitive closure
  (\fref{sec:spec}),

\item We define the concrete data structures that the algorithm
  manipulates (\fref{sec:matrices}),

\item We derive Valiant's algorithm from part of that specification
  (\fref{sec:derivation}), 

\item We conclude by showing that the algorithm satisfies the rest of
  the specification (\fref{sec:smallest}).
\end{itemize}
\providecommand{\structure}[2]{C#1 & \ref{structure:#1}  & #2 & \pageref{structure:#1}}
\begin{figure}
  \centering
\begin{tabular}{lllr}
  \parbox[b]{3em}{Code\\ section} & \parbox[b]{3em}{Paper\\ section} & Heading & Page
\\\hline
  \structure{1}{\ensuremath{\Conid{SemiNearRing}}}
\\\structure{1.1}{Carriers, operators}
\\\structure{1.2}{Commutative monoid \ensuremath{(\Varid{+,0})}}
\\\structure{1.3}{Distributive, idempotent, \ldots}
\\\structure{1.4}{Exporting commutative monoid operations}
\\\structure{1.5}{Setoid, \ldots}
\\\structure{1.6}{Lower bounds}
\\\hline
  \structure{2}{\ensuremath{\Conid{SemiNearRing2}}}
\\\structure{2.1}{Plus and times for \ensuremath{\Varid{u}}, \ldots}
\\\structure{2.2}{Linear equation \ensuremath{\Conid{L}}}
\\\structure{2.3}{Properties of \ensuremath{\Conid{L}}}
\\\hline
  \structure{3}{\ensuremath{\Conid{ClosedSemiNearRing}}}
\\\structure{3.1}{Quadratic equation \ensuremath{\Conid{Q}} + properties}
\\\structure{3.2}{Closure function and correctness}
\\\structure{3.3}{Function for \ensuremath{\Conid{L}} and its correctness}
\\\structure{3.4}{Ordering properties of \ensuremath{\Conid{L}} and \ensuremath{\Conid{Q}}}
\\\hline
  \structure{4}{2-by-2 block matrix, preserving \ensuremath{\Conid{ClosedSemiNearRing}}}
\\\structure{4.1}{Square matrix}
\\\structure{4.2}{Upper triangular matrix}
\\\structure{4.3}{Laws}
\\\structure{4.4}{Lifting \ensuremath{\Conid{Q}} and its proof}
\\\structure{4.5}{Lifting orders and their properties}
\\\structure{4.6}{Lifting the proof of \ensuremath{\Conid{L}}}
\\\structure{4.7}{Proofs for ordering \ensuremath{\Conid{L}}-solutions}
\\\hline
  \structure{5}{One-by-one matrix}
\\\structure{5.1}{Base case for \ensuremath{\Conid{L}}}
\\\structure{5.2}{Base case for least \ensuremath{\Conid{Q}} }
\\\hline
  \structure{6}{Top level recursion for square matrices}
\\\structure{6.1}{Top level algorithm extraction}
\end{tabular}
\caption{Mapping from code sections to paper sections.}
\label{fig:codemapping}
\end{figure}
\renewcommand{\structure}[2]{\label{structure:#1}C#1: #2}
In \fref{fig:codemapping} we present the mapping between the order of
presentation in the paper and in the Agda development.
The paper starts from a simplified presentation of the development
including the top level algorithm and only then builds up to include
all the properties and proofs needed for the full formalization.
This makes the proof easier to follow and lets us explain step by step
the full algebra needed.
The Agda code, on the other hand, introduces the full algebra earlier
and only gets to the top level algorithm on the last line of the file.


\section{Algebra}\label{sec:algebra}

%
We begin by defining a record type called \ensuremath{\Conid{SemiNearRing}}, whose fields
and helper definitions capture the algebraic structure that we need for the
algorithm development.
First we introduce a carrier set \ensuremath{\Varid{s}} with an equivalence relation, a
zero, addition and multiplication.
\savecolumns[SemiNearRing]
\savecolumns[SemiNearRingA]
\begin{hscode}\SaveRestoreHook
\column{B}{@{}>{\hspre}l<{\hspost}@{}}%
\column{3}{@{}>{\hspre}l<{\hspost}@{}}%
\column{6}{@{}>{\hspre}l<{\hspost}@{}}%
\column{12}{@{}>{\hspre}l<{\hspost}@{}}%
\column{51}{@{}>{\hspre}l<{\hspost}@{}}%
\column{E}{@{}>{\hspre}l<{\hspost}@{}}%
\>[B]{}\kw{record}\;\Conid{SemiNearRing}\;\!:\!\;\Conid{Set₁}\;\kw{where}\;{}\<[51]%
\>[51]{}\mbox{\onelinecomment  \structure{1}{\ensuremath{\Conid{SemiNearRing}}}}{}\<[E]%
\\
\>[B]{}\hsindent{3}{}\<[3]%
\>[3]{}\kw{field}\;{}\<[51]%
\>[51]{}\mbox{\onelinecomment  \structure{1.1}{Carriers, operators}}{}\<[E]%
\\
\>[3]{}\hsindent{3}{}\<[6]%
\>[6]{}\Varid{s}\;\!:\!\;\Conid{Set}{}\<[E]%
\\
\>[3]{}\hsindent{3}{}\<[6]%
\>[6]{}\unopun{\mathrel{≃_s}}\;{}\<[12]%
\>[12]{}\!:\!\;\Varid{s}\;\Varid{→}\;\Varid{s}\;\Varid{→}\;\Conid{Set}{}\<[E]%
\\
\>[3]{}\hsindent{3}{}\<[6]%
\>[6]{}0_s\;{}\<[12]%
\>[12]{}\!:\!\;\Varid{s}{}\<[E]%
\\
\>[3]{}\hsindent{3}{}\<[6]%
\>[6]{}\unopun{\mathbin{+_s}}\;{}\<[12]%
\>[12]{}\!:\!\;\Varid{s}\;\Varid{→}\;\Varid{s}\;\Varid{→}\;\Varid{s}{}\<[E]%
\\
\>[3]{}\hsindent{3}{}\<[6]%
\>[6]{}\unopun{\mathbin{·_s}}\;{}\<[12]%
\>[12]{}\!:\!\;\Varid{s}\;\Varid{→}\;\Varid{s}\;\Varid{→}\;\Varid{s}{}\<[E]%
\ColumnHook
\end{hscode}\resethooks
Sets of non-terminals form a \ensuremath{\Conid{SemiNearRing}} with finite sets of
non-terminals for \ensuremath{\Varid{s}} and its operations for the other fields (the
usual equality, the empty set as zero, set union as addition and
``cross product filtered by the grammar'' as multiplication).

Further, the charts (from \fref{sec:charts}), also form a
\ensuremath{\Conid{SemiNearRing}}.
Indeed, lifting the operations on matrices preserve the
\ensuremath{\Conid{SemiNearRing}} structure, as we formally prove in \fref{sec:matrices}.

At this stage they are ``raw'' operations without laws.
Here is a summary of the laws needed.
We require that ($0$,$+$) forms a commutative monoid.
We also require that $0$ is absorbing for $(\cdot)$, that
$(\cdot)$ distributes over $(+)$ and that $(+)$ is idempotent.
Note that the product is not necessarily associative; in fact,
if it were, computing the transitive closure would be much easier and parsing would not be an application.
%
%
%
\begin{align*}
  x + 0 &= x =  0 + x    &  x + y &= y + x   &  x + (y + z) &= (x + y) + z   & \mbox{(now)}\\
  x · 0 &= 0 =  0 · x    &  x + x &= x       &  x · (y + z) &= x · y + x · z & \mbox{(later)}
\end{align*}
(It is easy to check that the definitions given in \fref{sec:charts} have these properties.)
We could specify these properties individually, but instead we take
advantage of ``specification building blocks'' from the \ensuremath{\Conid{Algebra}}
modules in the Agda standard library \citep{danielsson_agda_2013}.
More specifically we use the record \ensuremath{\Conid{IsCommutativeMonoid}} from the
library module \ensuremath{\Conid{Structures}} and we use the left and right zero laws
from the parametrised module \ensuremath{\Conid{FunctionProperties}} specialised to the
underlying equivalence (\ensuremath{\unopun{\mathrel{≃_s}}}) on our carrier set \ensuremath{\Varid{s}}.
\restorecolumns[SemiNearRingA]
\begin{hscode}\SaveRestoreHook
\column{B}{@{}>{\hspre}l<{\hspost}@{}}%
\column{3}{@{}>{\hspre}l<{\hspost}@{}}%
\column{42}{@{}>{\hspre}l<{\hspost}@{}}%
\column{E}{@{}>{\hspre}l<{\hspost}@{}}%
\>[3]{}\kw{open}\;\Conid{Algebra.Structures}\;{}\<[42]%
\>[42]{}\kw{using}\;(\Conid{IsCommutativeMonoid}){}\<[E]%
\\
\>[3]{}\kw{open}\;\Conid{Algebra.FunctionProperties}\;\unopun{\mathrel{≃_s}}\;{}\<[42]%
\>[42]{}\kw{using}\;(\Conid{LeftZero};\Conid{RightZero}){}\<[E]%
\ColumnHook
\end{hscode}\resethooks
Armed with these properties we now continue the \ensuremath{\Conid{SemiNearRing}} record
type by specifying (as new fields in the record) the laws we require
of our operations:
\ensuremath{\unopun{\mathbin{+_s}}} is a commutative monoid with \ensuremath{0_s} as the unit,
\ensuremath{0_s} is also a multiplicative zero of \ensuremath{\unopun{\mathbin{·_s}}} and multiplication preserves equivalence.
\restorecolumns[SemiNearRingA]
\begin{hscode}\SaveRestoreHook
\column{B}{@{}>{\hspre}l<{\hspost}@{}}%
\column{3}{@{}>{\hspre}l<{\hspost}@{}}%
\column{5}{@{}>{\hspre}l<{\hspost}@{}}%
\column{12}{@{}>{\hspre}l<{\hspost}@{}}%
\column{25}{@{}>{\hspre}l<{\hspost}@{}}%
\column{29}{@{}>{\hspre}l<{\hspost}@{}}%
\column{31}{@{}>{\hspre}l<{\hspost}@{}}%
\column{38}{@{}>{\hspre}l<{\hspost}@{}}%
\column{39}{@{}>{\hspre}l<{\hspost}@{}}%
\column{42}{@{}>{\hspre}l<{\hspost}@{}}%
\column{51}{@{}>{\hspre}l<{\hspost}@{}}%
\column{52}{@{}>{\hspre}l<{\hspost}@{}}%
\column{55}{@{}>{\hspre}l<{\hspost}@{}}%
\column{65}{@{}>{\hspre}l<{\hspost}@{}}%
\column{70}{@{}>{\hspre}l<{\hspost}@{}}%
\column{E}{@{}>{\hspre}l<{\hspost}@{}}%
\>[3]{}\kw{field}\;{}\<[51]%
\>[51]{}\mbox{\onelinecomment  \structure{1.2}{Commutative monoid \ensuremath{(\Varid{+,0})}}}{}\<[E]%
\\
\>[3]{}\hsindent{2}{}\<[5]%
\>[5]{}\Varid{isCommMon}\;\!:\!\;\Conid{IsCommutativeMonoid}\;\unopun{\mathrel{≃_s}}\;\unopun{\mathbin{+_s}}\;0_s{}\<[E]%
\\[\blanklineskip]%
\>[3]{}\hsindent{2}{}\<[5]%
\>[5]{}\Varid{zeroˡ}\;{}\<[12]%
\>[12]{}\!:\!\;\Conid{LeftZero}\;{}\<[25]%
\>[25]{}0_s\;{}\<[31]%
\>[31]{}\unopun{\mathbin{·_s}}{}\<[38]%
\>[38]{}\mbox{\onelinecomment  expands to \ensuremath{\Varid{∀}\;\Varid{x}\;\Varid{→}\;(0_s\;\mathbin{·_s}\;\Varid{x})\;\mathrel{≃_s}\;0_s}}{}\<[E]%
\\
\>[3]{}\hsindent{2}{}\<[5]%
\>[5]{}\Varid{zeroʳ}\;{}\<[12]%
\>[12]{}\!:\!\;\Conid{RightZero}\;{}\<[25]%
\>[25]{}0_s\;{}\<[31]%
\>[31]{}\unopun{\mathbin{·_s}}{}\<[38]%
\>[38]{}\mbox{\onelinecomment  expands to \ensuremath{\Varid{∀}\;\Varid{x}\;\Varid{→}\;(\Varid{x}\;\mathbin{·_s}\;0_s)\;\mathrel{≃_s}\;0_s}}{}\<[E]%
\\
\>[3]{}\hsindent{2}{}\<[5]%
\>[5]{}\unopun{\mathbin{<\!\!\!·\!\!\!>}}\;{}\<[12]%
\>[12]{}\!:\!\;\Varid{∀}\;\{\mskip1.5mu \Varid{x}\;\Varid{y}\;\Varid{u}\;\Varid{v}\mskip1.5mu\}\;\Varid{→}\;{}\<[29]%
\>[29]{}(\Varid{x}\;\mathrel{≃_s}\;\Varid{y})\;{}\<[39]%
\>[39]{}\Varid{→}\;{}\<[42]%
\>[42]{}(\Varid{u}\;\mathrel{≃_s}\;\Varid{v})\;{}\<[52]%
\>[52]{}\Varid{→}\;{}\<[55]%
\>[55]{}(\Varid{x}\;\mathbin{·_s}\;\Varid{u}\;{}\<[65]%
\>[65]{}\mathrel{≃_s}\;{}\<[70]%
\>[70]{}\Varid{y}\;\mathbin{·_s}\;\Varid{v}){}\<[E]%
\ColumnHook
\end{hscode}\resethooks
The rest of the record type consists of helper declarations which are
useful to have in scope when working with the specification. They will
be put in scope whenever we access an instance of the record in the
definition of the algorithm or in its proof.
Also inside the record type we specify the precedence of operators
using the following declarations:
\restorecolumns[SemiNearRingA]
\begin{hscode}\SaveRestoreHook
\column{B}{@{}>{\hspre}l<{\hspost}@{}}%
\column{3}{@{}>{\hspre}l<{\hspost}@{}}%
\column{E}{@{}>{\hspre}l<{\hspost}@{}}%
\>[3]{}\kw{infix}\;\Varid{4}\;\unopun{\mathrel{≃_s}};\kw{infixl}\;\Varid{6}\;\unopun{\mathbin{+_s}};\kw{infixl}\;\Varid{7}\;\unopun{\mathbin{·_s}}{}\<[E]%
\ColumnHook
\end{hscode}\resethooks

\subsubsection*{Exporting inner names}
As we mentioned in \fref{sec:prel} the module \ensuremath{\Conid{Algebra.Structures}}
includes a record (\ensuremath{\Conid{IsCommutativeMonoid}}) which contains the
commutative monoid laws, and doubles up as a parametrised module.
In the \ensuremath{\Conid{SemiNearRing}} record type that we are defining there is
already a field \ensuremath{\Varid{isCommMon}} which, in turn, contains the proofs of the
monoid laws so that a user with a value \ensuremath{\Varid{snr}\;\!:\!\;\Conid{SemiNearRing}} can
access all these proofs by indexing through the two record layers.
But two (and later more) levels of records is inconvenient to use so
we include short hand names for the inner record fields as follows.
\restorecolumns[SemiNearRingA]
\begin{hscode}\SaveRestoreHook
\column{B}{@{}>{\hspre}l<{\hspost}@{}}%
\column{3}{@{}>{\hspre}l<{\hspost}@{}}%
\column{5}{@{}>{\hspre}l<{\hspost}@{}}%
\column{6}{@{}>{\hspre}c<{\hspost}@{}}%
\column{6E}{@{}l@{}}%
\column{9}{@{}>{\hspre}l<{\hspost}@{}}%
\column{24}{@{}>{\hspre}l<{\hspost}@{}}%
\column{51}{@{}>{\hspre}l<{\hspost}@{}}%
\column{E}{@{}>{\hspre}l<{\hspost}@{}}%
\>[51]{}\mbox{\onelinecomment  \structure{1.4}{Exporting commutative monoid operations}}{}\<[E]%
\\
\>[3]{}\kw{open}\;\Conid{Algebra.Structures.IsCommutativeMonoid}\;\Varid{isCommMon}\;\kw{public}{}\<[E]%
\\
\>[3]{}\hsindent{2}{}\<[5]%
\>[5]{}\kw{hiding}\;(\Varid{refl}){}\<[E]%
\\
\>[3]{}\hsindent{2}{}\<[5]%
\>[5]{}\kw{renaming}{}\<[E]%
\\
\>[5]{}\hsindent{1}{}\<[6]%
\>[6]{}({}\<[6E]%
\>[9]{}\Varid{isEquivalence}\;{}\<[24]%
\>[24]{}\kw{to}\;\Varid{isEquiv}_s{}\<[E]%
\\
\>[5]{}\hsindent{1}{}\<[6]%
\>[6]{};{}\<[6E]%
\>[9]{}\Varid{assoc}\;{}\<[24]%
\>[24]{}\kw{to}\;\Varid{assoc}_s{}\<[E]%
\\
\>[5]{}\hsindent{1}{}\<[6]%
\>[6]{};{}\<[6E]%
\>[9]{}\Varid{comm}\;{}\<[24]%
\>[24]{}\kw{to}\;\Varid{comm}_s{}\<[E]%
\\
\>[5]{}\hsindent{1}{}\<[6]%
\>[6]{};{}\<[6E]%
\>[9]{}\text{•-}\Varid{cong}\;{}\<[24]%
\>[24]{}\kw{to}\;\unopun{\mathbin{<\!\!\!+\!\!\!>}}{}\<[E]%
\\
\>[5]{}\hsindent{1}{}\<[6]%
\>[6]{};{}\<[6E]%
\>[9]{}\Varid{identityˡ}\;{}\<[24]%
\>[24]{}\kw{to}\;\Varid{identityˡ}_s{}\<[E]%
\\
\>[5]{}\hsindent{1}{}\<[6]%
\>[6]{}){}\<[6E]%
\\
\>[3]{}\Varid{identityʳ}_s\;\mathrel{=}\;\Varid{proj₂}\;\Varid{identity}{}\<[E]%
\ColumnHook
\end{hscode}\resethooks
Recall that \ensuremath{\kw{open}} means that the new names are brought into scope for later
definitions inside the record type and \ensuremath{\kw{public}} means the new names
are also exported (publicly visible).
We rename (with a subscript $s$-suffix) to avoid clashes when we add most of the
same for another set \ensuremath{\Varid{u}} later.
The infix notation \ensuremath{\unopun{\mathbin{<\!\!\!+\!\!\!>}}} for congruence is useful in equality
proofs: if we have proofs
\ensuremath{\Varid{ap}\;\!:\!\;\Varid{a}\;\mathrel{≃_s}\;\Varid{a'}} and
\ensuremath{\Varid{bp}\;\!:\!\;\Varid{b}\;\mathrel{≃_s}\;\Varid{b'}} we get a proof of \ensuremath{(\Varid{a}\;\mathbin{+_s}\;\Varid{b})\;\mathrel{≃_s}\;(\Varid{a'}\;\mathbin{+_s}\;\Varid{b'})} which can be layed out nicely:
\restorecolumns[SemiNearRing]
\aligncolumn{13}{@{}>{\hspre}c<{\hspost}@{}}
\begin{hscode}\SaveRestoreHook
\column{B}{@{}>{\hspre}l<{\hspost}@{}}%
\column{5}{@{}>{\hspre}l<{\hspost}@{}}%
\column{9}{@{}>{\hspre}l<{\hspost}@{}}%
\column{13}{@{}>{\hspre}l<{\hspost}@{}}%
\column{18}{@{}>{\hspre}l<{\hspost}@{}}%
\column{22}{@{}>{\hspre}l<{\hspost}@{}}%
\column{E}{@{}>{\hspre}l<{\hspost}@{}}%
\>[5]{}\Varid{begin}\;{}\<[E]%
\\
\>[5]{}\hsindent{4}{}\<[9]%
\>[9]{}\Varid{a}\;{}\<[13]%
\>[13]{}\mathbin{+_s}\;{}\<[18]%
\>[18]{}\Varid{b}{}\<[E]%
\\
\>[5]{}≈\!\!⟨\;{}\<[9]%
\>[9]{}\Varid{ap}\;{}\<[13]%
\>[13]{}\mathbin{<\!\!\!+\!\!\!>}\;{}\<[18]%
\>[18]{}\Varid{bp}\;{}\<[22]%
\>[22]{}\Varid{⟩}\;{}\<[E]%
\\
\>[9]{}\Varid{a'}\;{}\<[13]%
\>[13]{}\mathbin{+_s}\;{}\<[18]%
\>[18]{}\Varid{b'}{}\<[E]%
\\
\>[5]{}\Varid{∎}{}\<[E]%
\ColumnHook
\end{hscode}\resethooks
%

Finally \ensuremath{\Varid{sSetoid}} packages up the carrier set \ensuremath{\Varid{s}}, the relation
\ensuremath{\unopun{\mathrel{≃_s}}} and the proof that it is an equivalence.
This is useful not only for documentation purposes (\ensuremath{\Varid{s}} \emph{is} a
setoid), but certain parts of the standard library require properties
to be packaged in such a manner, for example the module giving
convenient syntax for equality proofs, which we use later.
We write \ensuremath{0_L} for universe level \ensuremath{\Varid{0}}.
\restorecolumns[SemiNearRingA]
\begin{hscode}\SaveRestoreHook
\column{B}{@{}>{\hspre}l<{\hspost}@{}}%
\column{3}{@{}>{\hspre}l<{\hspost}@{}}%
\column{23}{@{}>{\hspre}l<{\hspost}@{}}%
\column{38}{@{}>{\hspre}l<{\hspost}@{}}%
\column{51}{@{}>{\hspre}l<{\hspost}@{}}%
\column{E}{@{}>{\hspre}l<{\hspost}@{}}%
\>[3]{}\Varid{sSetoid}\;\!:\!\;\Conid{Setoid}\;0_L\;0_L{}\<[51]%
\>[51]{}\mbox{\onelinecomment  \structure{1.5}{Setoid, \ldots}}{}\<[E]%
\\
\>[3]{}\Varid{sSetoid}\;\mathrel{=}\;\kw{record}\;\{\mskip1.5mu {}\<[23]%
\>[23]{}\Conid{Carrier}\;{}\<[38]%
\>[38]{}\mathrel{=}\;\Varid{s};{}\<[E]%
\\
\>[23]{}\unopun{\mathrel{≈}}\;{}\<[38]%
\>[38]{}\mathrel{=}\;\unopun{\mathrel{≃_s}};{}\<[E]%
\\
\>[23]{}\Varid{isEquivalence}\;{}\<[38]%
\>[38]{}\mathrel{=}\;\Varid{isEquiv}_s\mskip1.5mu\}{}\<[E]%
\ColumnHook
\end{hscode}\resethooks
\restorecolumns[SemiNearRingA]
\begin{hscode}\SaveRestoreHook
\column{B}{@{}>{\hspre}l<{\hspost}@{}}%
\column{3}{@{}>{\hspre}l<{\hspost}@{}}%
\column{5}{@{}>{\hspre}l<{\hspost}@{}}%
\column{14}{@{}>{\hspre}l<{\hspost}@{}}%
\column{E}{@{}>{\hspre}l<{\hspost}@{}}%
\>[3]{}\kw{open}\;\Conid{IsEquivalence}\;\Varid{isEquiv}_s\;\kw{public}{}\<[E]%
\\
\>[3]{}\hsindent{2}{}\<[5]%
\>[5]{}\kw{hiding}\;{}\<[14]%
\>[14]{}(\Varid{reflexive})\;\kw{renaming}\;(\Varid{refl}\;\kw{to}\;\Varid{refl}_s;\Varid{sym}\;\kw{to}\;\Varid{sym}_s;\Varid{trans}\;\kw{to}\;\Varid{trans}_s){}\<[E]%
\ColumnHook
\end{hscode}\resethooks
The \ensuremath{\kw{open}} \ensuremath{\kw{public}} of \ensuremath{\Varid{isEquiv}_s} is (again) to avoid the need for
multiple layers of record projections.

To summarise, a record value \ensuremath{\Varid{snr}\;\!:\!\;\Conid{SemiNearRing}} contains as fields a
carrier set \ensuremath{\Varid{s}}, the operations (\ensuremath{\unopun{\mathrel{≃_s}}}, \ensuremath{0_s}, \ensuremath{\unopun{\mathbin{+_s}}}, \ensuremath{\unopun{\mathbin{·_s}}}) and
the proofs of the properties (\ensuremath{\Varid{isCommMon}}, \ensuremath{\Varid{zeroˡ}}, \ensuremath{\Varid{zeroʳ}}, \ensuremath{\unopun{\mathbin{<\!\!\!·\!\!\!>}}}).
We will later return to this record type and add a few more fields (in
\fref{sec:spec}) and helpers (in \fref{sec:completion} and
\fref{sec:smallest}) to capture the full specification of the closure
algorithm, but for now this will do.
\subsection{Matrix Algebra(s)}
Valiant's algorithm works on square matrices.
We carry on and define the algebraic structures required on matrices
for Valiant's algorithm to work.
Some of the structure of these matrices is the same as that required
of elements, and we will later (in \fref{sec:matrices}) show that the
additional properties follow from just a \ensuremath{\Conid{SemiNearRing}} structure on
the elements.

The name of the carrier set (\ensuremath{\Varid{s}}) defined earlier was in reference to
\ensuremath{\Varid{s}}quare matrices.
However, Valiant's algorithm works only on (strictly) upper triangular
matrices--- that is, those whose elements at and below the diagonal
are zero.
We could have defined upper triangular matrices as the type of square
matrices satisfying the triangularity predicate. 
However, such a definition yields a large amount of tedious work to
manipulate the extra predicate.

Instead, we enforce the property of upper-triangularity axiomatically,
by defining a separate type \ensuremath{\Varid{u}} (for upper triangular matrix) and
require an embedding from \ensuremath{\Varid{u}} to \ensuremath{\Varid{s}} (square matrix).
We package the types \ensuremath{\Varid{u}} and \ensuremath{\Varid{s}}, together with all their properties,
in a single record \ensuremath{\Conid{SemiNearRing2}}.
This packaging helps, because we later build structures by induction,
and the inductive case for \ensuremath{\Varid{u}} depends on the induction hypothesis for
\ensuremath{\Varid{s}} and \textit{vice-versa}.
(Yet we keep \ensuremath{\Conid{SemiNearRing}} as a separate entity, as it is an adequate
specification of the elements of the matrices.)
%
\savecolumns[SemiNearRing2]
\savecolumns[SemiNearRing2A]
\begin{hscode}\SaveRestoreHook
\column{B}{@{}>{\hspre}l<{\hspost}@{}}%
\column{3}{@{}>{\hspre}l<{\hspost}@{}}%
\column{5}{@{}>{\hspre}l<{\hspost}@{}}%
\column{6}{@{}>{\hspre}l<{\hspost}@{}}%
\column{12}{@{}>{\hspre}l<{\hspost}@{}}%
\column{34}{@{}>{\hspre}l<{\hspost}@{}}%
\column{51}{@{}>{\hspre}l<{\hspost}@{}}%
\column{E}{@{}>{\hspre}l<{\hspost}@{}}%
\>[B]{}\kw{record}\;\Conid{SemiNearRing2}\;\!:\!\;\Conid{Set₁}\;\kw{where}\;{}\<[51]%
\>[51]{}\mbox{\onelinecomment  \structure{2}{\ensuremath{\Conid{SemiNearRing2}}}}{}\<[E]%
\\
\>[B]{}\hsindent{3}{}\<[3]%
\>[3]{}\kw{field}{}\<[E]%
\\
\>[3]{}\hsindent{2}{}\<[5]%
\>[5]{}\Varid{snr}\;\!:\!\;\Conid{SemiNearRing}{}\<[E]%
\\
\>[B]{}\hsindent{3}{}\<[3]%
\>[3]{}\kw{open}\;\Conid{SemiNearRing}\;\Varid{snr}\;\kw{public}\;{}\<[34]%
\>[34]{}\mbox{\onelinecomment  public = export the "local" names from \ensuremath{\Conid{SemiNearRing}}}{}\<[E]%
\\
\>[B]{}\hsindent{3}{}\<[3]%
\>[3]{}\kw{field}\;{}\<[51]%
\>[51]{}\mbox{\onelinecomment  \structure{2.1}{Plus and times for \ensuremath{\Varid{u}}, \ldots}}{}\<[E]%
\\
\>[3]{}\hsindent{3}{}\<[6]%
\>[6]{}\Varid{u}\;\!:\!\;\Conid{Set}{}\<[E]%
\\
\>[3]{}\hsindent{3}{}\<[6]%
\>[6]{}\unopun{\mathbin{+_u}}\;{}\<[12]%
\>[12]{}\!:\!\;\Varid{u}\;\Varid{→}\;\Varid{u}\;\Varid{→}\;\Varid{u}{}\<[E]%
\\
\>[3]{}\hsindent{3}{}\<[6]%
\>[6]{}\unopun{\mathbin{·_u}}\;{}\<[12]%
\>[12]{}\!:\!\;\Varid{u}\;\Varid{→}\;\Varid{u}\;\Varid{→}\;\Varid{u}{}\<[E]%
\\
\>[3]{}\hsindent{3}{}\<[6]%
\>[6]{}\Varid{u2s}\;{}\<[12]%
\>[12]{}\!:\!\;\Varid{u}\;\Varid{→}\;\Varid{s}{}\<[E]%
\ColumnHook
\end{hscode}\resethooks
Beside the type \ensuremath{\Varid{u}} and an embedding into \ensuremath{\Varid{s}}, we require addition
and multiplication over \ensuremath{\Varid{u}}: this is a simple specification of the
property that those operations preserve upper-triangularity.
Note that we require no relation between \ensuremath{\unopun{\mathbin{+_u}}} and \ensuremath{\unopun{\mathbin{+_s}}}, no
relation between \ensuremath{\unopun{\mathbin{·_u}}} and \ensuremath{\unopun{\mathbin{·_s}}} and no algebraic properties of the
\ensuremath{\Varid{u}}-operations.
Most of the link between \ensuremath{\Varid{u}} and \ensuremath{\Varid{s}} will become manifest when we see
the recursive structure tying them together.
The rest of the record contains helper definitions to lift the
structure of \ensuremath{\Varid{s}} onto \ensuremath{\Varid{u}} using \ensuremath{\Varid{u2s}}.
The lifting of binary relations is provided for us by the standard
library via the infix operator \ensuremath{\Conid{Function.on}}.
\restorecolumns[SemiNearRing2A]
\begin{hscode}\SaveRestoreHook
\column{B}{@{}>{\hspre}l<{\hspost}@{}}%
\column{3}{@{}>{\hspre}l<{\hspost}@{}}%
\column{9}{@{}>{\hspre}l<{\hspost}@{}}%
\column{17}{@{}>{\hspre}l<{\hspost}@{}}%
\column{20}{@{}>{\hspre}l<{\hspost}@{}}%
\column{24}{@{}>{\hspre}l<{\hspost}@{}}%
\column{28}{@{}>{\hspre}l<{\hspost}@{}}%
\column{E}{@{}>{\hspre}l<{\hspost}@{}}%
\>[3]{}\unopun{\mathrel{≃_u}}\;\!:\!\;\Varid{u}\;\Varid{→}\;\Varid{u}\;\Varid{→}\;\Conid{Set}{}\<[E]%
\\
\>[3]{}\unopun{\mathrel{≃_u}}\;{}\<[9]%
\>[9]{}\mathrel{=}\;\unopun{\mathrel{≃_s}}\;\Conid{Function.on}\;\Varid{u2s}{}\<[E]%
\\[\blanklineskip]%
\>[3]{}\unopun{\mathbin{{_u}\!·_s}}\;\!:\!\;\Varid{u}\;\Varid{→}\;\Varid{s}\;\Varid{→}\;\Varid{s}{}\<[E]%
\\
\>[3]{}\unopun{\mathbin{{_u}\!·_s}}\;\Varid{u}\;\Varid{s}\;\mathrel{=}\;{}\<[17]%
\>[17]{}\Varid{u2s}\;\Varid{u}\;{}\<[24]%
\>[24]{}\mathbin{·_s}\;{}\<[28]%
\>[28]{}\Varid{s}{}\<[E]%
\\[\blanklineskip]%
\>[3]{}\unopun{\mathbin{{_s}\!·_u}}\;\!:\!\;\Varid{s}\;\Varid{→}\;\Varid{u}\;\Varid{→}\;\Varid{s}{}\<[E]%
\\
\>[3]{}\unopun{\mathbin{{_s}\!·_u}}\;\Varid{s}\;\Varid{u}\;\mathrel{=}\;{}\<[17]%
\>[17]{}\Varid{s}\;{}\<[20]%
\>[20]{}\mathbin{·_s}\;{}\<[24]%
\>[24]{}\Varid{u2s}\;\Varid{u}{}\<[E]%
\ColumnHook
\end{hscode}\resethooks
The operator precedences and a \ensuremath{\Conid{Setoid}} instance follow, for completeness.
\pagebreak 
\restorecolumns[SemiNearRing2A]
\begin{hscode}\SaveRestoreHook
\column{B}{@{}>{\hspre}l<{\hspost}@{}}%
\column{3}{@{}>{\hspre}l<{\hspost}@{}}%
\column{11}{@{}>{\hspre}l<{\hspost}@{}}%
\column{14}{@{}>{\hspre}l<{\hspost}@{}}%
\column{21}{@{}>{\hspre}l<{\hspost}@{}}%
\column{29}{@{}>{\hspre}l<{\hspost}@{}}%
\column{32}{@{}>{\hspre}l<{\hspost}@{}}%
\column{39}{@{}>{\hspre}l<{\hspost}@{}}%
\column{47}{@{}>{\hspre}l<{\hspost}@{}}%
\column{50}{@{}>{\hspre}l<{\hspost}@{}}%
\column{E}{@{}>{\hspre}l<{\hspost}@{}}%
\>[3]{}\kw{infix}\;{}\<[11]%
\>[11]{}\Varid{4}\;{}\<[14]%
\>[14]{}\unopun{\mathrel{≃_u}};{}\<[21]%
\>[21]{}\kw{infixl}\;{}\<[29]%
\>[29]{}\Varid{6}\;{}\<[32]%
\>[32]{}\unopun{\mathbin{+_u}};{}\<[39]%
\>[39]{}\kw{infixl}\;{}\<[47]%
\>[47]{}\Varid{7}\;{}\<[50]%
\>[50]{}\unopun{\mathbin{·_u}}\;\unopun{\mathbin{{_u}\!·_s}}\;\unopun{\mathbin{{_s}\!·_u}}{}\<[E]%
\\[\blanklineskip]%
\>[3]{}\Varid{uSetoid}\;\!:\!\;\Conid{Setoid}\;0_L\;0_L{}\<[E]%
\\
\>[3]{}\Varid{uSetoid}\;\mathrel{=}\;\kw{record}\;\{\mskip1.5mu \Varid{isEquivalence}\;\mathrel{=}\;\Conid{Relation.Binary.On.isEquivalence}\;\Varid{u2s}\;\Varid{isEquiv}_s\mskip1.5mu\}{}\<[E]%
\ColumnHook
\end{hscode}\resethooks
%
%
To summarise, a record value \ensuremath{\Varid{snr2}\;\!:\!\;\Conid{SemiNearRing2}} contains the
\ensuremath{\Varid{s}}-operations from a subrecord \ensuremath{\Varid{snr}}, the corresponding operations
for \ensuremath{\Varid{u}} (but no laws) and the embedding \ensuremath{\Varid{u2s}\;\!:\!\;\Varid{u}\;\to \;\Varid{s}}.

\section{Specification}\label{sec:spec}

Recall that Valiant's algorithm aims at finding the smallest matrix
$C$ such that $ C = W + C · C $.
Our goal is to find a function computing $C$, and to prove it correct.
The first step is to formally transcribe the definition of transitive
closure.
Hence we revisit and extend the \ensuremath{\Conid{SemiNearRing}} structure with a few more
components needed to define the closure relation.
%
First we need an order to define ``smallest''.
Remember that, for parsing, a value of type \ensuremath{\Varid{s}} is a set of non-terminals, or a matrix thereof.
Hence, parsing uses set inclusion for the
preorder and set union for addition and in this context \ensuremath{\Varid{a}\;\Varid{⊆}\;\Varid{b}} iff \ensuremath{\Varid{a}\;\Varid{∪}\;\Varid{b}\;\mathrel{=}\;\Varid{b}}.
A natural generalisation is to define a partial order where \ensuremath{\Varid{x}\;\mathrel{≤_s}\;\Varid{y}}
iff \ensuremath{\Varid{x}\;\mathbin{+_s}\;\Varid{y}\;\mathrel{≃_s}\;\Varid{y}}.
For the relation \ensuremath{\unopun{\mathrel{≤_s}}} to be a preorder we need idempotence of
addition (which implies reflexivity and transitivity).
With this order addition is automatically monotonous, but for
multiplication to be monotonous we need distribution laws: that \ensuremath{\unopun{\mathbin{·_s}}}
distributes over \ensuremath{\unopun{\mathbin{+_s}}} (see \fref{fig:monoFromDist}).
\begin{figure}
  \centering
\begin{hscode}\SaveRestoreHook
\column{B}{@{}>{\hspre}l<{\hspost}@{}}%
\column{3}{@{}>{\hspre}l<{\hspost}@{}}%
\column{5}{@{}>{\hspre}l<{\hspost}@{}}%
\column{7}{@{}>{\hspre}l<{\hspost}@{}}%
\column{8}{@{}>{\hspre}l<{\hspost}@{}}%
\column{14}{@{}>{\hspre}l<{\hspost}@{}}%
\column{25}{@{}>{\hspre}l<{\hspost}@{}}%
\column{45}{@{}>{\hspre}l<{\hspost}@{}}%
\column{56}{@{}>{\hspre}l<{\hspost}@{}}%
\column{60}{@{}>{\hspre}l<{\hspost}@{}}%
\column{E}{@{}>{\hspre}l<{\hspost}@{}}%
\>[B]{}\Varid{monoTimesLeft}\;\!:\!\;\Varid{∀}\;\{\mskip1.5mu \Varid{a}\mskip1.5mu\}\;\{\mskip1.5mu \Varid{b}\mskip1.5mu\}\;\{\mskip1.5mu \Varid{c}\mskip1.5mu\}\;\Varid{→}\;(\Varid{a}\;\mathrel{≤_s}\;\Varid{b})\;\Varid{→}\;{}\<[45]%
\>[45]{}((\Varid{a}\;\mathbin{·_s}\;\Varid{c})\;{}\<[56]%
\>[56]{}\mathrel{≤_s}\;{}\<[60]%
\>[60]{}(\Varid{b}\;\mathbin{·_s}\;\Varid{c})){}\<[E]%
\\
\>[B]{}\Varid{monoTimesLeft}\;\{\mskip1.5mu \Varid{a}\mskip1.5mu\}\;\{\mskip1.5mu \Varid{b}\mskip1.5mu\}\;\{\mskip1.5mu \Varid{c}\mskip1.5mu\}\;\Varid{a}\!\!\mathrel{≤}\!\!\Varid{b}\;\mathrel{=}\;{}\<[E]%
\\
\>[B]{}\hsindent{3}{}\<[3]%
\>[3]{}\Varid{begin}\;{}\<[E]%
\\
\>[3]{}\hsindent{2}{}\<[5]%
\>[5]{}(\Varid{a}\;\mathbin{·_s}\;\Varid{c})\;\mathbin{+_s}\;(\Varid{b}\;\mathbin{·_s}\;\Varid{c})\;{}\<[E]%
\\
\>[B]{}\hsindent{3}{}\<[3]%
\>[3]{}≈\!\!⟨\;\Varid{sym}\;(\Varid{distr}\;\anonymous \;\anonymous \;\anonymous )\;\Varid{⟩}\;{}\<[E]%
\\
\>[3]{}\hsindent{2}{}\<[5]%
\>[5]{}(\Varid{a}\;\mathbin{+_s}\;\Varid{b})\;\mathbin{·_s}\;\Varid{c}\;{}\<[E]%
\\
\>[B]{}\hsindent{3}{}\<[3]%
\>[3]{}≈\!\!⟨\;{}\<[7]%
\>[7]{}\Varid{a}\!\!\mathrel{≤}\!\!\Varid{b}\;{}\<[14]%
\>[14]{}\mathbin{<\!\!\!·\!\!\!>}\;\Varid{refl}_s\;{}\<[25]%
\>[25]{}\Varid{⟩}\;\mbox{\onelinecomment  \ensuremath{\Varid{a}\;\mathrel{≤}\;\Varid{b}} means \ensuremath{\Varid{a}\;\Varid{+}\;\Varid{b}\;\mathrel{=}\;\Varid{b}}}{}\<[E]%
\\
\>[3]{}\hsindent{2}{}\<[5]%
\>[5]{}\Varid{b}\;{}\<[8]%
\>[8]{}\mathbin{·_s}\;\Varid{c}\;{}\<[E]%
\\
\>[B]{}\hsindent{3}{}\<[3]%
\>[3]{}\Varid{∎}{}\<[E]%
\ColumnHook
\end{hscode}\resethooks
  \caption{Monotonicity for multiplication follows from distributivity.}
  \label{fig:monoFromDist}
\end{figure}
\restorecolumns[SemiNearRing]
\begin{hscode}\SaveRestoreHook
\column{B}{@{}>{\hspre}l<{\hspost}@{}}%
\column{3}{@{}>{\hspre}l<{\hspost}@{}}%
\column{5}{@{}>{\hspre}l<{\hspost}@{}}%
\column{6}{@{}>{\hspre}l<{\hspost}@{}}%
\column{8}{@{}>{\hspre}l<{\hspost}@{}}%
\column{10}{@{}>{\hspre}l<{\hspost}@{}}%
\column{13}{@{}>{\hspre}l<{\hspost}@{}}%
\column{51}{@{}>{\hspre}l<{\hspost}@{}}%
\column{E}{@{}>{\hspre}l<{\hspost}@{}}%
\>[3]{}\kw{open}\;\Conid{Algebra.FunctionProperties}\;\unopun{\mathrel{≃_s}}{}\<[E]%
\\
\>[3]{}\hsindent{2}{}\<[5]%
\>[5]{}\kw{using}\;(\Conid{Idempotent};\Varid{\char95 DistributesOverˡ\char95 };\Varid{\char95 DistributesOverʳ\char95 }){}\<[E]%
\\[\blanklineskip]%
\>[3]{}\kw{field}\;{}\<[51]%
\>[51]{}\mbox{\onelinecomment  \structure{1.3}{Distributive, idempotent, \ldots}}{}\<[E]%
\\
\>[3]{}\hsindent{3}{}\<[6]%
\>[6]{}\Varid{idem}\;{}\<[13]%
\>[13]{}\!:\!\;\Conid{Idempotent}\;\unopun{\mathbin{+_s}}{}\<[E]%
\\[\blanklineskip]%
\>[3]{}\hsindent{3}{}\<[6]%
\>[6]{}\Varid{distl}\;{}\<[13]%
\>[13]{}\!:\!\;\unopun{\mathbin{·_s}}\;\Conid{DistributesOverˡ}\;\unopun{\mathbin{+_s}}{}\<[E]%
\\
\>[3]{}\hsindent{3}{}\<[6]%
\>[6]{}\Varid{distr}\;{}\<[13]%
\>[13]{}\!:\!\;\unopun{\mathbin{·_s}}\;\Conid{DistributesOverʳ}\;\unopun{\mathbin{+_s}}{}\<[E]%
\\
\>[6]{}\hsindent{2}{}\<[8]%
\>[8]{}\mbox{\onelinecomment  expands to \ensuremath{\Varid{∀}\;\Varid{a}\;\Varid{b}\;\Varid{c}\;\Varid{→}\;(\Varid{a}\;\mathbin{+_s}\;\Varid{b})\;\mathbin{·_s}\;\Varid{c}\;\mathrel{≃_s}\;(\Varid{a}\;\mathbin{·_s}\;\Varid{c})\;\mathbin{+_s}\;(\Varid{b}\;\mathbin{·_s}\;\Varid{c})}}{}\<[E]%
\\[\blanklineskip]%
\>[3]{}\kw{infix}\;{}\<[10]%
\>[10]{}\Varid{4}\;\unopun{\mathrel{≤_s}}{}\<[E]%
\\
\>[3]{}\unopun{\mathrel{≤_s}}\;\!:\!\;\Varid{s}\;\to \;\Varid{s}\;\to \;\Conid{Set}{}\<[E]%
\\
\>[3]{}\Varid{x}\;\mathrel{≤_s}\;\Varid{y}\;\mathrel{=}\;{}\<[13]%
\>[13]{}\Varid{x}\;\mathbin{+_s}\;\Varid{y}\;\mathrel{≃_s}\;\Varid{y}{}\<[E]%
\ColumnHook
\end{hscode}\resethooks
The algebraic structure we have specified so far in this record type
is nearly, but not quite, an idempotent semiring.
We just lack (and don't want) associativity (and unit) of our
multiplication.
(Remember that for our motivating example, parsing, the multiplication
operation is normally not associative.)

As the last addition to the \ensuremath{\Conid{SemiNearRing}} record type we define lower
bound with respect to this order:
\restorecolumns[SemiNearRing]
\begin{hscode}\SaveRestoreHook
\column{B}{@{}>{\hspre}l<{\hspost}@{}}%
\column{3}{@{}>{\hspre}l<{\hspost}@{}}%
\column{16}{@{}>{\hspre}l<{\hspost}@{}}%
\column{30}{@{}>{\hspre}l<{\hspost}@{}}%
\column{51}{@{}>{\hspre}l<{\hspost}@{}}%
\column{E}{@{}>{\hspre}l<{\hspost}@{}}%
\>[3]{}\Conid{LowerBound}_s\;{}\<[16]%
\>[16]{}\mathrel{=}\;\Conid{LowerBound}\;{}\<[30]%
\>[30]{}\unopun{\mathrel{≤_s}}{}\<[51]%
\>[51]{}\mbox{\onelinecomment  \structure{1.6}{Lower bounds}}{}\<[E]%
\ColumnHook
\end{hscode}\resethooks
In \ensuremath{\Conid{SemiNearRing2}} we obtain a suitable ordering on \ensuremath{\Varid{u}} by lifting the ordering on~\ensuremath{\Varid{s}}.
\restorecolumns[SemiNearRing2A]
\begin{hscode}\SaveRestoreHook
\column{B}{@{}>{\hspre}l<{\hspost}@{}}%
\column{3}{@{}>{\hspre}l<{\hspost}@{}}%
\column{9}{@{}>{\hspre}l<{\hspost}@{}}%
\column{E}{@{}>{\hspre}l<{\hspost}@{}}%
\>[3]{}\unopun{\mathrel{≤_u}}\;{}\<[9]%
\>[9]{}\!:\!\;\Varid{u}\;\Varid{→}\;\Varid{u}\;\Varid{→}\;\Conid{Set}{}\<[E]%
\\
\>[3]{}\unopun{\mathrel{≤_u}}\;{}\<[9]%
\>[9]{}\mathrel{=}\;\unopun{\mathrel{≤_s}}\;\Conid{Function.on}\;\Varid{u2s}{}\<[E]%
\ColumnHook
\end{hscode}\resethooks
%

\subsubsection*{Closure}
We have now seen the first layer \ensuremath{\Conid{SemiNearRing}} specifying the
underlying carrier set \ensuremath{\Varid{s}} for square matrices and the second layer
\ensuremath{\Conid{SemiNearRing2}} specifying the set \ensuremath{\Varid{u}} for upper triangular matrices.
The third layer \ensuremath{\Conid{ClosedSemiNearRing}} specifies the transitive closure
and Valiant's algorithm for computing it.
\savecolumns[ClosedSemiNearRing]
\begin{hscode}\SaveRestoreHook
\column{B}{@{}>{\hspre}l<{\hspost}@{}}%
\column{3}{@{}>{\hspre}l<{\hspost}@{}}%
\column{5}{@{}>{\hspre}l<{\hspost}@{}}%
\column{27}{@{}>{\hspre}l<{\hspost}@{}}%
\column{51}{@{}>{\hspre}l<{\hspost}@{}}%
\column{E}{@{}>{\hspre}l<{\hspost}@{}}%
\>[B]{}\kw{record}\;\Conid{ClosedSemiNearRing}\;\!:\!\;\Conid{Set₁}\;\kw{where}\;{}\<[51]%
\>[51]{}\mbox{\onelinecomment  \structure{3}{\ensuremath{\Conid{ClosedSemiNearRing}}}}{}\<[E]%
\\[\blanklineskip]%
\>[B]{}\hsindent{3}{}\<[3]%
\>[3]{}\kw{field}{}\<[E]%
\\
\>[3]{}\hsindent{2}{}\<[5]%
\>[5]{}\Varid{snr2}\;\!:\!\;\Conid{SemiNearRing2}{}\<[27]%
\>[27]{}\mbox{\onelinecomment  includes \ensuremath{\Varid{s}}, \ensuremath{\Varid{u}} and corresponding operations}{}\<[E]%
\\[\blanklineskip]%
\>[B]{}\hsindent{3}{}\<[3]%
\>[3]{}\kw{open}\;\Conid{SemiNearRing2}\;\Varid{snr2}{}\<[E]%
\ColumnHook
\end{hscode}\resethooks
We can now finally give the specification and we do it in three steps:
the relation \ensuremath{\Conid{Q}} capturing the ``quadratic'' equation, the relation
\ensuremath{\Conid{Closure}} capturing only the ``least'' solutions of \ensuremath{\Conid{Q}} and finally
\ensuremath{\Conid{Entire}\;\Conid{Q}} which says that there is a total function inside the
relation.
\restorecolumns[ClosedSemiNearRing]
\begin{hscode}\SaveRestoreHook
\column{B}{@{}>{\hspre}l<{\hspost}@{}}%
\column{3}{@{}>{\hspre}l<{\hspost}@{}}%
\column{5}{@{}>{\hspre}l<{\hspost}@{}}%
\column{14}{@{}>{\hspre}l<{\hspost}@{}}%
\column{18}{@{}>{\hspre}l<{\hspost}@{}}%
\column{26}{@{}>{\hspre}l<{\hspost}@{}}%
\column{30}{@{}>{\hspre}l<{\hspost}@{}}%
\column{51}{@{}>{\hspre}l<{\hspost}@{}}%
\column{E}{@{}>{\hspre}l<{\hspost}@{}}%
\>[3]{}\Conid{Q}\;\!:\!\;\Varid{u}\;\Varid{→}\;\Varid{u}\;\Varid{→}\;\Conid{Set}{}\<[51]%
\>[51]{}\mbox{\onelinecomment  \structure{3.1}{Quadratic equation \ensuremath{\Conid{Q}} + properties}}{}\<[E]%
\\
\>[3]{}\Conid{Q}\;\Varid{w}\;\Varid{c}\;\mathrel{=}\;\Varid{w}\;{}\<[14]%
\>[14]{}\mathbin{+_u}\;{}\<[18]%
\>[18]{}\Varid{c}\;\mathbin{·_u}\;\Varid{c}\;{}\<[26]%
\>[26]{}\mathrel{≃_u}\;{}\<[30]%
\>[30]{}\Varid{c}{}\<[E]%
\\[\blanklineskip]%
\>[3]{}\Conid{Closure}\;\!:\!\;\Varid{u}\;\to \;\Varid{u}\;\to \;\Conid{Set}{}\<[E]%
\\
\>[3]{}\Conid{Closure}\;\Varid{w}\;\Varid{c}\;\mathrel{=}\;\Conid{Least}\;\unopun{\mathrel{≤_u}}\;(\Conid{Q}\;\Varid{w})\;\Varid{c}{}\<[E]%
\\[\blanklineskip]%
\>[3]{}\kw{field}{}\<[E]%
\\
\>[3]{}\hsindent{2}{}\<[5]%
\>[5]{}\Varid{entireQ}\;\!:\!\;\Conid{Entire}\;\Conid{Q}{}\<[E]%
\ColumnHook
\end{hscode}\resethooks
From the \ensuremath{\Varid{entireQ}} field (which we will populate later) we can extract
the closure algorithm (\ensuremath{\Varid{closure}}) and part of its correctness proof
(\ensuremath{\Varid{closureHasAll}}).
\label{def:closure}
\restorecolumns[ClosedSemiNearRing]
\begin{hscode}\SaveRestoreHook
\column{B}{@{}>{\hspre}l<{\hspost}@{}}%
\column{3}{@{}>{\hspre}l<{\hspost}@{}}%
\column{18}{@{}>{\hspre}l<{\hspost}@{}}%
\column{51}{@{}>{\hspre}l<{\hspost}@{}}%
\column{E}{@{}>{\hspre}l<{\hspost}@{}}%
\>[3]{}\Varid{closure}\;\!:\!\;\Varid{u}\;\Varid{→}\;\Varid{u}{}\<[51]%
\>[51]{}\mbox{\onelinecomment  \structure{3.2}{Closure function and correctness}}{}\<[E]%
\\
\>[3]{}\Varid{closure}\;\mathrel{=}\;\Varid{fun}\;\Varid{entireQ}{}\<[E]%
\\[\blanklineskip]%
\>[3]{}\Varid{closureHasAll}\;{}\<[18]%
\>[18]{}\!:\!\;\Varid{∀}\;\{\mskip1.5mu \Varid{w}\;\!:\!\;\Varid{u}\mskip1.5mu\}\;\Varid{→}\;\Conid{Q}\;\Varid{w}\;(\Varid{closure}\;\Varid{w}){}\<[E]%
\\
\>[3]{}\Varid{closureHasAll}\;{}\<[18]%
\>[18]{}\mathrel{=}\;\Varid{correct}\;\Varid{entireQ}{}\<[E]%
\ColumnHook
\end{hscode}\resethooks
Proving \ensuremath{\Conid{Entire}\;\Conid{Closure}} means to show that \ensuremath{\Varid{closure}} yields the
smallest possible solution, and is deferred until \fref{sec:smallest}.
\restorecolumns[ClosedSemiNearRing]
\begin{hscode}\SaveRestoreHook
\column{B}{@{}>{\hspre}l<{\hspost}@{}}%
\column{3}{@{}>{\hspre}l<{\hspost}@{}}%
\column{E}{@{}>{\hspre}l<{\hspost}@{}}%
\>[3]{}\kw{open}\;\Conid{SemiNearRing2}\;\Varid{snr2}\;\kw{public}{}\<[E]%
\ColumnHook
\end{hscode}\resethooks

\section{Matrices}\label{sec:matrices}

In the specification and implementation we  make heavy use of
square matrices of size \ensuremath{\Varid{m}\;\mathbin{\!\times\!}\;\Varid{m}} for different values of \ensuremath{\Varid{m}}.
For concision we make the simplifying assumption that \ensuremath{\Varid{m}} is a power
of two (\ensuremath{\Varid{m}\;\mathrel{=}\;2^\Varid{n}}) --- handling the general case involves indexing
matrices with their sizes, which is straightforward but clutters the
development (\fref{sec:general-size}).

There are many different ways to represent matrices.
As usual when working with dependent types, some definitions yield a
concise presentation, while others require large amounts of
boilerplate, obscuring the intent.
A judicious choice is thus in order. Possibilities include:
\begin{itemize}
\item A vector of vectors of elements, where a vector is
  \begin{hscode}\SaveRestoreHook
\column{B}{@{}>{\hspre}l<{\hspost}@{}}%
\column{5}{@{}>{\hspre}l<{\hspost}@{}}%
\column{7}{@{}>{\hspre}l<{\hspost}@{}}%
\column{13}{@{}>{\hspre}l<{\hspost}@{}}%
\column{E}{@{}>{\hspre}l<{\hspost}@{}}%
\>[5]{}\kw{data}\;\Conid{Vec}\;\Varid{a}\;\Varid{n}\;\kw{where}{}\<[E]%
\\
\>[5]{}\hsindent{2}{}\<[7]%
\>[7]{}\Conid{Nil}\;{}\<[13]%
\>[13]{}\!:\!\;\Conid{Vec}\;\Varid{a}\;\Varid{0}{}\<[E]%
\\
\>[5]{}\hsindent{2}{}\<[7]%
\>[7]{}\Conid{Cons}\;{}\<[13]%
\>[13]{}\!:\!\;\Varid{a}\;\to \;\Conid{Vec}\;\Varid{a}\;\Varid{n}\;\to \;\Conid{Vec}\;\Varid{a}\;\Varid{n}\;\to \;\Conid{Vec}\;\Varid{a}\;(\Varid{1}\;\Varid{+}\;\Varid{n}){}\<[E]%
\ColumnHook
\end{hscode}\resethooks
\item A function from (two) indices to elements \ensuremath{\Conid{Bin}\;\Varid{n}\;\to \;\Conid{Bin}\;\Varid{n}\;\to \;\Varid{a}} (where \ensuremath{\Conid{Bin}} stands for a binary number of \ensuremath{\Varid{n}} bits)
\item A recursive data-type (a quad tree):
  \begin{hscode}\SaveRestoreHook
\column{B}{@{}>{\hspre}l<{\hspost}@{}}%
\column{5}{@{}>{\hspre}l<{\hspost}@{}}%
\column{7}{@{}>{\hspre}l<{\hspost}@{}}%
\column{13}{@{}>{\hspre}l<{\hspost}@{}}%
\column{E}{@{}>{\hspre}l<{\hspost}@{}}%
\>[5]{}\kw{data}\;\Conid{Mat}\;\Varid{a}\;\Varid{n}\;\kw{where}{}\<[E]%
\\
\>[5]{}\hsindent{2}{}\<[7]%
\>[7]{}\Conid{Unit}\;{}\<[13]%
\>[13]{}\!:\!\;\Varid{a}\;\to \;\Conid{Mat}\;\Varid{a}\;\Varid{0}{}\<[E]%
\\
\>[5]{}\hsindent{2}{}\<[7]%
\>[7]{}\Conid{Quad}\;{}\<[13]%
\>[13]{}\!:\!\;\Conid{Mat}\;\Varid{a}\;\Varid{n}\;\to \;\Conid{Mat}\;\Varid{a}\;\Varid{n}\;\to \;\Conid{Mat}\;\Varid{a}\;\Varid{n}\;\to \;\Conid{Mat}\;\Varid{a}\;\Varid{n}\;\to \;\Conid{Mat}\;\Varid{a}\;(\Varid{1}\;\Varid{+}\;\Varid{n}){}\<[E]%
\ColumnHook
\end{hscode}\resethooks
\item A function from \ensuremath{\Varid{n}} to \ensuremath{\Conid{Set}}:
  \begin{hscode}\SaveRestoreHook
\column{B}{@{}>{\hspre}l<{\hspost}@{}}%
\column{5}{@{}>{\hspre}l<{\hspost}@{}}%
\column{7}{@{}>{\hspre}l<{\hspost}@{}}%
\column{20}{@{}>{\hspre}l<{\hspost}@{}}%
\column{E}{@{}>{\hspre}l<{\hspost}@{}}%
\>[5]{}\Conid{Mat}\;\Varid{a}\;\Varid{0}\;{}\<[20]%
\>[20]{}\mathrel{=}\;\Conid{OneByOne}\;\Varid{a}{}\<[E]%
\\
\>[5]{}\hsindent{2}{}\<[7]%
\>[7]{}\kw{where}\;\Conid{OneByOne}\;\Varid{a}\;\mathrel{=}\;\Varid{a}{}\<[E]%
\\
\>[5]{}\Conid{Mat}\;\Varid{a}\;(\Varid{suc}\;\Varid{n})\;{}\<[20]%
\>[20]{}\mathrel{=}\;\Conid{Square}\;(\Conid{Mat}\;\Varid{a}\;\Varid{n}){}\<[E]%
\\
\>[5]{}\hsindent{2}{}\<[7]%
\>[7]{}\kw{where}\;\Conid{Square}\;\Varid{t}\;\mathrel{=}\;\Varid{t}\;\mathbin{\!\times\!}\;\Varid{t}\;\mathbin{\!\times\!}\;\Varid{t}\;\mathbin{\!\times\!}\;\Varid{t}{}\<[E]%
\ColumnHook
\end{hscode}\resethooks
\end{itemize}
Here we choose the last approach, which is specifically tailored to
the problem at hand. Indeed, the advantage of doing so is that
\ensuremath{\Conid{Square}} is a functor.  This approach allows to extend the \ensuremath{\Conid{Square}}
functor with all the axiomatisation we need, including ring-like
structures and up to the full closure specification.
In fact, most of the rest of the paper is devoted to the
definition of that functor: when it is fully defined we have a
proven-correct implementation of Valiant's algorithm.
In this section we show the abstraction over \ensuremath{\Conid{SemiNearRing}} and the
construction of matrices, up to the full definition of \ensuremath{\Conid{Mat}}.
In the base case \ensuremath{\Conid{OneByOne}} (shown later in \fref{sec:OneByOne} on page \pageref{sec:OneByOne}) we
lift a \ensuremath{\Conid{SemiNearRing}} to a \ensuremath{\Conid{ClosedSemiNearRing}} and in the inductive
case and we apply \ensuremath{\Conid{Square}} which preserves the \ensuremath{\Conid{ClosedSemiNearRing}}
structure.

\subsubsection*{The main functor: \ensuremath{\Conid{Square}}}
We attack the recursive case right away by defining the function
\ensuremath{\Conid{Square}} which lifts our algebraic structure (the types \ensuremath{\Varid{s}} and \ensuremath{\Varid{u}},
the operations on them and all the laws) from elements to 2-by-2
(block) matrices.
It can be seen as an implementation of a (first-class, but otherwise)
\ensuremath{\Conid{ML}}-style functor.
\savecolumns[Square]
\savecolumns[SquareAgain]
\savecolumns[SquareThird]
\begin{hscode}\SaveRestoreHook
\column{B}{@{}>{\hspre}l<{\hspost}@{}}%
\column{3}{@{}>{\hspre}l<{\hspost}@{}}%
\column{51}{@{}>{\hspre}l<{\hspost}@{}}%
\column{E}{@{}>{\hspre}l<{\hspost}@{}}%
\>[51]{}\mbox{\onelinecomment  \structure{4}{2-by-2 block matrix, preserving \ensuremath{\Conid{ClosedSemiNearRing}}}}{}\<[E]%
\\
\>[B]{}\Conid{Square}\;\!:\!\;\Conid{ClosedSemiNearRing}\;\to \;\Conid{ClosedSemiNearRing}{}\<[E]%
\\
\>[B]{}\Conid{Square}\;\Varid{csnr}\;\mathrel{=}\;\Conid{CSNR}\;\kw{where}{}\<[E]%
\\
\>[B]{}\hsindent{3}{}\<[3]%
\>[3]{}\kw{open}\;\Conid{ClosedSemiNearRing}\;\Varid{csnr}{}\<[E]%
\ColumnHook
\end{hscode}\resethooks
The rest of this section is the body of this \ensuremath{\kw{where}} clause where all
operations and laws from \ensuremath{\Varid{csnr}} are in scope.

\subsubsection*{Lifting types}
We start by defining two types \ensuremath{\Conid{U}} and \ensuremath{\Conid{S}}, instantiating the \ensuremath{\Varid{u}} and
\ensuremath{\Varid{s}} types for two-by-two (block) matrices.
Throughout the rest of the paper, we use the convention of using
capitals for the values of the fields of the record created (the
target of the functor).

For type \ensuremath{\Conid{S}} we could use \ensuremath{\Conid{S}\;\mathrel{=}\;\Varid{s}\;\mathbin{\!\times\!}\;\Varid{s}\;\mathbin{\!\times\!}\;\Varid{s}\;\mathbin{\!\times\!}\;\Varid{s}} but we prefer to use a
record type for clarity. The constructor name uses angle brackets to lift any ambiguity.
\restorecolumns[Square]
\aligncolumn{18}{@{\quad}>{\hspre}c<{\hspost}@{}}
\begin{hscode}\SaveRestoreHook
\column{B}{@{}>{\hspre}l<{\hspost}@{}}%
\column{3}{@{}>{\hspre}l<{\hspost}@{}}%
\column{5}{@{}>{\hspre}l<{\hspost}@{}}%
\column{7}{@{}>{\hspre}l<{\hspost}@{}}%
\column{12}{@{}>{\hspre}l<{\hspost}@{}}%
\column{18}{@{}>{\hspre}l<{\hspost}@{}}%
\column{23}{@{}>{\hspre}l<{\hspost}@{}}%
\column{51}{@{}>{\hspre}l<{\hspost}@{}}%
\column{E}{@{}>{\hspre}l<{\hspost}@{}}%
\>[3]{}\kw{record}\;\Conid{S}\;\!:\!\;\Conid{Set}\;\kw{where}\;{}\<[51]%
\>[51]{}\mbox{\onelinecomment  \structure{4.1}{Square matrix}}{}\<[E]%
\\
\>[3]{}\hsindent{2}{}\<[5]%
\>[5]{}\Varid{constructor}\;\Varid{⟨\char95 ,\char95 ,\char95 ,\char95 ⟩}{}\<[E]%
\\
\>[3]{}\hsindent{2}{}\<[5]%
\>[5]{}\kw{field}{}\<[E]%
\\
\>[5]{}\hsindent{2}{}\<[7]%
\>[7]{}\Varid{s}_{00}\;{}\<[12]%
\>[12]{}\!:\!\;\Varid{s};{}\<[18]%
\>[18]{}\Varid{s}_{01}\;{}\<[23]%
\>[23]{}\!:\!\;\Varid{s}{}\<[E]%
\\
\>[5]{}\hsindent{2}{}\<[7]%
\>[7]{}\Varid{s}_{10}\;{}\<[12]%
\>[12]{}\!:\!\;\Varid{s};{}\<[18]%
\>[18]{}\Varid{s}_{11}\;{}\<[23]%
\>[23]{}\!:\!\;\Varid{s}{}\<[E]%
\\
\>[3]{}\kw{infix}\;\Varid{4}\;\Varid{⟨\char95 ,\char95 ,\char95 ,\char95 ⟩}{}\<[E]%
\ColumnHook
\end{hscode}\resethooks

The two-by-two upper triangular matrix is composed of two
(smaller) upper-triangular matrices, and a square matrix for the
top-right corner. (The bottom-left corner is zero.)  Basically, \ensuremath{\Conid{U}\;\mathrel{=}\;\Varid{u}\;\mathbin{\!\times\!}\;\Varid{s}\;\mathbin{\!\times\!}\;\Varid{u}} but again, regular nested pairs obscure the intent. Here
we add a box inside the name of the record constructor, as a reminder that the bottom left corner is empty.
\restorecolumns[Square]
\aligncolumn{18}{@{\quad}>{\hspre}c<{\hspost}@{}}
\begin{hscode}\SaveRestoreHook
\column{B}{@{}>{\hspre}l<{\hspost}@{}}%
\column{3}{@{}>{\hspre}l<{\hspost}@{}}%
\column{5}{@{}>{\hspre}l<{\hspost}@{}}%
\column{7}{@{}>{\hspre}l<{\hspost}@{}}%
\column{18}{@{}>{\hspre}l<{\hspost}@{}}%
\column{24}{@{}>{\hspre}l<{\hspost}@{}}%
\column{51}{@{}>{\hspre}l<{\hspost}@{}}%
\column{E}{@{}>{\hspre}l<{\hspost}@{}}%
\>[3]{}\kw{record}\;\Conid{U}\;\!:\!\;\Conid{Set}\;\kw{where}\;{}\<[51]%
\>[51]{}\mbox{\onelinecomment  \structure{4.2}{Upper triangular matrix}}{}\<[E]%
\\
\>[3]{}\hsindent{2}{}\<[5]%
\>[5]{}\Varid{constructor}\;\Varid{⟨\char95 ,\char95 ,•,\char95 ⟩}{}\<[E]%
\\
\>[3]{}\hsindent{2}{}\<[5]%
\>[5]{}\kw{field}{}\<[E]%
\\
\>[5]{}\hsindent{2}{}\<[7]%
\>[7]{}\Varid{uu}_{00}\;\!:\!\;\Varid{u};{}\<[18]%
\>[18]{}\Varid{us}_{01}\;{}\<[24]%
\>[24]{}\!:\!\;\Varid{s};{}\<[E]%
\\
\>[18]{}\Varid{uu}_{11}\;{}\<[24]%
\>[24]{}\!:\!\;\Varid{u}{}\<[E]%
\\
\>[3]{}\kw{infix}\;\Varid{4}\;\Varid{⟨\char95 ,\char95 ,•,\char95 ⟩}{}\<[E]%
\ColumnHook
\end{hscode}\resethooks

\subsubsection*{Lifting operations}
We lift the operations $0,+,·$ from the underlying semi-near-ring to
matrices over that semi-near-ring, in the usual manner.
\restorecolumns[SquareAgain]
\aligncolumn{21}{@{}>{\hspre}c<{\hspost}@{\quad}}
\aligncolumn{34}{@{}>{\hspre}c<{\hspost}@{\quad}}
\aligncolumn{47}{@{}>{\hspre}c<{\hspost}@{\quad}}
\begin{hscode}\SaveRestoreHook
\column{B}{@{}>{\hspre}l<{\hspost}@{}}%
\column{3}{@{}>{\hspre}l<{\hspost}@{}}%
\column{9}{@{}>{\hspre}l<{\hspost}@{}}%
\column{16}{@{}>{\hspre}l<{\hspost}@{}}%
\column{17}{@{}>{\hspre}l<{\hspost}@{}}%
\column{21}{@{}>{\hspre}l<{\hspost}@{}}%
\column{24}{@{}>{\hspre}l<{\hspost}@{}}%
\column{30}{@{}>{\hspre}l<{\hspost}@{}}%
\column{34}{@{}>{\hspre}l<{\hspost}@{}}%
\column{37}{@{}>{\hspre}l<{\hspost}@{}}%
\column{43}{@{}>{\hspre}l<{\hspost}@{}}%
\column{47}{@{}>{\hspre}l<{\hspost}@{}}%
\column{50}{@{}>{\hspre}l<{\hspost}@{}}%
\column{56}{@{}>{\hspre}l<{\hspost}@{}}%
\column{60}{@{}>{\hspre}l<{\hspost}@{}}%
\column{E}{@{}>{\hspre}l<{\hspost}@{}}%
\>[3]{}\unopun{\mathbin{+\!_S}}\;{}\<[9]%
\>[9]{}\!:\!\;\Conid{S}\;\Varid{→}\;{}\<[16]%
\>[16]{}\Conid{S}\;\Varid{→}\;\Conid{S}{}\<[E]%
\\
\>[3]{}\unopun{\mathbin{+\!_S}}\;{}\<[9]%
\>[9]{}\Varid{⟨}\;\Varid{a}\;{}\<[21]%
\>[21]{}\Varid{,}\;{}\<[24]%
\>[24]{}\Varid{b}\;{}\<[34]%
\>[34]{}\Varid{,}\;{}\<[37]%
\>[37]{}\Varid{c}\;{}\<[47]%
\>[47]{}\Varid{,}\;{}\<[50]%
\>[50]{}\Varid{d}\;{}\<[60]%
\>[60]{}\Varid{⟩}\;{}\<[E]%
\\
\>[9]{}\Varid{⟨}\;{}\<[17]%
\>[17]{}\Varid{a'}\;{}\<[21]%
\>[21]{}\Varid{,}\;{}\<[30]%
\>[30]{}\Varid{b'}\;{}\<[34]%
\>[34]{}\Varid{,}\;{}\<[43]%
\>[43]{}\Varid{c'}\;{}\<[47]%
\>[47]{}\Varid{,}\;{}\<[56]%
\>[56]{}\Varid{d'}\;{}\<[60]%
\>[60]{}\Varid{⟩}\;\mathrel{=}\;{}\<[E]%
\\
\>[9]{}\Varid{⟨}\;\Varid{a}\;\mathbin{+_s}\;{}\<[17]%
\>[17]{}\Varid{a'}\;{}\<[21]%
\>[21]{}\Varid{,}\;{}\<[24]%
\>[24]{}\Varid{b}\;\mathbin{+_s}\;{}\<[30]%
\>[30]{}\Varid{b'}\;{}\<[34]%
\>[34]{}\Varid{,}\;{}\<[37]%
\>[37]{}\Varid{c}\;\mathbin{+_s}\;{}\<[43]%
\>[43]{}\Varid{c'}\;{}\<[47]%
\>[47]{}\Varid{,}\;{}\<[50]%
\>[50]{}\Varid{d}\;\mathbin{+_s}\;{}\<[56]%
\>[56]{}\Varid{d'}\;{}\<[60]%
\>[60]{}\Varid{⟩}{}\<[E]%
\ColumnHook
\end{hscode}\resethooks
Matrix multiplication is defined as expected.
\restorecolumns[SquareThird]
\aligncolumn{70}{@{}>{\hspre}c<{\hspost}@{\quad}}
\begin{hscode}\SaveRestoreHook
\column{B}{@{}>{\hspre}l<{\hspost}@{}}%
\column{3}{@{}>{\hspre}l<{\hspost}@{}}%
\column{9}{@{}>{\hspre}l<{\hspost}@{}}%
\column{11}{@{}>{\hspre}l<{\hspost}@{}}%
\column{12}{@{}>{\hspre}l<{\hspost}@{}}%
\column{14}{@{}>{\hspre}l<{\hspost}@{}}%
\column{15}{@{}>{\hspre}l<{\hspost}@{}}%
\column{22}{@{}>{\hspre}l<{\hspost}@{}}%
\column{25}{@{}>{\hspre}l<{\hspost}@{}}%
\column{29}{@{}>{\hspre}l<{\hspost}@{}}%
\column{35}{@{}>{\hspre}l<{\hspost}@{}}%
\column{38}{@{}>{\hspre}l<{\hspost}@{}}%
\column{41}{@{}>{\hspre}l<{\hspost}@{}}%
\column{44}{@{}>{\hspre}l<{\hspost}@{}}%
\column{55}{@{}>{\hspre}l<{\hspost}@{}}%
\column{59}{@{}>{\hspre}l<{\hspost}@{}}%
\column{70}{@{}>{\hspre}l<{\hspost}@{}}%
\column{73}{@{}>{\hspre}l<{\hspost}@{}}%
\column{84}{@{}>{\hspre}l<{\hspost}@{}}%
\column{88}{@{}>{\hspre}l<{\hspost}@{}}%
\column{E}{@{}>{\hspre}l<{\hspost}@{}}%
\>[3]{}\unopun{\mathbin{·\!_S}}\;{}\<[9]%
\>[9]{}\!:\!\;\Conid{S}\;\Varid{→}\;\Conid{S}\;\Varid{→}\;\Conid{S}{}\<[E]%
\\
\>[3]{}\unopun{\mathbin{·\!_S}}\;{}\<[9]%
\>[9]{}\Varid{⟨}\;{}\<[12]%
\>[12]{}\Varid{a}\;{}\<[15]%
\>[15]{}\Varid{,}\;\Varid{b}\;\Varid{,}\;{}\<[E]%
\\
\>[12]{}\Varid{c}\;{}\<[15]%
\>[15]{}\Varid{,}\;\Varid{d}\;\Varid{⟩}\;{}\<[22]%
\>[22]{}\Varid{⟨}\;{}\<[25]%
\>[25]{}\Varid{a'}\;{}\<[29]%
\>[29]{}\Varid{,}\;\Varid{b'}\;{}\<[35]%
\>[35]{}\Varid{,}\;{}\<[E]%
\\
\>[25]{}\Varid{c'}\;{}\<[29]%
\>[29]{}\Varid{,}\;\Varid{d'}\;{}\<[35]%
\>[35]{}\Varid{⟩}\;{}\<[38]%
\>[38]{}\mathrel{=}\;{}\<[41]%
\>[41]{}\Varid{⟨}\;{}\<[44]%
\>[44]{}(\Varid{a}\;\mathbin{·_s}\;\Varid{a'})\;{}\<[55]%
\>[55]{}\mathbin{+_s}\;{}\<[59]%
\>[59]{}(\Varid{b}\;\mathbin{·_s}\;\Varid{c'})\;{}\<[70]%
\>[70]{}\Varid{,}\;{}\<[73]%
\>[73]{}(\Varid{a}\;\mathbin{·_s}\;\Varid{b'})\;{}\<[84]%
\>[84]{}\mathbin{+_s}\;{}\<[88]%
\>[88]{}(\Varid{b}\;\mathbin{·_s}\;\Varid{d'})\;{}\<[E]%
\\
\>[41]{}\Varid{,}\;{}\<[44]%
\>[44]{}(\Varid{c}\;\mathbin{·_s}\;\Varid{a'})\;{}\<[55]%
\>[55]{}\mathbin{+_s}\;{}\<[59]%
\>[59]{}(\Varid{d}\;\mathbin{·_s}\;\Varid{c'})\;{}\<[70]%
\>[70]{}\Varid{,}\;{}\<[73]%
\>[73]{}(\Varid{c}\;\mathbin{·_s}\;\Varid{b'})\;{}\<[84]%
\>[84]{}\mathbin{+_s}\;{}\<[88]%
\>[88]{}(\Varid{d}\;\mathbin{·_s}\;\Varid{d'})\;\Varid{⟩}{}\<[E]%
\\[\blanklineskip]%
\>[3]{}\kw{infixl}\;{}\<[11]%
\>[11]{}\Varid{6}\;{}\<[14]%
\>[14]{}\unopun{\mathbin{+\!_S}}{}\<[E]%
\\
\>[3]{}\kw{infixl}\;{}\<[11]%
\>[11]{}\Varid{7}\;{}\<[14]%
\>[14]{}\unopun{\mathbin{·\!_S}}{}\<[E]%
\ColumnHook
\end{hscode}\resethooks
Zero is defined similarly.
\restorecolumns[SquareThird]
\begin{hscode}\SaveRestoreHook
\column{B}{@{}>{\hspre}l<{\hspost}@{}}%
\column{3}{@{}>{\hspre}l<{\hspost}@{}}%
\column{13}{@{}>{\hspre}l<{\hspost}@{}}%
\column{E}{@{}>{\hspre}l<{\hspost}@{}}%
\>[3]{}\Varid{zerS}\;\!:\!\;\Conid{S}{}\<[E]%
\\
\>[3]{}\Varid{zerS}\;\mathrel{=}\;\Varid{⟨}\;{}\<[13]%
\>[13]{}0_s\;\Varid{,}\;0_s\;\Varid{,}\;{}\<[E]%
\\
\>[13]{}0_s\;\Varid{,}\;0_s\;\Varid{⟩}{}\<[E]%
\ColumnHook
\end{hscode}\resethooks

We can then define the operations on upper-triangular matrices.
In particular, we give full definitions of point-wise upper triangular
matrix addition and multiplication.
Having to define operations both for \ensuremath{\Conid{U}} and \ensuremath{\Conid{S}} is an consequence of
our choice of separating \ensuremath{\Varid{u}} and \ensuremath{\Varid{s}} in the algebraic structure.
This may look redundant, but in fact the definitions for \ensuremath{\Conid{U}} serve the
purpose of proving that upper-triangularity is preserved by those
operations.
In particular, not that \ensuremath{\Varid{u}}-operations are used on the (block-)
diagonal and \ensuremath{\Varid{s}}-operations are used in the the ``square corner''.
\restorecolumns[SquareThird]
\aligncolumn{57}{@{}>{\hspre}c<{\hspost}@{\quad}}
\begin{hscode}\SaveRestoreHook
\column{B}{@{}>{\hspre}l<{\hspost}@{}}%
\column{3}{@{}>{\hspre}l<{\hspost}@{}}%
\column{9}{@{}>{\hspre}l<{\hspost}@{}}%
\column{12}{@{}>{\hspre}l<{\hspost}@{}}%
\column{16}{@{}>{\hspre}l<{\hspost}@{}}%
\column{19}{@{}>{\hspre}l<{\hspost}@{}}%
\column{25}{@{}>{\hspre}l<{\hspost}@{}}%
\column{28}{@{}>{\hspre}l<{\hspost}@{}}%
\column{35}{@{}>{\hspre}l<{\hspost}@{}}%
\column{41}{@{}>{\hspre}l<{\hspost}@{}}%
\column{44}{@{}>{\hspre}l<{\hspost}@{}}%
\column{47}{@{}>{\hspre}l<{\hspost}@{}}%
\column{57}{@{}>{\hspre}l<{\hspost}@{}}%
\column{61}{@{}>{\hspre}l<{\hspost}@{}}%
\column{72}{@{}>{\hspre}l<{\hspost}@{}}%
\column{76}{@{}>{\hspre}l<{\hspost}@{}}%
\column{87}{@{}>{\hspre}l<{\hspost}@{}}%
\column{E}{@{}>{\hspre}l<{\hspost}@{}}%
\>[3]{}\unopun{\mathbin{+_U\!}\!}\;{}\<[9]%
\>[9]{}\!:\!\;\Conid{U}\;\Varid{→}\;\Conid{U}\;\Varid{→}\;\Conid{U}{}\<[E]%
\\
\>[3]{}\unopun{\mathbin{+_U\!}\!}\;\Varid{⟨}\;\Varid{xl}\;\Varid{,}\;\Varid{xm}\;\Varid{,•,}\;\Varid{xr}\;\Varid{⟩}\;\Varid{⟨}\;\Varid{yl}\;\Varid{,}\;\Varid{ym}\;\Varid{,•,}\;\Varid{yr}\;\Varid{⟩}\;\mathrel{=}\;\Varid{⟨}\;\Varid{xl}\;\mathbin{+_u}\;\Varid{yl}\;\Varid{,}\;\Varid{xm}\;\mathbin{+_s}\;\Varid{ym}\;\Varid{,•,}\;\Varid{xr}\;\mathbin{+_u}\;\Varid{yr}\;\Varid{⟩}{}\<[E]%
\\[\blanklineskip]%
\>[3]{}\unopun{\mathbin{·_U}\!}\;\!:\!\;\Conid{U}\;\Varid{→}\;\Conid{U}\;\Varid{→}\;\Conid{U}{}\<[E]%
\\
\>[3]{}\unopun{\mathbin{·_U}\!}\;{}\<[9]%
\>[9]{}\Varid{⟨}\;{}\<[12]%
\>[12]{}\Varid{xl}\;{}\<[16]%
\>[16]{}\Varid{,}\;{}\<[19]%
\>[19]{}\Varid{xm}\;{}\<[E]%
\\
\>[12]{}\Varid{,•,}\;{}\<[19]%
\>[19]{}\Varid{xr}\;\Varid{⟩}\;{}\<[25]%
\>[25]{}\Varid{⟨}\;{}\<[28]%
\>[28]{}\Varid{yl}\;\Varid{,}\;{}\<[35]%
\>[35]{}\Varid{ym}\;{}\<[E]%
\\
\>[28]{}\Varid{,•,}\;{}\<[35]%
\>[35]{}\Varid{yr}\;\Varid{⟩}\;{}\<[41]%
\>[41]{}\mathrel{=}\;{}\<[44]%
\>[44]{}\Varid{⟨}\;{}\<[47]%
\>[47]{}\Varid{xl}\;\mathbin{·_u}\;\Varid{yl}\;{}\<[57]%
\>[57]{}\Varid{,}\;{}\<[61]%
\>[61]{}\Varid{xl}\;\mathbin{{_u}\!·_s}\;\Varid{ym}\;{}\<[72]%
\>[72]{}\mathbin{+_s}\;{}\<[76]%
\>[76]{}\Varid{xm}\;\mathbin{{_s}\!·_u}\;\Varid{yr}\;{}\<[E]%
\\
\>[47]{}\Varid{,•,}\;{}\<[61]%
\>[61]{}\Varid{xr}\;\mathbin{·_u}\;\Varid{yr}\;{}\<[87]%
\>[87]{}\Varid{⟩}{}\<[E]%
\ColumnHook
\end{hscode}\resethooks
Equivalence is structural: two matrices are equivalent if all the elements are equivalent.
\restorecolumns[SquareThird]
\begin{hscode}\SaveRestoreHook
\column{B}{@{}>{\hspre}l<{\hspost}@{}}%
\column{3}{@{}>{\hspre}l<{\hspost}@{}}%
\column{49}{@{}>{\hspre}l<{\hspost}@{}}%
\column{52}{@{}>{\hspre}l<{\hspost}@{}}%
\column{63}{@{}>{\hspre}l<{\hspost}@{}}%
\column{66}{@{}>{\hspre}l<{\hspost}@{}}%
\column{E}{@{}>{\hspre}l<{\hspost}@{}}%
\>[3]{}\unopun{\mathrel{≃_S}}\;\!:\!\;\Conid{S}\;\Varid{→}\;\Conid{S}\;\Varid{→}\;\Conid{Set}{}\<[E]%
\\
\>[3]{}\unopun{\mathrel{≃_S}}\;\Varid{⟨}\;\Varid{a}\;\Varid{,}\;\Varid{b}\;\Varid{,}\;\Varid{c}\;\Varid{,}\;\Varid{d}\;\Varid{⟩}\;\Varid{⟨}\;\Varid{a'}\;\Varid{,}\;\Varid{b'}\;\Varid{,}\;\Varid{c'}\;\Varid{,}\;\Varid{d'}\;\Varid{⟩}\;{}\<[49]%
\>[49]{}\mathrel{=}\;{}\<[52]%
\>[52]{}(\Varid{a}\;\mathrel{≃_s}\;\Varid{a'})\;{}\<[63]%
\>[63]{}\mathbin{\!\times\!}\;{}\<[66]%
\>[66]{}(\Varid{b}\;\mathrel{≃_s}\;\Varid{b'})\;{}\<[E]%
\\
\>[49]{}\mathbin{\!\times\!}\;{}\<[52]%
\>[52]{}(\Varid{c}\;\mathrel{≃_s}\;\Varid{c'})\;{}\<[63]%
\>[63]{}\mathbin{\!\times\!}\;{}\<[66]%
\>[66]{}(\Varid{d}\;\mathrel{≃_s}\;\Varid{d'}){}\<[E]%
\\[\blanklineskip]%
\>[3]{}\kw{infix}\;\Varid{4}\;\unopun{\mathrel{≃_S}}{}\<[E]%
\ColumnHook
\end{hscode}\resethooks
Embedding upper triangular matrices as (regular) matrices is straightforward:
\restorecolumns[SquareAgain]
\aligncolumn{48}{@{}>{\hspre}c<{\hspost}@{\quad}}
\begin{hscode}\SaveRestoreHook
\column{B}{@{}>{\hspre}l<{\hspost}@{}}%
\column{3}{@{}>{\hspre}l<{\hspost}@{}}%
\column{35}{@{}>{\hspre}l<{\hspost}@{}}%
\column{38}{@{}>{\hspre}l<{\hspost}@{}}%
\column{48}{@{}>{\hspre}l<{\hspost}@{}}%
\column{51}{@{}>{\hspre}l<{\hspost}@{}}%
\column{61}{@{}>{\hspre}l<{\hspost}@{}}%
\column{62}{@{}>{\hspre}l<{\hspost}@{}}%
\column{E}{@{}>{\hspre}l<{\hspost}@{}}%
\>[3]{}\Conid{U2S}\;\!:\!\;\Conid{U}\;\Varid{→}\;\Conid{S}{}\<[E]%
\\
\>[3]{}\Conid{U2S}\;\Varid{⟨}\;\Varid{uu}_{00}\;\Varid{,}\;\Varid{us}_{01}\;\Varid{,•,}\;\Varid{uu}_{11}\;\Varid{⟩}\;\mathrel{=}\;{}\<[35]%
\>[35]{}\Varid{⟨}\;{}\<[38]%
\>[38]{}\Varid{u2s}\;\Varid{uu}_{00}\;{}\<[48]%
\>[48]{}\Varid{,}\;{}\<[51]%
\>[51]{}\Varid{us}_{01}\;{}\<[61]%
\>[61]{}\Varid{,}\;{}\<[E]%
\\
\>[38]{}0_s\;{}\<[48]%
\>[48]{}\Varid{,}\;{}\<[51]%
\>[51]{}\Varid{u2s}\;\Varid{uu}_{11}\;{}\<[62]%
\>[62]{}\Varid{⟩}{}\<[E]%
\ColumnHook
\end{hscode}\resethooks
\subsubsection*{Lifting laws}
At this point we must verify that the above constructions preserve the
laws seen so far.
Most of this verification (of \ensuremath{\Varid{symS}}, \ensuremath{\Varid{transS}}, \ensuremath{\Varid{congS}}, etc.) is by
straightforward, and omitted, lifting of the proofs from the
underlying structure, for example \ensuremath{\Varid{reflS}}:
\restorecolumns[Square]
\begin{hscode}\SaveRestoreHook
\column{B}{@{}>{\hspre}l<{\hspost}@{}}%
\column{3}{@{}>{\hspre}l<{\hspost}@{}}%
\column{12}{@{}>{\hspre}l<{\hspost}@{}}%
\column{51}{@{}>{\hspre}l<{\hspost}@{}}%
\column{E}{@{}>{\hspre}l<{\hspost}@{}}%
\>[3]{}\Varid{reflS}\;\!:\!\;\{\mskip1.5mu \Varid{x}\;\!:\!\;\Conid{S}\mskip1.5mu\}\;\Varid{→}\;\Varid{x}\;\mathrel{≃_S}\;\Varid{x}{}\<[51]%
\>[51]{}\mbox{\onelinecomment  \structure{4.3}{Laws}}{}\<[E]%
\\
\>[3]{}\Varid{reflS}\;\mathrel{=}\;{}\<[12]%
\>[12]{}\Varid{refl}_s\;\Varid{,}\;\Varid{refl}_s\;\Varid{,}\;{}\<[E]%
\\
\>[12]{}\Varid{refl}_s\;\Varid{,}\;\Varid{refl}_s{}\<[E]%
\ColumnHook
\end{hscode}\resethooks
To prove the preservation of distributivity we use the following lemma
about commutativity (proving it is an easy equational reasoning
exercise in Agda).
\restorecolumns[Square]
\begin{hscode}\SaveRestoreHook
\column{B}{@{}>{\hspre}l<{\hspost}@{}}%
\column{3}{@{}>{\hspre}l<{\hspost}@{}}%
\column{26}{@{}>{\hspre}l<{\hspost}@{}}%
\column{30}{@{}>{\hspre}l<{\hspost}@{}}%
\column{40}{@{}>{\hspre}l<{\hspost}@{}}%
\column{44}{@{}>{\hspre}l<{\hspost}@{}}%
\column{E}{@{}>{\hspre}l<{\hspost}@{}}%
\>[3]{}\Varid{swapMid}\;\!:\!\;\Varid{∀}\;\{\mskip1.5mu \Varid{a}\;\Varid{b}\;\Varid{c}\;\Varid{d}\mskip1.5mu\}\;{}\<[26]%
\>[26]{}\Varid{→}\;{}\<[30]%
\>[30]{}(\Varid{a}\;\mathbin{+_s}\;\Varid{b})\;{}\<[40]%
\>[40]{}\mathbin{+_s}\;{}\<[44]%
\>[44]{}(\Varid{c}\;\mathbin{+_s}\;\Varid{d})\;{}\<[E]%
\\
\>[26]{}\mathrel{≃_s}\;{}\<[30]%
\>[30]{}(\Varid{a}\;\mathbin{+_s}\;\Varid{c})\;{}\<[40]%
\>[40]{}\mathbin{+_s}\;{}\<[44]%
\>[44]{}(\Varid{b}\;\mathbin{+_s}\;\Varid{d}){}\<[E]%
\ColumnHook
\end{hscode}\resethooks

\restorecolumns[Square]
\begin{hscode}\SaveRestoreHook
\column{B}{@{}>{\hspre}l<{\hspost}@{}}%
\column{3}{@{}>{\hspre}l<{\hspost}@{}}%
\column{6}{@{}>{\hspre}l<{\hspost}@{}}%
\column{13}{@{}>{\hspre}l<{\hspost}@{}}%
\column{19}{@{}>{\hspre}l<{\hspost}@{}}%
\column{42}{@{}>{\hspre}l<{\hspost}@{}}%
\column{46}{@{}>{\hspre}l<{\hspost}@{}}%
\column{66}{@{}>{\hspre}l<{\hspost}@{}}%
\column{70}{@{}>{\hspre}l<{\hspost}@{}}%
\column{E}{@{}>{\hspre}l<{\hspost}@{}}%
\>[3]{}\Varid{distlS}\;\!:\!\;(\Varid{x}\;\Varid{y}\;\Varid{z}\;\!:\!\;\Conid{S})\;\Varid{→}\;\Varid{x}\;\mathbin{·\!_S}\;(\Varid{y}\;\mathbin{+\!_S}\;\Varid{z})\;\mathrel{≃_S}\;\Varid{x}\;\mathbin{·\!_S}\;\Varid{y}\;\mathbin{+\!_S}\;\Varid{x}\;\mathbin{·\!_S}\;\Varid{z}{}\<[E]%
\\
\>[3]{}\Varid{distlS}\;\anonymous \;\anonymous \;\anonymous \;\mathrel{=}\;{}\<[19]%
\>[19]{}\Varid{distrHelp}\;\Varid{,}\;\Varid{distrHelp}\;\Varid{,}\;\Varid{distrHelp}\;\Varid{,}\;\Varid{distrHelp}{}\<[E]%
\\
\>[3]{}\hsindent{3}{}\<[6]%
\>[6]{}\kw{where}\;{}\<[13]%
\>[13]{}\Varid{distrHelp}\;\!:\!\;\Varid{∀}\;\{\mskip1.5mu \Varid{a}\;\Varid{b}\;\Varid{c}\;\Varid{d}\;\Varid{e}\;\Varid{f}\mskip1.5mu\}\;{}\<[42]%
\>[42]{}\Varid{→}\;{}\<[46]%
\>[46]{}\Varid{a}\;\mathbin{·_s}\;(\Varid{b}\;\mathbin{+_s}\;\Varid{c})\;{}\<[66]%
\>[66]{}\mathbin{+_s}\;{}\<[70]%
\>[70]{}\Varid{d}\;\mathbin{·_s}\;(\Varid{e}\;\mathbin{+_s}\;\Varid{f})\;{}\<[E]%
\\
\>[42]{}\mathrel{≃_s}\;{}\<[46]%
\>[46]{}(\Varid{a}\;\mathbin{·_s}\;\Varid{b}\;\mathbin{+_s}\;\Varid{d}\;\mathbin{·_s}\;\Varid{e})\;{}\<[66]%
\>[66]{}\mathbin{+_s}\;(\Varid{a}\;\mathbin{·_s}\;\Varid{c}\;\mathbin{+_s}\;\Varid{d}\;\mathbin{·_s}\;\Varid{f}){}\<[E]%
\\
\>[13]{}\Varid{distrHelp}\;\mathrel{=}\;\Varid{trans}_s\;(\Varid{distl}\;\anonymous \;\anonymous \;\anonymous \;\mathbin{<\!\!\!+\!\!\!>}\;\Varid{distl}\;\anonymous \;\anonymous \;\anonymous )\;\Varid{swapMid}{}\<[E]%
\ColumnHook
\end{hscode}\resethooks

The record value of type \ensuremath{\Conid{Square}} should eventually contain all
the fields, including the proof of closure --- but we leave the rest
to the upcoming sections.
Thus we leave the definition of the functor \ensuremath{\Conid{Square}} for now and
instead give the the base case of the inductive structure.
\subsubsection*{Base case: 1-by-1 matrices}
\label{sec:OneByOne}
Any strictly upper triangular matrix is by definition zero on (and below) the diagonal.
At size 1-by-1 that means the only element is zero so need not store any element.
We represent this case by the unit type (\ensuremath{\Varid{tt}\;\!:\!\;\Varid{⊤}}) and it is therefore
trivially equipped with closure.
\begin{hscode}\SaveRestoreHook
\column{B}{@{}>{\hspre}l<{\hspost}@{}}%
\column{3}{@{}>{\hspre}l<{\hspost}@{}}%
\column{5}{@{}>{\hspre}l<{\hspost}@{}}%
\column{12}{@{}>{\hspre}l<{\hspost}@{}}%
\column{26}{@{}>{\hspre}l<{\hspost}@{}}%
\column{45}{@{}>{\hspre}l<{\hspost}@{}}%
\column{51}{@{}>{\hspre}l<{\hspost}@{}}%
\column{64}{@{}>{\hspre}l<{\hspost}@{}}%
\column{70}{@{}>{\hspre}l<{\hspost}@{}}%
\column{E}{@{}>{\hspre}l<{\hspost}@{}}%
\>[3]{}\Conid{OneByOne}\;\!:\!\;\Conid{SemiNearRing}\;\to \;\Conid{ClosedSemiNearRing}{}\<[51]%
\>[51]{}\mbox{\onelinecomment  \structure{5}{One-by-one matrix}}{}\<[E]%
\\
\>[3]{}\Conid{OneByOne}\;\Varid{snr}\;\mathrel{=}\;\kw{record}\;{}\<[26]%
\>[26]{}\{\mskip1.5mu \Varid{snr2}\;\mathrel{=}\;\kw{record}\;\{\mskip1.5mu {}\<[45]%
\>[45]{}\Varid{snr}\;\mathrel{=}\;\Varid{snr};\Varid{u}\;\mathrel{=}\;\Varid{⊤};{}\<[64]%
\>[64]{}\unopun{\mathbin{+_u}}\;{}\<[70]%
\>[70]{}\mathrel{=}\;\lambda \;\anonymous \;\anonymous \;\Varid{→}\;\Varid{tt};{}\<[E]%
\\
\>[45]{}\Varid{u2s}\;\mathrel{=}\;\lambda \;\anonymous \;\Varid{→}\;0_s;{}\<[64]%
\>[64]{}\unopun{\mathbin{·_u}}\;{}\<[70]%
\>[70]{}\mathrel{=}\;\lambda \;\anonymous \;\anonymous \;\Varid{→}\;\Varid{tt}\mskip1.5mu\}{}\<[E]%
\\
\>[26]{}\mskip1.5mu\}{}\<[E]%
\\
\>[3]{}\hsindent{2}{}\<[5]%
\>[5]{}\kw{where}\;{}\<[12]%
\>[12]{}\kw{open}\;\Conid{SemiNearRing}\;\Varid{snr}\;\kw{using}\;(0_s){}\<[E]%
\ColumnHook
\end{hscode}\resethooks
Finally we give the recursion schema, following the pattern that we hinted at the beginning of the section:
we define \ensuremath{\Conid{Mat}\;\Varid{n}} as the semi-near-ring structures lifted \ensuremath{\Varid{n+1}} times.
\begin{hscode}\SaveRestoreHook
\column{B}{@{}>{\hspre}l<{\hspost}@{}}%
\column{3}{@{}>{\hspre}l<{\hspost}@{}}%
\column{16}{@{}>{\hspre}l<{\hspost}@{}}%
\column{51}{@{}>{\hspre}l<{\hspost}@{}}%
\column{E}{@{}>{\hspre}l<{\hspost}@{}}%
\>[3]{}\Conid{Mat}\;\!:\!\;\Conid{ℕ}\;\Varid{→}\;\Conid{SemiNearRing}\;\Varid{→}\;\Conid{ClosedSemiNearRing}{}\<[51]%
\>[51]{}\mbox{\onelinecomment  \structure{6}{Top level recursion for matrices}}{}\<[E]%
\\
\>[3]{}\Conid{Mat}\;\Varid{zero}\;{}\<[16]%
\>[16]{}\Varid{el}\;\mathrel{=}\;\Conid{OneByOne}\;\Varid{el}{}\<[E]%
\\
\>[3]{}\Conid{Mat}\;(\Varid{suc}\;\Varid{n})\;{}\<[16]%
\>[16]{}\Varid{el}\;\mathrel{=}\;\Conid{Square}\;(\Conid{Mat}\;\Varid{n}\;\Varid{el}){}\<[E]%
\ColumnHook
\end{hscode}\resethooks
The fields of \ensuremath{\Conid{Mat}\;\Varid{n}\;\Varid{el}} are of special interest to us.
In particular the \ensuremath{\Varid{u}} field is the type of (strictly upper triangular)
matrices of size \ensuremath{2^\Varid{n}}, which we will soon show how to equip
with a closure operation.
\begin{hscode}\SaveRestoreHook
\column{B}{@{}>{\hspre}l<{\hspost}@{}}%
\column{3}{@{}>{\hspre}l<{\hspost}@{}}%
\column{E}{@{}>{\hspre}l<{\hspost}@{}}%
\>[3]{}\Conid{Upper}\;\!:\!\;\Conid{ℕ}\;\Varid{→}\;\Conid{SemiNearRing}\;\Varid{→}\;\Conid{Set}{}\<[E]%
\\
\>[3]{}\Conid{Upper}\;\Varid{n}\;\Varid{el}\;\mathrel{=}\;\Conid{ClosedSemiNearRing.u}\;(\Conid{Mat}\;\Varid{n}\;\Varid{el}){}\<[E]%
\ColumnHook
\end{hscode}\resethooks
(The expression \ensuremath{\Conid{ClosedSemiNearRing.u}\;\Varid{csnr}} extracts the field \ensuremath{\Varid{u}} from
the record \ensuremath{\Varid{csnr}} of type \ensuremath{\Conid{ClosedSemiNearRing}}.)

\section{Transitive closure: derivation}\label{sec:derivation}

We have now completed the definition of all the structures necessary
to develop our proof.
In this section we describe (and partially prove) the closure
algorithm.

We do so by deriving it from the specification \ensuremath{\Conid{Closure}\;\Varid{w}\;\Varid{c}\;\mathrel{=}\;\Conid{Least}\;\unopun{\mathrel{≤_u}}\;(\Conid{Q}\;\Varid{w})\;\Varid{c}},
where
\ensuremath{\Conid{Q}\;\Varid{w}\;\Varid{c}\;\mathrel{=}\;\Varid{w}\;\mathbin{+_u}\;\Varid{c}\;\mathbin{·_u}\;\Varid{c}\;\mathrel{≃_u}\;\Varid{c}}
and we start with the equational part, the \ensuremath{\Conid{Entire}\;\Conid{Q}} requirement, and
return to \ensuremath{\Conid{Least}} in \fref{sec:smallest}.
Technically, we give a definition for the \ensuremath{\Varid{entireQ}\;\!:\!\;\Conid{Entire}\;\Conid{Q}} field
in both the \ensuremath{\Conid{OneByOne}} and \ensuremath{\Conid{Square}} cases of the \ensuremath{\Conid{Mat}} function.

We first proceed semi-formally, to show what a non-certified proof of
the algorithm looks like, and to be able to compare it with the fully
formal, certified Agda proof that we subsequently present.

\subsection{Closure of triangular matrices}
Recall that our task is to find a function $\_^+$ which maps an upper
triangular matrix $W$ to its transitive closure $C = \closure W$.

If $W$ is a 1 by 1 matrix, $C = W = 0$.
Otherwise, let us divide $W$ and $C$ in blocks as follows:
\begin{align*}
   W &= \tritwo A Y B \hspace{2cm} &  C &= \tritwo {A'} {X'} {B'}
\end{align*}
Then the condition that $C$ satisfies \ensuremath{\Conid{Q}} becomes:
$$ \tritwo A Y B + \tritwo {A'} {X'} {B'} ·
                   \tritwo {A'} {X'} {B'} = \tritwo {A'} {X'} {B'} $$
Applying matrix multiplication and addition block-wise:
\begin{align*}
 A + A' A'         &=  A'\\
 Y + A' X'+X' B'   &=  X'\\
 B + B' B'         &=  B'
\end{align*}
Because $A$ and $B$ are smaller matrices than $W$ (and still upper triangular),
we know how to compute $A'$ and $B'$ recursively ($A' = \closure A$,
$B' = \closure B$). Before showing how $X'$ is computed, we show how to formalise
the above reasoning in Agda. The main job is to populate the \ensuremath{\Varid{entireQ}} field
in the \ensuremath{\Conid{ClosedSemiNearRing}} record. We have to do so both for the base case and the
recursive case of the \ensuremath{\Conid{Mat}} construction. The base case being trivial, we show
here the inductive case.

We first define the \ensuremath{\Conid{Q}} relation for 2-by-2 (block-)matrices:
\restorecolumns[Square]
\begin{hscode}\SaveRestoreHook
\column{B}{@{}>{\hspre}l<{\hspost}@{}}%
\column{3}{@{}>{\hspre}l<{\hspost}@{}}%
\column{21}{@{}>{\hspre}l<{\hspost}@{}}%
\column{25}{@{}>{\hspre}l<{\hspost}@{}}%
\column{51}{@{}>{\hspre}l<{\hspost}@{}}%
\column{E}{@{}>{\hspre}l<{\hspost}@{}}%
\>[3]{}\Conid{QU}\;\mathrel{=}\;\lambda \;\Conid{W}\;\Conid{C}\;\to \;(\Conid{W}\;{}\<[21]%
\>[21]{}\mathbin{+_U\!}\;{}\<[25]%
\>[25]{}(\Conid{C}\;\mathbin{·_U}\;\Conid{C}))\;\mathrel{≃_U}\;\Conid{C}{}\<[51]%
\>[51]{}\mbox{\onelinecomment  \structure{4.4}{Lifting \ensuremath{\Conid{Q}} and its proof}}{}\<[E]%
\ColumnHook
\end{hscode}\resethooks
and then we proceed with the proof that it is \ensuremath{\Conid{Entire}}: for any input
matrix \ensuremath{\Conid{W}} we construct another matrix \ensuremath{\Conid{C}} and a \ensuremath{\Varid{proof}} that it is a
closure (strictly speaking, so far only that it satisfies \ensuremath{\Conid{QU}\;\Conid{W}\;\Conid{C}}).
One remarkable feature is that Agda can infer the solution matrix (\ensuremath{\Conid{C}\;\mathrel{=}\;\anonymous }), from the \ensuremath{\Varid{proof}}. 
The proof follows exactly the semi-formal development given at the
beginning of the subsection.
\savecolumns[proof]
\begin{hscode}\SaveRestoreHook
\column{B}{@{}>{\hspre}l<{\hspost}@{}}%
\column{3}{@{}>{\hspre}l<{\hspost}@{}}%
\column{10}{@{}>{\hspre}l<{\hspost}@{}}%
\column{19}{@{}>{\hspre}l<{\hspost}@{}}%
\column{21}{@{}>{\hspre}l<{\hspost}@{}}%
\column{22}{@{}>{\hspre}l<{\hspost}@{}}%
\column{24}{@{}>{\hspre}l<{\hspost}@{}}%
\column{25}{@{}>{\hspre}l<{\hspost}@{}}%
\column{26}{@{}>{\hspre}l<{\hspost}@{}}%
\column{27}{@{}>{\hspre}l<{\hspost}@{}}%
\column{29}{@{}>{\hspre}l<{\hspost}@{}}%
\column{30}{@{}>{\hspre}l<{\hspost}@{}}%
\column{33}{@{}>{\hspre}l<{\hspost}@{}}%
\column{37}{@{}>{\hspre}l<{\hspost}@{}}%
\column{38}{@{}>{\hspre}l<{\hspost}@{}}%
\column{39}{@{}>{\hspre}l<{\hspost}@{}}%
\column{41}{@{}>{\hspre}l<{\hspost}@{}}%
\column{43}{@{}>{\hspre}l<{\hspost}@{}}%
\column{48}{@{}>{\hspre}l<{\hspost}@{}}%
\column{49}{@{}>{\hspre}l<{\hspost}@{}}%
\column{55}{@{}>{\hspre}l<{\hspost}@{}}%
\column{58}{@{}>{\hspre}l<{\hspost}@{}}%
\column{60}{@{}>{\hspre}l<{\hspost}@{}}%
\column{64}{@{}>{\hspre}l<{\hspost}@{}}%
\column{66}{@{}>{\hspre}l<{\hspost}@{}}%
\column{70}{@{}>{\hspre}l<{\hspost}@{}}%
\column{E}{@{}>{\hspre}l<{\hspost}@{}}%
\>[3]{}\Varid{entireQStep}\;\!:\!\;\Varid{∀}\;\Conid{W}\;\to \;\Varid{∃}\;(\Conid{QU}\;\Conid{W}){}\<[E]%
\\
\>[3]{}\Varid{entireQStep}\;\Conid{W}\;\mathrel{=}\;\Conid{C}\;\Varid{,}\;\Varid{proof}\;\kw{where}{}\<[E]%
\\
\>[3]{}\hsindent{7}{}\<[10]%
\>[10]{}\Conid{C}\;\!:\!\;\Conid{U}{}\<[E]%
\\
\>[3]{}\hsindent{7}{}\<[10]%
\>[10]{}\Conid{C}\;\mathrel{=}\;\anonymous {}\<[E]%
\\[\blanklineskip]%
\>[3]{}\hsindent{7}{}\<[10]%
\>[10]{}\kw{open}\;\Conid{EqReasoning}\;(\Conid{SemiNearRing2.uSetoid}\;\Conid{SNR2}){}\<[E]%
\\
\>[3]{}\hsindent{7}{}\<[10]%
\>[10]{}\Varid{proof}\;\!:\!\;(\Conid{W}\;{}\<[22]%
\>[22]{}\mathbin{+_U\!}\;{}\<[26]%
\>[26]{}(\Conid{C}\;\mathbin{·_U}\;\Conid{C}))\;{}\<[37]%
\>[37]{}\mathrel{≃_U}\;{}\<[41]%
\>[41]{}\Conid{C}{}\<[E]%
\\
\>[3]{}\hsindent{7}{}\<[10]%
\>[10]{}\Varid{proof}\;\mathrel{=}\;{}\<[19]%
\>[19]{}\Varid{begin}\;{}\<[E]%
\\
\>[19]{}\hsindent{2}{}\<[21]%
\>[21]{}(\Conid{W}\;{}\<[25]%
\>[25]{}\mathbin{+_U\!}\;{}\<[29]%
\>[29]{}(\Conid{C}\;\mathbin{·_U}\;\Conid{C}))\;{}\<[E]%
\\
\>[19]{}≡\!\!⟨\;\Varid{refl}\;\Varid{⟩}{}\<[30]%
\>[30]{}\mbox{\onelinecomment  expand matrix components}{}\<[E]%
\\
\>[19]{}\hsindent{2}{}\<[21]%
\>[21]{}\kw{let}\;{}\<[27]%
\>[27]{}\Varid{⟨}\;\Conid{A}\;{}\<[33]%
\>[33]{}\Varid{,}\;\Conid{Y}\;{}\<[39]%
\>[39]{}\Varid{,•,}\;\Conid{B}\;{}\<[48]%
\>[48]{}\Varid{⟩}\;\mathrel{=}\;\Conid{W}{}\<[E]%
\\
\>[27]{}\Varid{⟨}\;\Conid{A'}\;{}\<[33]%
\>[33]{}\Varid{,}\;\Conid{Y'}\;{}\<[39]%
\>[39]{}\Varid{,•,}\;\Conid{B'}\;{}\<[48]%
\>[48]{}\Varid{⟩}\;\mathrel{=}\;\Conid{C}\;\kw{in}{}\<[E]%
\\
\>[19]{}\hsindent{2}{}\<[21]%
\>[21]{}\Varid{⟨}\;\Conid{A}\;\Varid{,}\;\Conid{Y}\;\Varid{,•,}\;\Conid{B}\;\Varid{⟩}\;{}\<[38]%
\>[38]{}\mathbin{+_U\!}\;(\Varid{⟨}\;\Conid{A'}\;\Varid{,}\;\Conid{Y'}\;\Varid{,•,}\;\Conid{B'}\;\Varid{⟩}\;\mathbin{·_U}\;\Varid{⟨}\;\Conid{A'}\;\Varid{,}\;\Conid{Y'}\;\Varid{,•,}\;\Conid{B'}\;\Varid{⟩}){}\<[E]%
\\
\>[19]{}≡\!\!⟨\;\Varid{refl}\;\Varid{⟩}\;{}\<[30]%
\>[30]{}\mbox{\onelinecomment  expand definition of \ensuremath{\mathbin{·_U}}}{}\<[E]%
\\
\>[19]{}\hsindent{2}{}\<[21]%
\>[21]{}\Varid{⟨}\;\Conid{A}\;\Varid{,}\;\Conid{Y}\;\Varid{,•,}\;\Conid{B}\;\Varid{⟩}\;{}\<[38]%
\>[38]{}\mathbin{+_U\!}\;{}\<[43]%
\>[43]{}\Varid{⟨}\;\Conid{A'}\;\mathbin{·_u}\;\Conid{A'}\;{}\<[55]%
\>[55]{}\Varid{,}\;{}\<[58]%
\>[58]{}(\Conid{A'}\;\mathbin{{_u}\!·_s}\;\Conid{Y'}\;{}\<[70]%
\>[70]{}\mathbin{+_s}\;\Conid{Y'}\;\mathbin{{_s}\!·_u}\;\Conid{B'})\;{}\<[E]%
\\
\>[55]{}\Varid{,•,}\;{}\<[60]%
\>[60]{}\Conid{B'}\;\mathbin{·_u}\;\Conid{B'}\;\Varid{⟩}{}\<[E]%
\\
\>[19]{}≡\!\!⟨\;\Varid{refl}\;\Varid{⟩}\;{}\<[30]%
\>[30]{}\mbox{\onelinecomment  by def. of \ensuremath{\mathbin{+_U\!}}}{}\<[E]%
\\
\>[19]{}\hsindent{2}{}\<[21]%
\>[21]{}\Varid{⟨}\;{}\<[24]%
\>[24]{}\Conid{A}\;\mathbin{+_u}\;\Conid{A'}\;\mathbin{·_u}\;\Conid{A'}\;{}\<[43]%
\>[43]{}\Varid{,}\;{}\<[49]%
\>[49]{}\Conid{Y}\;\mathbin{+_s}\;(\Conid{A'}\;\mathbin{{_u}\!·_s}\;\Conid{Y'}\;{}\<[66]%
\>[66]{}\mathbin{+_s}\;{}\<[70]%
\>[70]{}\Conid{Y'}\;\mathbin{{_s}\!·_u}\;\Conid{B'})\;{}\<[E]%
\\
\>[43]{}\Varid{,•,}\;{}\<[49]%
\>[49]{}\Conid{B}\;\mathbin{+_u}\;\Conid{B'}\;\mathbin{·_u}\;\Conid{B'}\;{}\<[64]%
\>[64]{}\Varid{⟩}{}\<[E]%
\\
\>[19]{}≈\!\!⟨\;\Varid{congU}\;\Varid{closureHasAll}\;\Varid{completionHasAll}\;\Varid{closureHasAll}\;\Varid{⟩}\;{}\<[E]%
\\
\>[19]{}\hsindent{2}{}\<[21]%
\>[21]{}\Varid{⟨}\;\Conid{A'}\;\Varid{,}\;\Conid{Y'}\;\Varid{,•,}\;\Conid{B'}\;\Varid{⟩}{}\<[E]%
\\
\>[19]{}≡\!\!⟨\;\Varid{refl}\;\Varid{⟩}\;{}\<[E]%
\\
\>[19]{}\hsindent{3}{}\<[22]%
\>[22]{}\Conid{C}{}\<[E]%
\\
\>[19]{}\Varid{∎}{}\<[E]%
\ColumnHook
\end{hscode}\resethooks

The definition \ensuremath{\Conid{C}\;\mathrel{=}\;\anonymous } works because Agda unifies the type of the
proof that it infers with the type that we give (\ensuremath{(\Conid{W}\;\mathbin{+_U\!}\;(\Conid{C}\;\mathbin{·_U}\;\Conid{C}))\;\mathrel{≃_U}\;\Conid{C}}). The left-hand-side of the equivalence inferred type is given by
each step in the proof: \ensuremath{\Conid{C}\;\mathrel{=}\;\Varid{⟨}\;\Conid{A'}\;\Varid{,}\;\Conid{Y'}\;\Varid{,•,}\;\Conid{B'}\;\Varid{⟩}}, and so on. If we
expand those steps this is the core algorithm:

\restorecolumns[proof]
\begin{hscode}\SaveRestoreHook
\column{B}{@{}>{\hspre}l<{\hspost}@{}}%
\column{10}{@{}>{\hspre}l<{\hspost}@{}}%
\column{15}{@{}>{\hspre}l<{\hspost}@{}}%
\column{21}{@{}>{\hspre}l<{\hspost}@{}}%
\column{38}{@{}>{\hspre}l<{\hspost}@{}}%
\column{E}{@{}>{\hspre}l<{\hspost}@{}}%
\>[10]{}\Conid{C}\;\mathrel{=}\;{}\<[15]%
\>[15]{}\kw{let}\;{}\<[21]%
\>[21]{}\Varid{⟨}\;\Conid{A}\;\Varid{,}\;\Conid{Y}\;\Varid{,•,}\;\Conid{B}\;\Varid{⟩}\;{}\<[38]%
\>[38]{}\mathrel{=}\;\Conid{W}{}\<[E]%
\\
\>[21]{}(\Conid{A'}\;\Varid{,}\;\Varid{proofA})\;{}\<[38]%
\>[38]{}\mathrel{=}\;\Varid{entireQ}\;\Conid{A}{}\<[E]%
\\
\>[21]{}(\Conid{B'}\;\Varid{,}\;\Varid{proofB})\;{}\<[38]%
\>[38]{}\mathrel{=}\;\Varid{entireQ}\;\Conid{B}{}\<[E]%
\\
\>[21]{}(\Conid{Y'}\;\Varid{,}\;\Varid{proofY})\;{}\<[38]%
\>[38]{}\mathrel{=}\;\Varid{entireL}\;\Conid{A'}\;\Conid{Y}\;\Conid{B'}{}\<[E]%
\\
\>[15]{}\kw{in}\;{}\<[21]%
\>[21]{}\Varid{⟨}\;\Conid{A'}\;\Varid{,}\;\Conid{Y'}\;\Varid{,•,}\;\Conid{B'}\;\Varid{⟩}{}\<[E]%
\ColumnHook
\end{hscode}\resethooks
The use of equational reasoning in \ensuremath{\Varid{proof}} shows another very useful
feature of the proof notation: using a normal \ensuremath{\kw{let}}-\ensuremath{\kw{in}} expression
together with the distfix transitivity operator \ensuremath{\ensuremath{\un{≡}⟨\un⟩\un}} we can do
what is often done in paper-proofs: introduce new names in the middle
of the reasoning chain.
The new names (here \ensuremath{\Conid{A,}\;\Conid{Y,}\;\Conid{B}} etc.) are in scope in the rest of the
\ensuremath{\Varid{proof}}.
Another interesting feature of this proof is that many of the
steps can be justified simply by \ensuremath{\Varid{refl}\;\!:\!\;\Varid{x}\;\Varid{≡}\;\Varid{x}}. Indeed, Agda automatically expands
definitions during type-checking, and thus automatically expands the definitions
of operators on block matrices.
In fact, because \ensuremath{\Varid{refl}} is ``the unit of transitivity'', Agda would be just as
happy with only:
\restorecolumns[proof]
\begin{hscode}\SaveRestoreHook
\column{B}{@{}>{\hspre}l<{\hspost}@{}}%
\column{10}{@{}>{\hspre}l<{\hspost}@{}}%
\column{E}{@{}>{\hspre}l<{\hspost}@{}}%
\>[10]{}\Varid{proof}\;\mathrel{=}\;\Varid{congU}\;\Varid{closureHasAll}\;\Varid{completionHasAll}\;\Varid{closureHasAll}{}\<[E]%
\ColumnHook
\end{hscode}\resethooks
Yet, this one line still holds the full proof of the inner
induction (\ensuremath{\Varid{completionHasAll}}) which is the topic of
\fref{sec:completion}.
(Remember that \ensuremath{\Varid{closureHasAll}\;\mathrel{=}\;\Varid{correct}\;\Varid{entireQ}} from \fref{sec:spec}
(page \pageref{def:closure}) plays the role of induction hypothesis
for the closure of triangular matrices.)

The base case is completely trivial; a 1-by-1 upper triangular matrix
contains no non-zero element and is represented by the unit type.
Thus \ensuremath{\Varid{entireQBase}\;\Varid{tt}\;\mathrel{=}\;\Varid{tt}\;\Varid{,}\;\Varid{refl}_s}.

\subsection{Completion of square matrices}\label{sec:completion}
\newcommand{\reccall}[4] {
  \fill[black, very nearly transparent] (#1,#1) -- (#1,#4) -- (#4,#4) -- (#3,#3) -- (#2,#3) -- (#2,#2) -- cycle;
  \fill[black, very nearly transparent] (#1,#3) -- (#1,#4) -- (#2,#4) -- (#2,#3) -- cycle;
  \fill[black, very nearly transparent] (#1,#1) -- (#1,#2) -- (#2,#2) -- cycle;
  \fill[black, very nearly transparent] (#3,#3) -- (#3,#4) -- (#4,#4) -- cycle;
 }
\newcommand{\vpicture}[4]{
    \pgftransformrotate{-90}
    \pgftransformscale{1.1}
    \node at (0,-1) {Step #4:};
    \small
    \draw (0,0) -- (4,4);
    \subc 0 1 {$A_{11}$};
    \subc 1 2 {$A_{22}$};
    \subc 0 2 {$A_{12}$};
    \subc 2 4 {$B_{12}$};
    \subc 2 3 {$B_{11}$};
    \subc 3 4 {$B_{22}$};

    \subc 1 3 {$#1_{21}$};
    \subc 0 3 {$#2_{11}$};
    \subc 1 4 {$#3_{22}$};
    \subc 0 4 {$Y_{12}$};
}
\begin{figure}
  \centering
\begin{tikzpicture} \vpicture X Y Y 2 \reccall 0 1 2 3 \end{tikzpicture}
\begin{tikzpicture} \vpicture X X X 4 \reccall 0 1 3 4 \end{tikzpicture}
\\
\begin{tikzpicture} \vpicture Y Y Y 1 \reccall 1 2 2 3 \end{tikzpicture}
\begin{tikzpicture} \vpicture X X Y 3 \reccall 1 2 3 4 \end{tikzpicture}

  \caption{\small The recursive step of function $V$.
    The charts $A$ and $B$ are already complete.
    To complete the matrix $Y$, that is, compute $X = V(A,Y,B)$, one
    splits the matrices and performs 4 recursive calls.
    Each recursive call is depicted graphically.
    In each figure, to complete the dark-gray square, multiply the
    light-gray rectangles and add them to the dark-gray square,
    then do a recursive call on triangular matrix composed of
    the completed dark-gray square and the triangles.
%
  }
\label{fig:valiant}
\end{figure}
As in the previous subsection, we first proceed semi-formally.
The problem is to find a recursive function $V$ which maps $A$, $Y$
and $B$ to $X = V(A,Y,B)$, such that $X = Y + A·X + X·B$.
In terms of parsing, the function $V$ combines the chart $A$ of the
first part of the input with the chart $B$ of the second part of the
input, via a \emph{partial} chart $Y$ concerned only with strings
starting in $A$ and ending in $B$, and produces a full chart $X$.
We proceed as before and divide each matrix into blocks:
$$
\begin{array}{cccc}
  X = \twobytwo {X_{11}} {X_{12}} {X_{21}} {X_{22}} &
  Y = \twobytwo {Y_{11}} {Y_{12}} {Y_{21}} {Y_{22}} &
  A = \tritwo {A_{11}} {A_{12}} {A_{22}} &
  B = \tritwo {B_{11}} {B_{12}} {B_{22}}
\end{array}
$$
The condition on $X$ then becomes
\begin{align*}
\twobytwo {X_{11}} {X_{12}} {X_{21}} {X_{22}} =
\twobytwo {Y_{11}} {Y_{12}} {Y_{21}} {Y_{22}} +
\tritwo {A_{11}} {A_{12}} {A_{22}}  · \twobytwo {X_{11}} {X_{12}} {X_{21}} {X_{22}} +
\twobytwo {X_{11}} {X_{12}} {X_{21}} {X_{22}} · \tritwo {B_{11}} {B_{12}} {B_{22}}
\end{align*}
By applying matrix multiplication and addition block-wise:
$$
\begin{array}{l@{~=~}l@{ + }l@{ + }l@{ + }l@{ + }l}
X_{11} & Y_{11} & A_{11} X_{11} & A_{12} X_{21} & X_{11} B_{11} & 0               \\
X_{12} & Y_{12} & A_{11} X_{12} & A_{12} X_{22} & X_{11} B_{12} & X_{12} B_{22}    \\
X_{21} & Y_{21} & 0             & A_{22} X_{21} & X_{21} B_{11} & 0             \\
X_{22} & Y_{22} & 0             & A_{22} X_{22} & X_{21} B_{12} & X_{22} B_{22}   \\
\end{array}
$$
By commutativity of $(+)$ and 0 being its unit:
$$
\begin{array}{r@{~=~}l@{~+~}l@{~+~}l}
 X_{11} & Y_{11} + A_{12} X_{21}                 & A_{11}  X_{11}  & X_{11}   B_{11}\\
 X_{12} & Y_{12} + A_{12} X_{22} + X_{11} B_{12} & A_{11}  X_{12}    & X_{12}   B_{22}\\
 X_{21} & Y_{21}                                 & A_{22}  X_{21}&  X_{21}  B_{11}\\
 X_{22} & Y_{22} + X_{21} B_{12}                 & A_{22}  X_{22}  & X_{22}  B_{22}
\end{array}
$$
Now we have four equations, all of the form that $V$ can compute
solutions to.
Because each of the sub-matrices is smaller and because of the absence
of circular dependencies, $X$ can be computed recursively by $V$.
The internal dependencies dictate the order: start computing $X_{21}$,
use that to compute $X_{11}$ and $X_{22}$ (possibly in parallel) and
finally compute $X_{12}$:
$$
\begin{array}{r@{~= V (}l@{,~}l@{,~}l@{)}}
 X_{21} & A_{22}     & Y_{21}                                & B_{11}\\
 X_{11} & A_{11}     & Y_{11}   + A_{12} X_{21}                & B_{11}\\
 X_{22} & A_{22}     & Y_{22}   + X_{21} B_{12}                & B_{22} \\
 X_{12} & A_{11}     & Y_{12}   + A_{12} X_{22} + X_{11} B_{12}  & B_{22}\\
\end{array}
$$
A graphical summary is shown in fig. \ref{fig:valiant}.

We proceed to certify this proof step in Agda.
The problem is to solve the equation \ensuremath{\Conid{L}}, defined as follows.
(We pick the name \ensuremath{\Conid{L}} for linear equation as we used \ensuremath{\Conid{Q}} for quadratic earlier.)
%
%
\restorecolumns[SemiNearRing2A]
\begin{hscode}\SaveRestoreHook
\column{B}{@{}>{\hspre}l<{\hspost}@{}}%
\column{3}{@{}>{\hspre}l<{\hspost}@{}}%
\column{16}{@{}>{\hspre}l<{\hspost}@{}}%
\column{19}{@{}>{\hspre}l<{\hspost}@{}}%
\column{23}{@{}>{\hspre}l<{\hspost}@{}}%
\column{33}{@{}>{\hspre}l<{\hspost}@{}}%
\column{37}{@{}>{\hspre}l<{\hspost}@{}}%
\column{51}{@{}>{\hspre}l<{\hspost}@{}}%
\column{E}{@{}>{\hspre}l<{\hspost}@{}}%
\>[3]{}\Conid{L}\;\!:\!\;\Varid{u}\;\Varid{→}\;\Varid{s}\;\Varid{→}\;\Varid{u}\;\Varid{→}\;\Varid{s}\;\Varid{→}\;\Conid{Set}{}\<[51]%
\>[51]{}\mbox{\onelinecomment  \structure{2.2}{Linear equation \ensuremath{\Conid{L}}}}{}\<[E]%
\\
\>[3]{}\Conid{L}\;\Varid{a}\;\Varid{y}\;\Varid{b}\;\Varid{x}\;\mathrel{=}\;{}\<[16]%
\>[16]{}\Varid{y}\;{}\<[19]%
\>[19]{}\mathbin{+_s}\;{}\<[23]%
\>[23]{}(\Varid{a}\;\mathbin{{_u}\!·_s}\;\Varid{x}\;{}\<[33]%
\>[33]{}\mathbin{+_s}\;{}\<[37]%
\>[37]{}\Varid{x}\;\mathbin{{_s}\!·_u}\;\Varid{b})\;\mathrel{≃_s}\;\Varid{x}{}\<[E]%
\ColumnHook
\end{hscode}\resethooks
%
%
We must prove that the relation \ensuremath{\Conid{L}} is entire, and thus get an
algorithm (to compute the last argument) as well as its correctness
proof.
To this end, we add the appropriate field to the \ensuremath{\Conid{ClosedSemiNearRing}}
record.
We now reap the full benefits of using a record structure, which frees
us from repeating the recursion pattern.
\restorecolumns[ClosedSemiNearRing]
\begin{hscode}\SaveRestoreHook
\column{B}{@{}>{\hspre}l<{\hspost}@{}}%
\column{3}{@{}>{\hspre}l<{\hspost}@{}}%
\column{5}{@{}>{\hspre}l<{\hspost}@{}}%
\column{16}{@{}>{\hspre}l<{\hspost}@{}}%
\column{35}{@{}>{\hspre}l<{\hspost}@{}}%
\column{51}{@{}>{\hspre}l<{\hspost}@{}}%
\column{E}{@{}>{\hspre}l<{\hspost}@{}}%
\>[3]{}\kw{field}\;{}\<[51]%
\>[51]{}\mbox{\onelinecomment  \structure{3.3}{Function for \ensuremath{\Conid{L}} and its correctness}}{}\<[E]%
\\
\>[3]{}\hsindent{2}{}\<[5]%
\>[5]{}\Varid{entireL}\;{}\<[16]%
\>[16]{}\!:\!\;\Conid{Entire3}\;\Conid{L}{}\<[E]%
\\[\blanklineskip]%
\>[3]{}\Varid{completion}\;\!:\!\;\Varid{u}\;\Varid{→}\;\Varid{s}\;\Varid{→}\;\Varid{u}\;\Varid{→}\;\Varid{s}{}\<[E]%
\\
\>[3]{}\Varid{completion}\;\mathrel{=}\;\Varid{fun3}\;\Varid{entireL}{}\<[E]%
\\[\blanklineskip]%
\>[3]{}\Varid{completionHasAll}\;\!:\!\;\Varid{∀}\;\{\mskip1.5mu \Varid{a}\;\Varid{y}\;\Varid{b}\mskip1.5mu\}\;\Varid{→}\;{}\<[35]%
\>[35]{}\Conid{L}\;\Varid{a}\;\Varid{y}\;\Varid{b}\;(\Varid{completion}\;\Varid{a}\;\Varid{y}\;\Varid{b}){}\<[E]%
\\
\>[3]{}\Varid{completionHasAll}\;\mathrel{=}\;\Varid{correct3}\;\Varid{entireL}{}\<[E]%
\ColumnHook
\end{hscode}\resethooks
Again, the bulk of the proof is the recursive case (part of the
definition of \ensuremath{\Conid{Square}}), the lifting of \ensuremath{\Varid{completionHasAll}} to 2-by-2
block matrices.
Here as well, the semi-formal proof is faithfully represented in Agda.
\restorecolumns[ClosedSemiNearRing]
\begin{hscode}\SaveRestoreHook
\column{B}{@{}>{\hspre}l<{\hspost}@{}}%
\column{3}{@{}>{\hspre}l<{\hspost}@{}}%
\column{5}{@{}>{\hspre}l<{\hspost}@{}}%
\column{14}{@{}>{\hspre}l<{\hspost}@{}}%
\column{51}{@{}>{\hspre}l<{\hspost}@{}}%
\column{E}{@{}>{\hspre}l<{\hspost}@{}}%
\>[51]{}\mbox{\onelinecomment  \structure{4.6}{Lifting the proof of \ensuremath{\Conid{L}}}}{}\<[E]%
\\
\>[3]{}\Varid{entireLS}\;\!:\!\;\Varid{∀}\;(\Conid{A}\;\!:\!\;\Conid{U})\;(\Conid{Y}\;\!:\!\;\Conid{S})\;(\Conid{B}\;\!:\!\;\Conid{U})\;\Varid{→}\;\Varid{∃}\;(\lambda \;\Conid{X}\;\Varid{→}\;\Conid{Y}\;\mathbin{+\!_S}\;(\Conid{A}\;\mathbin{{_U}·_S}\;\Conid{X}\;\mathbin{+\!_S}\;\Conid{X}\;\mathbin{{_S}·_U}\;\Conid{B})\;\mathrel{≃_S}\;\Conid{X}){}\<[E]%
\\
\>[3]{}\Varid{entireLS}\;\Conid{A}\;\Conid{Y}\;\Conid{B}\;\mathrel{=}\;\Conid{X}\;\Varid{,}\;\Varid{proof}\;\kw{where}{}\<[E]%
\\
\>[3]{}\hsindent{2}{}\<[5]%
\>[5]{}\Conid{X}\;\!:\!\;\Conid{S}{}\<[E]%
\\
\>[3]{}\hsindent{2}{}\<[5]%
\>[5]{}\Conid{X}\;\mathrel{=}\;\anonymous {}\<[14]%
\>[14]{}\mbox{\onelinecomment  filled in by unification with \ensuremath{\Conid{Y}\;\mathbin{+\!_S}\;(\Conid{U2S}\;\Conid{A}\;\mathbin{·\!_S}\;\Conid{X}\;\mathbin{+\!_S}\;\Conid{X}\;\mathbin{·\!_S}\;\Conid{U2S}\;\Conid{B})}}{}\<[E]%
\\
\>[3]{}\hsindent{2}{}\<[5]%
\>[5]{}\Varid{proof}\;\!:\!\;\Conid{Y}\;\mathbin{+\!_S}\;(\Conid{U2S}\;\Conid{A}\;\mathbin{·\!_S}\;\Conid{X}\;\mathbin{+\!_S}\;\Conid{X}\;\mathbin{·\!_S}\;\Conid{U2S}\;\Conid{B})\;\mathrel{≃_S}\;\Conid{X}{}\<[E]%
\\
\>[3]{}\hsindent{2}{}\<[5]%
\>[5]{}\kw{open}\;\Conid{EqReasoning}\;(\Conid{SemiNearRing2.sSetoid}\;\Conid{SNR2}){}\<[E]%
\\
\>[3]{}\hsindent{2}{}\<[5]%
\>[5]{}\Varid{proof}\;\mathrel{=}{}\<[14]%
\>[14]{}\mbox{\onelinecomment  continued below to fit the width of the paper (it is still in the \ensuremath{\kw{where}} clause)}{}\<[E]%
\ColumnHook
\end{hscode}\resethooks
\begin{hscode}\SaveRestoreHook
\column{B}{@{}>{\hspre}l<{\hspost}@{}}%
\column{7}{@{}>{\hspre}l<{\hspost}@{}}%
\column{9}{@{}>{\hspre}l<{\hspost}@{}}%
\column{13}{@{}>{\hspre}l<{\hspost}@{}}%
\column{14}{@{}>{\hspre}l<{\hspost}@{}}%
\column{17}{@{}>{\hspre}l<{\hspost}@{}}%
\column{21}{@{}>{\hspre}l<{\hspost}@{}}%
\column{27}{@{}>{\hspre}l<{\hspost}@{}}%
\column{31}{@{}>{\hspre}l<{\hspost}@{}}%
\column{35}{@{}>{\hspre}l<{\hspost}@{}}%
\column{41}{@{}>{\hspre}l<{\hspost}@{}}%
\column{E}{@{}>{\hspre}l<{\hspost}@{}}%
\>[7]{}\Varid{begin}\;{}\<[E]%
\\
\>[7]{}\hsindent{2}{}\<[9]%
\>[9]{}(\Conid{Y}\;{}\<[13]%
\>[13]{}\mathbin{+\!_S}\;{}\<[17]%
\>[17]{}(\Conid{A}\;\mathbin{{_U}·_S}\;\Conid{X}\;{}\<[27]%
\>[27]{}\mathbin{+\!_S}\;{}\<[31]%
\>[31]{}\Conid{X}\;\mathbin{{_S}·_U}\;\Conid{B})){}\<[E]%
\\
\>[7]{}≡\!\!⟨\;\Varid{refl}\;\Varid{⟩}\mbox{\onelinecomment  name the components}{}\<[E]%
\\
\>[7]{}\hsindent{2}{}\<[9]%
\>[9]{}\kw{let}\;{}\<[14]%
\>[14]{}\Varid{⟨}\;\Varid{a}_{00}\;\Varid{,}\;\Varid{a}_{01}\;\Varid{,•,}\;\Varid{a}_{11}\;\Varid{⟩}\;{}\<[41]%
\>[41]{}\mathrel{=}\;\Conid{A}{}\<[E]%
\\
\>[14]{}\Varid{⟨}\;\Varid{b}_{00}\;\Varid{,}\;\Varid{b}_{01}\;\Varid{,•,}\;\Varid{b}_{11}\;\Varid{⟩}\;{}\<[41]%
\>[41]{}\mathrel{=}\;\Conid{B}{}\<[E]%
\\
\>[14]{}\Varid{⟨}\;\Varid{y}_{00}\;\Varid{,}\;\Varid{y}_{01}\;\Varid{,}\;\Varid{y}_{10}\;\Varid{,}\;\Varid{y}_{11}\;\Varid{⟩}\;{}\<[41]%
\>[41]{}\mathrel{=}\;\Conid{Y}{}\<[E]%
\\
\>[14]{}\Varid{⟨}\;\Varid{x}_{00}\;\Varid{,}\;\Varid{x}_{01}\;\Varid{,}\;\Varid{x}_{10}\;\Varid{,}\;\Varid{x}_{11}\;\Varid{⟩}\;{}\<[41]%
\>[41]{}\mathrel{=}\;\Conid{X}{}\<[E]%
\\
\>[7]{}\hsindent{2}{}\<[9]%
\>[9]{}\kw{in}\;{}\<[14]%
\>[14]{}\Conid{Y}\;{}\<[17]%
\>[17]{}\mathbin{+\!_S}\;{}\<[21]%
\>[21]{}(\Conid{A}\;\mathbin{{_U}·_S}\;\Conid{X}\;{}\<[31]%
\>[31]{}\mathbin{+\!_S}\;{}\<[35]%
\>[35]{}\Conid{X}\;\mathbin{{_S}·_U}\;\Conid{B}){}\<[E]%
\ColumnHook
\end{hscode}\resethooks
\begin{hscode}\SaveRestoreHook
\column{B}{@{}>{\hspre}l<{\hspost}@{}}%
\column{7}{@{}>{\hspre}l<{\hspost}@{}}%
\column{9}{@{}>{\hspre}l<{\hspost}@{}}%
\column{14}{@{}>{\hspre}l<{\hspost}@{}}%
\column{23}{@{}>{\hspre}l<{\hspost}@{}}%
\column{26}{@{}>{\hspre}l<{\hspost}@{}}%
\column{27}{@{}>{\hspre}l<{\hspost}@{}}%
\column{31}{@{}>{\hspre}l<{\hspost}@{}}%
\column{43}{@{}>{\hspre}l<{\hspost}@{}}%
\column{48}{@{}>{\hspre}l<{\hspost}@{}}%
\column{53}{@{}>{\hspre}l<{\hspost}@{}}%
\column{62}{@{}>{\hspre}l<{\hspost}@{}}%
\column{77}{@{}>{\hspre}l<{\hspost}@{}}%
\column{86}{@{}>{\hspre}l<{\hspost}@{}}%
\column{87}{@{}>{\hspre}l<{\hspost}@{}}%
\column{E}{@{}>{\hspre}l<{\hspost}@{}}%
\>[7]{}≡\!\!⟨\;\Varid{refl}\;\Varid{⟩}\mbox{\onelinecomment  expand \ensuremath{\mathbin{{_U}·_S}} and \ensuremath{\mathbin{{_S}·_U}} and use components}{}\<[E]%
\\
\>[7]{}\hsindent{2}{}\<[9]%
\>[9]{}\kw{let}\;{}\<[14]%
\>[14]{}\Conid{A}\Varid{·}\Conid{X}\;\mathrel{=}\;{}\<[23]%
\>[23]{}\Varid{⟨}\;{}\<[26]%
\>[26]{}\Varid{a}_{00}\;\mathbin{{_u}\!·_s}\;\Varid{x}_{00}\;{}\<[43]%
\>[43]{}\mathbin{+_s}\;{}\<[48]%
\>[48]{}\Varid{a}_{01}\;{}\<[53]%
\>[53]{}\mathbin{·_s}\;\Varid{x}_{10}\;{}\<[62]%
\>[62]{}\Varid{,}\;\Varid{a}_{00}\;\mathbin{{_u}\!·_s}\;\Varid{x}_{01}\;{}\<[77]%
\>[77]{}\mathbin{+_s}\;\Varid{a}_{01}\;{}\<[87]%
\>[87]{}\mathbin{·_s}\;\Varid{x}_{11}\;{}\<[E]%
\\
\>[23]{}\Varid{,}\;{}\<[26]%
\>[26]{}0_s\;\mathbin{·_s}\;\Varid{x}_{00}\;{}\<[43]%
\>[43]{}\mathbin{+_s}\;{}\<[48]%
\>[48]{}\Varid{a}_{11}\;\mathbin{{_u}\!·_s}\;\Varid{x}_{10}\;{}\<[62]%
\>[62]{}\Varid{,}\;0_s\;\mathbin{·_s}\;\Varid{x}_{01}\;{}\<[77]%
\>[77]{}\mathbin{+_s}\;\Varid{a}_{11}\;{}\<[86]%
\>[86]{}\mathbin{{_u}\!·_s}\;\Varid{x}_{11}\;\Varid{⟩}{}\<[E]%
\\
\>[14]{}\Conid{X}\Varid{·}\Conid{B}\;\mathrel{=}\;{}\<[23]%
\>[23]{}\Varid{⟨}\;{}\<[26]%
\>[26]{}\Varid{x}_{00}\;\mathbin{{_s}\!·_u}\;\Varid{b}_{00}\;{}\<[43]%
\>[43]{}\mathbin{+_s}\;{}\<[48]%
\>[48]{}\Varid{x}_{01}\;\mathbin{·_s}\;0_s\;{}\<[62]%
\>[62]{}\Varid{,}\;\Varid{x}_{00}\;\mathbin{·_s}\;\Varid{b}_{01}\;{}\<[77]%
\>[77]{}\mathbin{+_s}\;\Varid{x}_{01}\;{}\<[86]%
\>[86]{}\mathbin{{_s}\!·_u}\;\Varid{b}_{11}\;{}\<[E]%
\\
\>[23]{}\Varid{,}\;{}\<[26]%
\>[26]{}\Varid{x}_{10}\;\mathbin{{_s}\!·_u}\;\Varid{b}_{00}\;{}\<[43]%
\>[43]{}\mathbin{+_s}\;{}\<[48]%
\>[48]{}\Varid{x}_{11}\;\mathbin{·_s}\;0_s\;{}\<[62]%
\>[62]{}\Varid{,}\;\Varid{x}_{10}\;\mathbin{·_s}\;\Varid{b}_{01}\;{}\<[77]%
\>[77]{}\mathbin{+_s}\;\Varid{x}_{11}\;{}\<[86]%
\>[86]{}\mathbin{{_s}\!·_u}\;\Varid{b}_{11}\;\Varid{⟩}{}\<[E]%
\\
\>[7]{}\hsindent{2}{}\<[9]%
\>[9]{}\kw{in}\;{}\<[14]%
\>[14]{}\Conid{Y}\;\mathbin{+\!_S}\;(\Conid{A}\Varid{·}\Conid{X}\;{}\<[27]%
\>[27]{}\mathbin{+\!_S}\;{}\<[31]%
\>[31]{}\Conid{X}\Varid{·}\Conid{B}){}\<[E]%
\ColumnHook
\end{hscode}\resethooks
\begin{hscode}\SaveRestoreHook
\column{B}{@{}>{\hspre}l<{\hspost}@{}}%
\column{7}{@{}>{\hspre}l<{\hspost}@{}}%
\column{9}{@{}>{\hspre}l<{\hspost}@{}}%
\column{12}{@{}>{\hspre}l<{\hspost}@{}}%
\column{18}{@{}>{\hspre}l<{\hspost}@{}}%
\column{30}{@{}>{\hspre}l<{\hspost}@{}}%
\column{35}{@{}>{\hspre}l<{\hspost}@{}}%
\column{41}{@{}>{\hspre}l<{\hspost}@{}}%
\column{46}{@{}>{\hspre}l<{\hspost}@{}}%
\column{51}{@{}>{\hspre}l<{\hspost}@{}}%
\column{60}{@{}>{\hspre}l<{\hspost}@{}}%
\column{69}{@{}>{\hspre}l<{\hspost}@{}}%
\column{78}{@{}>{\hspre}l<{\hspost}@{}}%
\column{86}{@{}>{\hspre}l<{\hspost}@{}}%
\column{87}{@{}>{\hspre}l<{\hspost}@{}}%
\column{92}{@{}>{\hspre}l<{\hspost}@{}}%
\column{E}{@{}>{\hspre}l<{\hspost}@{}}%
\>[7]{}≡\!\!⟨\;\Varid{refl}\;\Varid{⟩}\;\mbox{\onelinecomment  Expand \ensuremath{\mathrel{≃_S}}, \ensuremath{\mathbin{+\!_S}} and collect components}{}\<[E]%
\\
\>[7]{}\hsindent{2}{}\<[9]%
\>[9]{}\Varid{⟨}\;{}\<[12]%
\>[12]{}\Varid{y}_{00}\;{}\<[18]%
\>[18]{}\mathbin{+_s}\;((\Varid{a}_{00}\;{}\<[30]%
\>[30]{}\mathbin{{_u}\!·_s}\;{}\<[35]%
\>[35]{}\Varid{x}_{00}\;{}\<[41]%
\>[41]{}\mathbin{+_s}\;{}\<[46]%
\>[46]{}\Varid{a}_{01}\;{}\<[51]%
\>[51]{}\mathbin{·_s}\;\Varid{x}_{10})\;{}\<[60]%
\>[60]{}\mathbin{+_s}\;(\Varid{x}_{00}\;\mathbin{{_s}\!·_u}\;\Varid{b}_{00}\;{}\<[78]%
\>[78]{}\mathbin{+_s}\;\Varid{x}_{01}\;{}\<[87]%
\>[87]{}\mathbin{·_s}\;{}\<[92]%
\>[92]{}0_s))\;{}\<[E]%
\\
\>[7]{}\hsindent{2}{}\<[9]%
\>[9]{}\Varid{,}\;{}\<[12]%
\>[12]{}\Varid{y}_{01}\;{}\<[18]%
\>[18]{}\mathbin{+_s}\;((\Varid{a}_{00}\;{}\<[30]%
\>[30]{}\mathbin{{_u}\!·_s}\;{}\<[35]%
\>[35]{}\Varid{x}_{01}\;{}\<[41]%
\>[41]{}\mathbin{+_s}\;{}\<[46]%
\>[46]{}\Varid{a}_{01}\;{}\<[51]%
\>[51]{}\mathbin{·_s}\;\Varid{x}_{11})\;{}\<[60]%
\>[60]{}\mathbin{+_s}\;(\Varid{x}_{00}\;{}\<[69]%
\>[69]{}\mathbin{·_s}\;\Varid{b}_{01}\;{}\<[78]%
\>[78]{}\mathbin{+_s}\;\Varid{x}_{01}\;{}\<[86]%
\>[86]{}\mathbin{{_s}\!·_u}\;{}\<[92]%
\>[92]{}\Varid{b}_{11}))\;{}\<[E]%
\\
\>[7]{}\hsindent{2}{}\<[9]%
\>[9]{}\Varid{,}\;{}\<[12]%
\>[12]{}\Varid{y}_{10}\;{}\<[18]%
\>[18]{}\mathbin{+_s}\;((0_s\;{}\<[30]%
\>[30]{}\mathbin{·_s}\;{}\<[35]%
\>[35]{}\Varid{x}_{00}\;{}\<[41]%
\>[41]{}\mathbin{+_s}\;{}\<[46]%
\>[46]{}\Varid{a}_{11}\;\mathbin{{_u}\!·_s}\;\Varid{x}_{10})\;{}\<[60]%
\>[60]{}\mathbin{+_s}\;(\Varid{x}_{10}\;\mathbin{{_s}\!·_u}\;\Varid{b}_{00}\;{}\<[78]%
\>[78]{}\mathbin{+_s}\;\Varid{x}_{11}\;{}\<[87]%
\>[87]{}\mathbin{·_s}\;{}\<[92]%
\>[92]{}0_s))\;{}\<[E]%
\\
\>[7]{}\hsindent{2}{}\<[9]%
\>[9]{}\Varid{,}\;{}\<[12]%
\>[12]{}\Varid{y}_{11}\;{}\<[18]%
\>[18]{}\mathbin{+_s}\;((0_s\;{}\<[30]%
\>[30]{}\mathbin{·_s}\;{}\<[35]%
\>[35]{}\Varid{x}_{01}\;{}\<[41]%
\>[41]{}\mathbin{+_s}\;{}\<[46]%
\>[46]{}\Varid{a}_{11}\;\mathbin{{_u}\!·_s}\;\Varid{x}_{11})\;{}\<[60]%
\>[60]{}\mathbin{+_s}\;(\Varid{x}_{10}\;{}\<[69]%
\>[69]{}\mathbin{·_s}\;\Varid{b}_{01}\;{}\<[78]%
\>[78]{}\mathbin{+_s}\;\Varid{x}_{11}\;{}\<[86]%
\>[86]{}\mathbin{{_s}\!·_u}\;{}\<[92]%
\>[92]{}\Varid{b}_{11}))\;\Varid{⟩}{}\<[E]%
\ColumnHook
\end{hscode}\resethooks
\begin{hscode}\SaveRestoreHook
\column{B}{@{}>{\hspre}l<{\hspost}@{}}%
\column{7}{@{}>{\hspre}l<{\hspost}@{}}%
\column{9}{@{}>{\hspre}l<{\hspost}@{}}%
\column{10}{@{}>{\hspre}l<{\hspost}@{}}%
\column{12}{@{}>{\hspre}l<{\hspost}@{}}%
\column{49}{@{}>{\hspre}l<{\hspost}@{}}%
\column{E}{@{}>{\hspre}l<{\hspost}@{}}%
\>[7]{}≈\!\!⟨\;\Varid{congS}\;\Varid{zeroLemma00}\;\Varid{zeroLemma01}\;\Varid{zeroLemma10}\;\Varid{zeroLemma11}\;\Varid{⟩}\;{}\<[E]%
\\
\>[7]{}\hsindent{3}{}\<[10]%
\>[10]{}\mbox{\onelinecomment  assoc. and comm. of +; zero absorption.}{}\<[E]%
\\
\>[7]{}\hsindent{2}{}\<[9]%
\>[9]{}\Varid{⟨}\;{}\<[12]%
\>[12]{}\Varid{y}_{00}\;\mathbin{+_s}\;\Varid{a}_{01}\;\mathbin{·_s}\;\Varid{x}_{10}\;\mathbin{+_s}\;{}\<[49]%
\>[49]{}(\Varid{a}_{00}\;\mathbin{{_u}\!·_s}\;\Varid{x}_{00}\;\mathbin{+_s}\;\Varid{x}_{00}\;\mathbin{{_s}\!·_u}\;\Varid{b}_{00})\;{}\<[E]%
\\
\>[7]{}\hsindent{2}{}\<[9]%
\>[9]{}\Varid{,}\;{}\<[12]%
\>[12]{}\Varid{y}_{01}\;\mathbin{+_s}\;\Varid{a}_{01}\;\mathbin{·_s}\;\Varid{x}_{11}\;\mathbin{+_s}\;\Varid{x}_{00}\;\mathbin{·_s}\;\Varid{b}_{01}\;\mathbin{+_s}\;{}\<[49]%
\>[49]{}(\Varid{a}_{00}\;\mathbin{{_u}\!·_s}\;\Varid{x}_{01}\;\mathbin{+_s}\;\Varid{x}_{01}\;\mathbin{{_s}\!·_u}\;\Varid{b}_{11})\;{}\<[E]%
\\
\>[7]{}\hsindent{2}{}\<[9]%
\>[9]{}\Varid{,}\;{}\<[12]%
\>[12]{}\Varid{y}_{10}\;\mathbin{+_s}\;{}\<[49]%
\>[49]{}(\Varid{a}_{11}\;\mathbin{{_u}\!·_s}\;\Varid{x}_{10}\;\mathbin{+_s}\;\Varid{x}_{10}\;\mathbin{{_s}\!·_u}\;\Varid{b}_{00})\;{}\<[E]%
\\
\>[7]{}\hsindent{2}{}\<[9]%
\>[9]{}\Varid{,}\;{}\<[12]%
\>[12]{}\Varid{y}_{11}\;\mathbin{+_s}\;\Varid{x}_{10}\;\mathbin{·_s}\;\Varid{b}_{01}\;\mathbin{+_s}\;{}\<[49]%
\>[49]{}(\Varid{a}_{11}\;\mathbin{{_u}\!·_s}\;\Varid{x}_{11}\;\mathbin{+_s}\;\Varid{x}_{11}\;\mathbin{{_s}\!·_u}\;\Varid{b}_{11})\;\Varid{⟩}{}\<[E]%
\ColumnHook
\end{hscode}\resethooks
\begin{hscode}\SaveRestoreHook
\column{B}{@{}>{\hspre}l<{\hspost}@{}}%
\column{7}{@{}>{\hspre}l<{\hspost}@{}}%
\column{9}{@{}>{\hspre}l<{\hspost}@{}}%
\column{12}{@{}>{\hspre}l<{\hspost}@{}}%
\column{17}{@{}>{\hspre}l<{\hspost}@{}}%
\column{35}{@{}>{\hspre}l<{\hspost}@{}}%
\column{69}{@{}>{\hspre}l<{\hspost}@{}}%
\column{E}{@{}>{\hspre}l<{\hspost}@{}}%
\>[7]{}≈\!\!⟨\;\Varid{congS}\;{}\<[17]%
\>[17]{}\Varid{completionHasAll}\;\Varid{completionHasAll}\;\Varid{completionHasAll}\;{}\<[69]%
\>[69]{}\Varid{completionHasAll}\;\Varid{⟩}\;{}\<[E]%
\\
\>[7]{}\hsindent{2}{}\<[9]%
\>[9]{}\Varid{⟨}\;{}\<[12]%
\>[12]{}\Varid{x}_{00}\;\Varid{,}\;\Varid{x}_{01}\;\Varid{,}\;\Varid{x}_{10}\;\Varid{,}\;\Varid{x}_{11}\;{}\<[35]%
\>[35]{}\Varid{⟩}{}\<[E]%
\\
\>[7]{}≡\!\!⟨\;\Varid{refl}\;\Varid{⟩}\;{}\<[E]%
\\
\>[7]{}\hsindent{2}{}\<[9]%
\>[9]{}\Conid{X}{}\<[E]%
\\
\>[7]{}\Varid{∎}{}\<[E]%
\ColumnHook
\end{hscode}\resethooks
The series of \ensuremath{\Varid{zeroLemma}}s use that zeros are absorbing, in addition to commutativity and associativity of addition.
One more time, the actual solution of the equation \ensuremath{\Conid{X}} can be inferred by Agda on the basis of the proof.
The base case is a mere application of the properties of zero.
\begin{hscode}\SaveRestoreHook
\column{B}{@{}>{\hspre}l<{\hspost}@{}}%
\column{12}{@{}>{\hspre}l<{\hspost}@{}}%
\column{51}{@{}>{\hspre}l<{\hspost}@{}}%
\column{E}{@{}>{\hspre}l<{\hspost}@{}}%
\>[12]{}\kw{open}\;\kw{import}\;\Conid{ZeroLemmas}\;\Varid{snr}{}\<[51]%
\>[51]{}\mbox{\onelinecomment  \structure{5.1}{Base case for \ensuremath{\Conid{L}}}}{}\<[E]%
\\
\>[12]{}\Varid{entireLBase}\;\!:\!\;(\Varid{a}\;\!:\!\;\Varid{⊤})\;(\Varid{y}\;\!:\!\;\Varid{s})\;(\Varid{b}\;\!:\!\;\Varid{⊤})\;\Varid{→}\;\Varid{∃}\;(\lambda \;\Varid{x}\;\Varid{→}\;\Varid{y}\;\mathbin{+_s}\;(\Varid{a}\;\mathbin{{_u}\!·_s}\;\Varid{x}\;\mathbin{+_s}\;\Varid{x}\;\mathbin{{_s}\!·_u}\;\Varid{b})\;\mathrel{≃_s}\;\Varid{x}){}\<[E]%
\\
\>[12]{}\Varid{entireLBase}\;\Varid{tt}\;\Varid{y}\;\Varid{tt}\;\mathrel{=}\;\Varid{y}\;\Varid{,}\;\Varid{zeroˡʳLemma}\;\Varid{y}\;\Varid{y}\;\Varid{y}{}\<[E]%
\ColumnHook
\end{hscode}\resethooks
We have now completely specified Valiant's algorithm, and it can be
accessed via the appropriate field of \ensuremath{\Conid{Mat}\;\Varid{n}}:
\begin{hscode}\SaveRestoreHook
\column{B}{@{}>{\hspre}l<{\hspost}@{}}%
\column{3}{@{}>{\hspre}l<{\hspost}@{}}%
\column{51}{@{}>{\hspre}l<{\hspost}@{}}%
\column{E}{@{}>{\hspre}l<{\hspost}@{}}%
\>[51]{}\mbox{\onelinecomment  \structure{6.1}{Top level algorithm extraction}}{}\<[E]%
\\
\>[3]{}\Varid{valiantAlgorithm}\;\!:\!\;(\Varid{el}\;\!:\!\;\Conid{SemiNearRing})\;\Varid{→}\;\Varid{∀}\;\Varid{n}\;\Varid{→}\;\Conid{Upper}\;\Varid{n}\;\Varid{el}\;\Varid{→}\;\Conid{Upper}\;\Varid{n}\;\Varid{el}{}\<[E]%
\\
\>[3]{}\Varid{valiantAlgorithm}\;\Varid{el}\;\Varid{n}\;\Varid{u}\;\mathrel{=}\;\Conid{ClosedSemiNearRing.closure}\;(\Conid{Mat}\;\Varid{n}\;\Varid{el})\;\Varid{u}{}\<[E]%
\ColumnHook
\end{hscode}\resethooks

\section{smallest}\label{sec:smallest}

Only one bit of proof remains to obtain full correctness: namely, we
should prove that the solution computed by our algorithm is a lower
bound of all solutions of the \ensuremath{\Conid{Q}} equation.
The proof is as follows. We have three lemmas. The first lemma is that the
\ensuremath{\Conid{L}} equation (recall \ensuremath{\Conid{L}\;\Varid{a}\;\Varid{y}\;\Varid{b}\;\Varid{x}\;\mathrel{=}\;\Varid{y}\;\mathbin{+_s}\;(\Varid{a}\;\mathbin{{_u}\!·_s}\;\Varid{x}\;\mathbin{+_s}\;\Varid{x}\;\mathbin{{_s}\!·_u}\;\Varid{b})\;\mathrel{≃_s}\;\Varid{x}})
admits a single solution for \ensuremath{\Varid{x}}. (This is not surprising, as it is
a linear equation.) Being unique, this solution is thus necessarily the
smallest.
The second lemma is that the \ensuremath{\Conid{L}} relation is a congruence in its second argument.
The third lemma is that the \ensuremath{\Varid{completion}} function is monotonous.
The theorem (lower bound) and the two lemmas are proved by induction, as before.


We start by stating the two first lemmas:
1. the relation \ensuremath{\Conid{L}} admits a unique solution in its last argument and
2. the relation \ensuremath{\Conid{L}} is a congruence in its second argument.
\restorecolumns[SemiNearRing2]
\begin{hscode}\SaveRestoreHook
\column{B}{@{}>{\hspre}l<{\hspost}@{}}%
\column{3}{@{}>{\hspre}l<{\hspost}@{}}%
\column{12}{@{}>{\hspre}l<{\hspost}@{}}%
\column{51}{@{}>{\hspre}l<{\hspost}@{}}%
\column{E}{@{}>{\hspre}l<{\hspost}@{}}%
\>[51]{}\mbox{\onelinecomment  \structure{2.3}{Properties of \ensuremath{\Conid{L}}}}{}\<[E]%
\\
\>[3]{}\Conid{UniqueL}\;{}\<[12]%
\>[12]{}\mathrel{=}\;\Varid{∀}\;\{\mskip1.5mu \Varid{a}\;\Varid{y}\;\Varid{b}\mskip1.5mu\}\;\Varid{→}\;\Conid{UniqueSolution}\;\unopun{\mathrel{≃_s}}\;(\Conid{L}\;\Varid{a}\;\Varid{y}\;\Varid{b}){}\<[E]%
\\
\>[3]{}\Conid{CongL}\;{}\<[12]%
\>[12]{}\mathrel{=}\;\Varid{∀}\;\{\mskip1.5mu \Varid{a}\;\Varid{x}\;\Varid{b}\mskip1.5mu\}\;\to \;\Varid{∀}\;\{\mskip1.5mu \Varid{y}\;\Varid{y'}\mskip1.5mu\}\;\to \;\Varid{y}\;\mathrel{≃_s}\;\Varid{y'}\;\to \;\Conid{L}\;\Varid{a}\;\Varid{y}\;\Varid{b}\;\Varid{x}\;\to \;\Conid{L}\;\Varid{a}\;\Varid{y'}\;\Varid{b}\;\Varid{x}{}\<[E]%
\ColumnHook
\end{hscode}\resethooks
%
We then prove those two lemmas, as well as monotonicity of
\ensuremath{\Varid{completion}} and the main theorem of this section (\ensuremath{\Varid{closureIsLeast}}),
and we do so by induction on the matrix structure.
Formally, we proceed as before: the induction pattern is encoded by
adding fields to our \ensuremath{\Conid{ClosedSemiNearRing}} record. The fields to add
are as follows:
\restorecolumns[ClosedSemiNearRing]
\begin{hscode}\SaveRestoreHook
\column{B}{@{}>{\hspre}l<{\hspost}@{}}%
\column{3}{@{}>{\hspre}l<{\hspost}@{}}%
\column{5}{@{}>{\hspre}l<{\hspost}@{}}%
\column{21}{@{}>{\hspre}l<{\hspost}@{}}%
\column{46}{@{}>{\hspre}l<{\hspost}@{}}%
\column{51}{@{}>{\hspre}l<{\hspost}@{}}%
\column{55}{@{}>{\hspre}l<{\hspost}@{}}%
\column{58}{@{}>{\hspre}l<{\hspost}@{}}%
\column{67}{@{}>{\hspre}l<{\hspost}@{}}%
\column{70}{@{}>{\hspre}l<{\hspost}@{}}%
\column{79}{@{}>{\hspre}l<{\hspost}@{}}%
\column{E}{@{}>{\hspre}l<{\hspost}@{}}%
\>[3]{}\kw{field}\;{}\<[51]%
\>[51]{}\mbox{\onelinecomment  \structure{3.4}{Ordering properties of \ensuremath{\Conid{L}} and \ensuremath{\Conid{Q}}}}{}\<[E]%
\\
\>[3]{}\hsindent{2}{}\<[5]%
\>[5]{}\Varid{uniqueL}\;{}\<[21]%
\>[21]{}\!:\!\;\Conid{UniqueL}{}\<[E]%
\\
\>[3]{}\hsindent{2}{}\<[5]%
\>[5]{}\Varid{congL}\;{}\<[21]%
\>[21]{}\!:\!\;\Conid{CongL}{}\<[E]%
\\
\>[3]{}\hsindent{2}{}\<[5]%
\>[5]{}\Varid{completionMono}\;{}\<[21]%
\>[21]{}\!:\!\;\Varid{∀}\;\{\mskip1.5mu \Varid{a}\;\Varid{a'}\;\Varid{y}\;\Varid{y'}\;\Varid{b}\;\Varid{b'}\mskip1.5mu\}\;\Varid{→}\;{}\<[46]%
\>[46]{}\Varid{a}\;\mathrel{≤_u}\;\Varid{a'}\;{}\<[55]%
\>[55]{}\Varid{→}\;{}\<[58]%
\>[58]{}\Varid{y}\;\mathrel{≤_s}\;\Varid{y'}\;{}\<[67]%
\>[67]{}\Varid{→}\;{}\<[70]%
\>[70]{}\Varid{b}\;\mathrel{≤_u}\;\Varid{b'}\;{}\<[79]%
\>[79]{}\Varid{→}\;{}\<[E]%
\\
\>[46]{}\Varid{completion}\;\Varid{a}\;\Varid{y}\;\Varid{b}\;\mathrel{≤_s}\;\Varid{completion}\;\Varid{a'}\;\Varid{y'}\;\Varid{b'}{}\<[E]%
\\
\>[3]{}\hsindent{2}{}\<[5]%
\>[5]{}\Varid{closureIsLeast}\;{}\<[21]%
\>[21]{}\!:\!\;\{\mskip1.5mu \Varid{w}\;\!:\!\;\Varid{u}\mskip1.5mu\}\;\to \;\Conid{LowerBound}\;\unopun{\mathrel{≤_u}}\;(\Conid{Q}\;\Varid{w})\;(\Varid{closure}\;\Varid{w}){}\<[E]%
\ColumnHook
\end{hscode}\resethooks
\noindent
We then formalize our above remark: the uniqueness of \ensuremath{\Conid{L}} immediately
implies that \ensuremath{\Varid{completion}} gives a least solution. We do this in the
\ensuremath{\Conid{ClosedSemiNearRing}} structure, in order to get this result for every
induction step.
\restorecolumns[ClosedSemiNearRing]
\begin{hscode}\SaveRestoreHook
\column{B}{@{}>{\hspre}l<{\hspost}@{}}%
\column{3}{@{}>{\hspre}l<{\hspost}@{}}%
\column{E}{@{}>{\hspre}l<{\hspost}@{}}%
\>[3]{}\kw{open}\;\Conid{OrderLemmas}\;\Varid{snr}\;\kw{public}{}\<[E]%
\\[\blanklineskip]%
\>[3]{}\Varid{completionIsLeast}\;\!:\!\;\Varid{∀}\;(\Varid{a}\;\!:\!\;\Varid{u})\;(\Varid{y}\;\!:\!\;\Varid{s})\;(\Varid{b}\;\!:\!\;\Varid{u})\;\to \;\Conid{LowerBound}_s\;(\Conid{L}\;\Varid{a}\;\Varid{y}\;\Varid{b})\;(\Varid{completion}\;\Varid{a}\;\Varid{y}\;\Varid{b}){}\<[E]%
\\
\>[3]{}\Varid{completionIsLeast}\;\Varid{a}\;\Varid{y}\;\Varid{b}\;\Varid{z}\;\Varid{p}\;\mathrel{=}\;{≃}\Varid{sTo}{≤}\Varid{s}\;(\Varid{uniqueL}\;\Varid{completionHasAll}\;\Varid{p}){}\<[E]%
\ColumnHook
\end{hscode}\resethooks
We then proceed with the induction proofs.
First, we prove the induction case of our main theorem, from the induction hypotheses.
As usual, the upper-right corner is the difficult case, and requires
the monotonicity lemma.
\restorecolumns[ClosedSemiNearRing]
\begin{hscode}\SaveRestoreHook
\column{B}{@{}>{\hspre}l<{\hspost}@{}}%
\column{3}{@{}>{\hspre}l<{\hspost}@{}}%
\column{5}{@{}>{\hspre}l<{\hspost}@{}}%
\column{9}{@{}>{\hspre}l<{\hspost}@{}}%
\column{10}{@{}>{\hspre}l<{\hspost}@{}}%
\column{12}{@{}>{\hspre}l<{\hspost}@{}}%
\column{17}{@{}>{\hspre}l<{\hspost}@{}}%
\column{19}{@{}>{\hspre}l<{\hspost}@{}}%
\column{20}{@{}>{\hspre}l<{\hspost}@{}}%
\column{24}{@{}>{\hspre}l<{\hspost}@{}}%
\column{30}{@{}>{\hspre}l<{\hspost}@{}}%
\column{31}{@{}>{\hspre}l<{\hspost}@{}}%
\column{38}{@{}>{\hspre}l<{\hspost}@{}}%
\column{47}{@{}>{\hspre}l<{\hspost}@{}}%
\column{51}{@{}>{\hspre}l<{\hspost}@{}}%
\column{E}{@{}>{\hspre}l<{\hspost}@{}}%
\>[3]{}\unopun{\mathrel{≤_U}}\;\mathrel{=}\;\Conid{SemiNearRing2.}\!\unopun{\mathrel{≤_u}}\;{}\<[30]%
\>[30]{}\Conid{SNR2}{}\<[51]%
\>[51]{}\mbox{\onelinecomment  \structure{4.5}{Lifting orders and their properties}}{}\<[E]%
\\
\>[3]{}\unopun{\mathrel{≤_S}}\;\mathrel{=}\;\Conid{SemiNearRing2.}\!\unopun{\mathrel{≤_s}}\;{}\<[30]%
\>[30]{}\Conid{SNR2}{}\<[E]%
\\
\>[3]{}\Varid{closureIsLeastS}\;\!:\!\;\Varid{∀}\;\{\mskip1.5mu \Conid{W}\mskip1.5mu\}\;\to \;\Conid{LowerBound}\;\unopun{\mathrel{≤_U}}\;(\Conid{QU}\;\Conid{W})\;(\Varid{fun}\;\Varid{entireQStep}\;\Conid{W}){}\<[E]%
\\
\>[3]{}\Varid{closureIsLeastS}\;{}\<[20]%
\>[20]{}\Conid{Z}\;{}\<[24]%
\>[24]{}\Conid{QUWZ}\;\mathrel{=}{}\<[E]%
\\
\>[3]{}\hsindent{2}{}\<[5]%
\>[5]{}\kw{let}\;{}\<[10]%
\>[10]{}\Varid{⟨}\;\Varid{z}_{00}\;{}\<[17]%
\>[17]{}\Varid{,}\;\Varid{z}_{01}\;\Varid{,•,}\;{}\<[31]%
\>[31]{}\Varid{z}_{11}\;\Varid{⟩}\;{}\<[38]%
\>[38]{}\mathrel{=}\;\Conid{Z}{}\<[47]%
\>[47]{}\mbox{\onelinecomment  every matrix \ensuremath{\Conid{Z}}}{}\<[E]%
\\
\>[10]{}(\Varid{p}_{00}\;{}\<[17]%
\>[17]{}\Varid{,}\;\Varid{p}_{01}\;\Varid{,}\;\anonymous \;\Varid{,}\;{}\<[31]%
\>[31]{}\Varid{p}_{11})\;{}\<[38]%
\>[38]{}\mathrel{=}\;\Conid{QUWZ}{}\<[47]%
\>[47]{}\mbox{\onelinecomment  which satisfies \ensuremath{(\Conid{QU}\;\Conid{W}\;\Conid{Z})}}{}\<[E]%
\\
\>[10]{}\Varid{q}_{10}\;{}\<[19]%
\>[19]{}\mathrel{=}\;\Varid{identityˡ}_s\;0_s{}\<[E]%
\\
\>[10]{}\Varid{q}_{00}\;{}\<[19]%
\>[19]{}\mathrel{=}\;\Varid{closureIsLeast}\;\Varid{z}_{00}\;\Varid{p}_{00}{}\<[47]%
\>[47]{}\mbox{\onelinecomment  is bigger than \ensuremath{\Conid{C}\;\mathrel{=}\;\Varid{fun}\;\Varid{entireQStep}\;\Conid{W}}}{}\<[E]%
\\
\>[10]{}\Varid{q}_{11}\;{}\<[19]%
\>[19]{}\mathrel{=}\;\Varid{closureIsLeast}\;\Varid{z}_{11}\;\Varid{p}_{11}{}\<[E]%
\\
\>[10]{}\Varid{vs01}\;{}\<[19]%
\>[19]{}\mathrel{=}\;\Varid{completionIsLeast}\;\anonymous \;\anonymous \;\anonymous \;\Varid{z}_{01}\;\Varid{p}_{01}{}\<[E]%
\\
\>[10]{}\Varid{mono01}\;{}\<[19]%
\>[19]{}\mathrel{=}\;\Varid{completionMono}\;\Varid{q}_{00}\;({≃}\Varid{sTo}{≤}\Varid{s}\;\Varid{refl}_s)\;\Varid{q}_{11}{}\<[E]%
\\
\>[3]{}\hsindent{2}{}\<[5]%
\>[5]{}\kw{in}\;{}\<[9]%
\>[9]{}({}\<[12]%
\>[12]{}\Varid{q}_{00}\;{}\<[17]%
\>[17]{}\Varid{,}\;\mathrel{≤}{\kern -4pt}\Varid{-trans}_s\;\Varid{mono01}\;\Varid{vs01}{}\<[E]%
\\
\>[9]{}\Varid{,}\;{}\<[12]%
\>[12]{}\Varid{q}_{10}\;{}\<[17]%
\>[17]{}\Varid{,}\;\Varid{q}_{11}){}\<[E]%
\ColumnHook
\end{hscode}\resethooks
We can then prove the induction case of the uniqueness of \ensuremath{\Conid{L}}. The proof uses the induction
hypotheses, replicating the structure derived in the previous
section. The only slight difficulty is to use the identity of \ensuremath{\Varid{0}} at
the appropriate places.

\begin{hscode}\SaveRestoreHook
\column{B}{@{}>{\hspre}l<{\hspost}@{}}%
\column{3}{@{}>{\hspre}l<{\hspost}@{}}%
\column{6}{@{}>{\hspre}l<{\hspost}@{}}%
\column{14}{@{}>{\hspre}l<{\hspost}@{}}%
\column{27}{@{}>{\hspre}l<{\hspost}@{}}%
\column{51}{@{}>{\hspre}l<{\hspost}@{}}%
\column{55}{@{}>{\hspre}l<{\hspost}@{}}%
\column{61}{@{}>{\hspre}l<{\hspost}@{}}%
\column{63}{@{}>{\hspre}l<{\hspost}@{}}%
\column{E}{@{}>{\hspre}l<{\hspost}@{}}%
\>[3]{}\Varid{uniqueLS}\;\!:\!\;\Conid{SemiNearRing2.UniqueL}\;\Conid{SNR2}{}\<[51]%
\>[51]{}\mbox{\onelinecomment  \structure{4.7}{Proofs for ordering \ensuremath{\Conid{L}}-solutions}}{}\<[E]%
\\
\>[3]{}\Varid{uniqueLS}\;(\Varid{p}_{00}\;\Varid{,}\;\Varid{p}_{01}\;\Varid{,}\;\Varid{p}_{10}\;\Varid{,}\;\Varid{p}_{11})\;(\Varid{q}_{00}\;\Varid{,}\;\Varid{q}_{01}\;\Varid{,}\;\Varid{q}_{10}\;\Varid{,}\;\Varid{q}_{11})\;\mathrel{=}\;{}\<[63]%
\>[63]{}\Varid{eq00}\;\Varid{,}\;\Varid{eq01}\;\Varid{,}\;\Varid{eq10}\;\Varid{,}\;\Varid{eq11}{}\<[E]%
\\
\>[3]{}\hsindent{3}{}\<[6]%
\>[6]{}\kw{where}\;\kw{mutual}{}\<[E]%
\\
\>[6]{}\hsindent{8}{}\<[14]%
\>[14]{}\Varid{s}_{00}\;\mathrel{=}\;\Varid{congL}\;(\Varid{refl}_s\;\mathbin{<\!\!\!+\!\!\!>}\;(\Varid{refl}_s\;\mathbin{<\!\!\!·\!\!\!>}\;\Varid{sym}\;\Varid{eq10})){}\<[E]%
\\
\>[6]{}\hsindent{8}{}\<[14]%
\>[14]{}\Varid{s}_{11}\;\mathrel{=}\;\Varid{congL}\;(\Varid{refl}_s\;\mathbin{<\!\!\!+\!\!\!>}\;(\Varid{sym}\;\Varid{eq10}\;\mathbin{<\!\!\!·\!\!\!>}\;\Varid{refl}_s)){}\<[E]%
\\
\>[6]{}\hsindent{8}{}\<[14]%
\>[14]{}\Varid{s}_{01}\;\mathrel{=}\;\Varid{congL}\;((\Varid{refl}_s\;\mathbin{<\!\!\!+\!\!\!>}\;(\Varid{refl}_s\;\mathbin{<\!\!\!·\!\!\!>}\;\Varid{sym}\;\Varid{eq11}))\;\mathbin{<\!\!\!+\!\!\!>}\;(\Varid{sym}\;\Varid{eq00}\;\mathbin{<\!\!\!·\!\!\!>}\;\Varid{refl}_s)){}\<[E]%
\\
\>[6]{}\hsindent{8}{}\<[14]%
\>[14]{}\Varid{r}_{10}\;\mathrel{=}\;{}\<[27]%
\>[27]{}\Varid{trans}\;(\Varid{sym}\;\Varid{zeroLemma10})\;\Varid{q}_{10}{}\<[E]%
\\
\>[6]{}\hsindent{8}{}\<[14]%
\>[14]{}\Varid{r}_{00}\;\mathrel{=}\;\Varid{s}_{00}\;({}\<[27]%
\>[27]{}\Varid{trans}\;(\Varid{sym}\;\Varid{zeroLemma00})\;\Varid{q}_{00}){}\<[E]%
\\
\>[6]{}\hsindent{8}{}\<[14]%
\>[14]{}\Varid{r}_{11}\;\mathrel{=}\;\Varid{s}_{11}\;({}\<[27]%
\>[27]{}\Varid{trans}\;(\Varid{sym}\;\Varid{zeroLemma11})\;\Varid{q}_{11}){}\<[E]%
\\
\>[6]{}\hsindent{8}{}\<[14]%
\>[14]{}\Varid{r}_{01}\;\mathrel{=}\;\Varid{s}_{01}\;({}\<[27]%
\>[27]{}\Varid{trans}\;(\Varid{sym}\;\Varid{zeroLemma01})\;\Varid{q}_{01}){}\<[E]%
\\
\>[6]{}\hsindent{8}{}\<[14]%
\>[14]{}\Varid{eq10}\;\mathrel{=}\;\Varid{uniqueL}\;(\Varid{trans}\;(\Varid{sym}\;\Varid{zeroLemma10})\;{}\<[55]%
\>[55]{}\Varid{p}_{10})\;{}\<[61]%
\>[61]{}\Varid{r}_{10}{}\<[E]%
\\
\>[6]{}\hsindent{8}{}\<[14]%
\>[14]{}\Varid{eq00}\;\mathrel{=}\;\Varid{uniqueL}\;(\Varid{trans}\;(\Varid{sym}\;\Varid{zeroLemma00})\;{}\<[55]%
\>[55]{}\Varid{p}_{00})\;{}\<[61]%
\>[61]{}\Varid{r}_{00}{}\<[E]%
\\
\>[6]{}\hsindent{8}{}\<[14]%
\>[14]{}\Varid{eq11}\;\mathrel{=}\;\Varid{uniqueL}\;(\Varid{trans}\;(\Varid{sym}\;\Varid{zeroLemma11})\;{}\<[55]%
\>[55]{}\Varid{p}_{11})\;{}\<[61]%
\>[61]{}\Varid{r}_{11}{}\<[E]%
\\
\>[6]{}\hsindent{8}{}\<[14]%
\>[14]{}\Varid{eq01}\;\mathrel{=}\;\Varid{uniqueL}\;(\Varid{trans}\;(\Varid{sym}\;\Varid{zeroLemma01})\;{}\<[55]%
\>[55]{}\Varid{p}_{01})\;{}\<[61]%
\>[61]{}\Varid{r}_{01}{}\<[E]%
\ColumnHook
\end{hscode}\resethooks
Completion monotonicity (induction case) is also straightforward, as is the base case.
(The operators \ensuremath{\Varid{[+]}} and \ensuremath{\Varid{[*]}} are monotonicity for sum and product.)
\begin{hscode}\SaveRestoreHook
\column{B}{@{}>{\hspre}l<{\hspost}@{}}%
\column{3}{@{}>{\hspre}l<{\hspost}@{}}%
\column{6}{@{}>{\hspre}l<{\hspost}@{}}%
\column{13}{@{}>{\hspre}l<{\hspost}@{}}%
\column{20}{@{}>{\hspre}l<{\hspost}@{}}%
\column{22}{@{}>{\hspre}l<{\hspost}@{}}%
\column{39}{@{}>{\hspre}l<{\hspost}@{}}%
\column{52}{@{}>{\hspre}l<{\hspost}@{}}%
\column{55}{@{}>{\hspre}l<{\hspost}@{}}%
\column{64}{@{}>{\hspre}l<{\hspost}@{}}%
\column{67}{@{}>{\hspre}l<{\hspost}@{}}%
\column{76}{@{}>{\hspre}l<{\hspost}@{}}%
\column{E}{@{}>{\hspre}l<{\hspost}@{}}%
\>[3]{}\Varid{completionMonoS}\;\!:\!\;{}\<[22]%
\>[22]{}\Varid{∀}\;\{\mskip1.5mu \Varid{a}\;\Varid{a'}\;\Varid{y}\;\Varid{y'}\;\Varid{b}\;\Varid{b'}\mskip1.5mu\}\;\Varid{→}\;\Varid{a}\;\mathrel{≤_U}\;\Varid{a'}\;{}\<[52]%
\>[52]{}\Varid{→}\;{}\<[55]%
\>[55]{}\Varid{y}\;\mathrel{≤_S}\;\Varid{y'}\;{}\<[64]%
\>[64]{}\Varid{→}\;{}\<[67]%
\>[67]{}\Varid{b}\;\mathrel{≤_U}\;\Varid{b'}\;{}\<[76]%
\>[76]{}\Varid{→}\;{}\<[E]%
\\
\>[22]{}\Varid{proj₁}\;(\Varid{entireLS}\;\Varid{a}\;\Varid{y}\;\Varid{b})\;\mathrel{≤_S}\;\Varid{proj₁}\;(\Varid{entireLS}\;\Varid{a'}\;\Varid{y'}\;\Varid{b'}){}\<[E]%
\\
\>[3]{}\Varid{completionMonoS}\;{}\<[20]%
\>[20]{}(\Varid{p}_{00}\;\Varid{,}\;\Varid{p}_{01}\;\Varid{,}\;\Varid{p}_{10}\;\Varid{,}\;\Varid{p}_{11})\;(\Varid{q}_{00}\;\Varid{,}\;\Varid{q}_{01}\;\Varid{,}\;\Varid{q}_{10}\;\Varid{,}\;\Varid{q}_{11})\;{}\<[E]%
\\
\>[20]{}(\Varid{r}_{00}\;\Varid{,}\;\Varid{r}_{01}\;\Varid{,}\;\Varid{r}_{10}\;\Varid{,}\;\Varid{r}_{11})\;\mathrel{=}\;\Varid{m}_{00}\;\Varid{,}\;\Varid{m}_{01}\;\Varid{,}\;\Varid{m}_{10}\;\Varid{,}\;\Varid{m}_{11}{}\<[E]%
\\
\>[3]{}\hsindent{3}{}\<[6]%
\>[6]{}\kw{where}\;{}\<[13]%
\>[13]{}\Varid{m}_{10}\;\mathrel{=}\;\Varid{completionMono}\;\Varid{p}_{11}\;{}\<[39]%
\>[39]{}\Varid{q}_{10}\;\Varid{r}_{00}{}\<[E]%
\\
\>[13]{}\Varid{m}_{00}\;\mathrel{=}\;\Varid{completionMono}\;\Varid{p}_{00}\;{}\<[39]%
\>[39]{}(\Varid{q}_{00}\;\Varid{[+]}\;\Varid{p}_{01}\;\Varid{[*]}\;\Varid{m}_{10})\;\Varid{r}_{00}{}\<[E]%
\\
\>[13]{}\Varid{m}_{11}\;\mathrel{=}\;\Varid{completionMono}\;\Varid{p}_{11}\;{}\<[39]%
\>[39]{}(\Varid{q}_{11}\;\Varid{[+]}\;\Varid{m}_{10}\;\Varid{[*]}\;\Varid{r}_{01})\;\Varid{r}_{11}{}\<[E]%
\\
\>[13]{}\Varid{m}_{01}\;\mathrel{=}\;\Varid{completionMono}\;\Varid{p}_{00}\;{}\<[39]%
\>[39]{}(\Varid{q}_{01}\;\Varid{[+]}\;\Varid{p}_{01}\;\Varid{[*]}\;\Varid{m}_{11}\;\Varid{[+]}\;\Varid{m}_{00}\;\Varid{[*]}\;\Varid{r}_{01})\;\Varid{r}_{11}{}\<[E]%
\\[\blanklineskip]%
\>[3]{}\Varid{congLS}\;\!:\!\;\Conid{SemiNearRing2.CongL}\;\Conid{SNR2}{}\<[E]%
\\
\>[3]{}\Varid{congLS}\;\Conid{P}\;\Conid{Q}\;\mathrel{=}\;\Varid{transS}\;(\Varid{symS}\;\Conid{P}\;\mathbin{<\!\!\!+\!_S\!\!>}\;\Varid{reflS})\;\Conid{Q}{}\<[E]%
\ColumnHook
\end{hscode}\resethooks
\begin{hscode}\SaveRestoreHook
\column{B}{@{}>{\hspre}l<{\hspost}@{}}%
\column{3}{@{}>{\hspre}l<{\hspost}@{}}%
\column{4}{@{}>{\hspre}l<{\hspost}@{}}%
\column{21}{@{}>{\hspre}l<{\hspost}@{}}%
\column{E}{@{}>{\hspre}l<{\hspost}@{}}%
\>[3]{}\Conid{CSNR}\;\!:\!\;\Conid{ClosedSemiNearRing}{}\<[E]%
\\
\>[3]{}\Conid{CSNR}\;\mathrel{=}\;\kw{record}\;\{\mskip1.5mu {}\<[E]%
\\
\>[3]{}\hsindent{1}{}\<[4]%
\>[4]{}\Varid{snr2}\;{}\<[21]%
\>[21]{}\mathrel{=}\;\Conid{SNR2};{}\<[E]%
\\
\>[3]{}\hsindent{1}{}\<[4]%
\>[4]{}\Varid{entireQ}\;{}\<[21]%
\>[21]{}\mathrel{=}\;\Varid{entireQStep};{}\<[E]%
\\
\>[3]{}\hsindent{1}{}\<[4]%
\>[4]{}\Varid{closureIsLeast}\;{}\<[21]%
\>[21]{}\mathrel{=}\;\Varid{closureIsLeastS};{}\<[E]%
\\
\>[3]{}\hsindent{1}{}\<[4]%
\>[4]{}\Varid{entireL}\;{}\<[21]%
\>[21]{}\mathrel{=}\;\Varid{entireLS};{}\<[E]%
\\
\>[3]{}\hsindent{1}{}\<[4]%
\>[4]{}\Varid{uniqueL}\;{}\<[21]%
\>[21]{}\mathrel{=}\;\Varid{uniqueLS};{}\<[E]%
\\
\>[3]{}\hsindent{1}{}\<[4]%
\>[4]{}\Varid{congL}\;{}\<[21]%
\>[21]{}\mathrel{=}\;\Varid{congLS};{}\<[E]%
\\
\>[3]{}\hsindent{1}{}\<[4]%
\>[4]{}\Varid{completionMono}\;{}\<[21]%
\>[21]{}\mathrel{=}\;\Varid{completionMonoS}\mskip1.5mu\}{}\<[E]%
\ColumnHook
\end{hscode}\resethooks
The base cases of lemmas are trivial and omitted. For the main theorem, we have:
\begin{hscode}\SaveRestoreHook
\column{B}{@{}>{\hspre}l<{\hspost}@{}}%
\column{12}{@{}>{\hspre}l<{\hspost}@{}}%
\column{51}{@{}>{\hspre}l<{\hspost}@{}}%
\column{E}{@{}>{\hspre}l<{\hspost}@{}}%
\>[12]{}\Varid{leastQBase}\;\mathrel{=}\;\lambda \;\anonymous \;\anonymous \;\Varid{→}\;\Varid{identityˡ}_s\;0_s{}\<[51]%
\>[51]{}\mbox{\onelinecomment  \structure{5.2}{Base case for least \ensuremath{\Conid{Q}} }}{}\<[E]%
\ColumnHook
\end{hscode}\resethooks

This concludes our proof.
The complete development is checked by Agda 2.4, and is available as
supplementary material for this paper.
The set of Agda files is written in literate programming style, and
doubles up as the input for the typesetting program.
\section{Related Work}
\label{sec:related}

\subsection{Efficient Parsing}

One of the main motivations for this work is the discovery that
Valiant's algorithm is not only interesting theoretically (as it gives
an upper bound on the complexity of context-free recognition), but
also practically.

Indeed \citet{bernardy_efficient_2013,bernardy_efficient_2015} have
recently shown that the divide and conquer structure of Valiant's
parsing algorithm yields an efficient parallel algorithm, given
commonly occurring conditions on the input. In sum, if the input is
organised hierarchically, then the conquer step is $O(\log^2 n)$,
instead of being as complex as matrix multiplication.

\subsection{Certified Parsing}

Several certified parsers exist.
\citet{firsov_certified_2013,coquand_decision_2011} have implemented
parsers for regular languages.  \citet{jourdan_validating_2012} have
implemented a parser for LR languages. Two certifications of full
context-free parsers have been produced while the present paper was in
submission.

\citet{ridge_simple_2014} has produced a novel, fully verified parser
that also has good practical performance. \citet{firsov_certified_2014} have
verified the CYK parsing algorithm (a precursor of Valiant's).

While Valiant's algorithm gives the best asymptotic bounds and is also
known to behave well in many practical situations
\citep{bernardy_efficient_2015}, we have not measured the practical
performance of (the Agda version of) our implementation.
As we write, we expect it to be very bad: the program extraction
mechanism of Agda is not mature and we cannot expect good results.

A previous formalisation of Valiant's algorithm was carried by
\citet{bth_sjblom_agda_2013} under the supervision of the authors of
this paper.  The present work is a redevelopment of the proof from
scratch. Indeed, the proof produced by
\citeauthor{bth_sjblom_agda_2013} is opaque: its structure does not
match the informal proof.  A close match between the formal and the
informal proof was enabled by two key design choices:
1. our representation of matrices as an (extensible) record with all
the necessary lemmas and
2. the use of two different types for square and upper triangular
matrices (instead of using sigma types).
In particular, using sigma types mean that every triangular matrix
would be composed of two fields (a square matrix and a proof of
triangularity).
In turn, using this structure requires to writes a proof whenever such
a matrix is produced.
Baking triangularity into the structure of types avoids this
complication.

\subsection{Certified Parsing Combinators}
In functional programming, one often uses parsing combinators as a
language formalism.  This means that the grammar is represented
directly as code in the host language, instead of data structures.
On the one hand, combinator parsing is generally less efficient than
context-free parsing: the latter technique has access to the full
grammar data, including its recursive structure, and thus more
intelligent processing of the input is possible. On the other hand,
combinators can in principle describe languages which are not context
free.

\citet{danielsson_total_2010} has formalised combinator parsing in
Agda.  The expressivity of Danielsson's formalism is maximal: it can
express all languages which can be decided by an Agda algorithm.

While the parser of \citep{ridge_simple_2014}, provides also a
combinator interface, it begins by converting the grammar to a first
order representation.

\subsection{Algebra of Programming}

The idea of deriving programs from specifications comes from the school of
``Algebra of Programming'' (AoP) \citep{birddemoor96}. While AoP uses
relational algebra, our setting is Agda. This paper is thus a direct
descendent of ``Algebra of Programming in Agda''
\citep{mukojansson08:mpc:dcc, MuKoJansson2009AoPA}, and as such leverages the power not only
of Agda, but of an extensive set of standard libraries, developed over
last decade mainly by \citet{danielsson_agda_2013}. These libraries
have already been used for several applications.

Yet, during the maturation of the present work we have found the AoP
canon too limited for our purposes, and have departed from it
significantly.
We have not derived the algorithm from the full specification, as AoP
prescribes, but only part of it (Equation \ensuremath{\Conid{Q}}).

Further, while the AoP school prescribes to perform derivations
outside the recursive structure (and thus makes extensive usage of
catamorphisms and their properties), we have first expanded the
recursive structure, and used equational derivations for the inductive
case. This primary unrolling of recursion significantly simplified our
proof, as every theorem (and lemma) proved using the same induction
pattern is simply added as a component of a record (coding up the
induction hypothesis).

\section{Extensions}
\label{sec:extensions}

\subsection{General size for matrices}
\label{sec:general-size}

In order to deal with matrices of general sizes (not only \ensuremath{2^\Varid{n}\;\mathbin{\!\times\!}\;2^\Varid{n}}), we need to
define the following type for the size of matrices:

\begin{hscode}\SaveRestoreHook
\column{B}{@{}>{\hspre}l<{\hspost}@{}}%
\column{3}{@{}>{\hspre}l<{\hspost}@{}}%
\column{36}{@{}>{\hspre}l<{\hspost}@{}}%
\column{E}{@{}>{\hspre}l<{\hspost}@{}}%
\>[B]{}\kw{data}\;\Conid{Shape}\;\!:\!\;\Conid{Set}\;\kw{where}{}\<[E]%
\\
\>[B]{}\hsindent{3}{}\<[3]%
\>[3]{}\Conid{Leaf}\;\!:\!\;\Conid{Shape}{}\<[36]%
\>[36]{}\mbox{\onelinecomment  One}{}\<[E]%
\\
\>[B]{}\hsindent{3}{}\<[3]%
\>[3]{}\Conid{Bin}\;\!:\!\;\Conid{Shape}\;\to \;\Conid{Shape}\;\to \;\Conid{Shape}{}\<[36]%
\>[36]{}\mbox{\onelinecomment  Sum of the two shapes}{}\<[E]%
\ColumnHook
\end{hscode}\resethooks

In addition to the size, the above type gives the way to divide a
matrix into submatrices. It is a good idea to use the above structure
instead of a (unary or binary) natural number, because the combination
of sizes requires no computation.

Then, every type for matrices needs to be indexed on such a \ensuremath{\Conid{Shape}},
and in particular the functor generating matrices becomes a
doubly-indexed functor, and the proof needs special cases when one of
the dimensions is \ensuremath{\Conid{One}}.
\pagebreak 

\begin{hscode}\SaveRestoreHook
\column{B}{@{}>{\hspre}l<{\hspost}@{}}%
\column{3}{@{}>{\hspre}l<{\hspost}@{}}%
\column{9}{@{}>{\hspre}l<{\hspost}@{}}%
\column{12}{@{}>{\hspre}l<{\hspost}@{}}%
\column{15}{@{}>{\hspre}l<{\hspost}@{}}%
\column{E}{@{}>{\hspre}l<{\hspost}@{}}%
\>[B]{}\kw{data}\;\Conid{Mat}\;\!:\!\;\Conid{Shape}\;\to \;\Conid{Shape}\;\to \;\Conid{Set}\;\to \;\Conid{Set}\;\kw{where}{}\<[E]%
\\
\>[B]{}\hsindent{3}{}\<[3]%
\>[3]{}\Conid{Quad}\;\!:\!\;{}\<[12]%
\>[12]{}\Conid{Mat}\;\Varid{x}_1\;\Varid{y}_1\;\Varid{a}\;\to \;\Conid{Mat}\;\Varid{x}_2\;\Varid{y}_1\;\Varid{a}\;\to \;{}\<[E]%
\\
\>[12]{}\Conid{Mat}\;\Varid{x}_1\;\Varid{y}_2\;\Varid{a}\;\to \;\Conid{Mat}\;\Varid{x}_2\;\Varid{y}_2\;\Varid{a}\;\to \;{}\<[E]%
\\
\>[12]{}\Conid{Mat}\;(\Conid{Bin}\;\Varid{x}_1\;\Varid{x}_2)\;(\Conid{Bin}\;\Varid{y}_1\;\Varid{y}_2)\;\Varid{a}{}\<[E]%
\\
\>[B]{}\hsindent{3}{}\<[3]%
\>[3]{}\Conid{OneByOne}\;\!:\!\;{}\<[15]%
\>[15]{}\Varid{a}\;\to \;\Conid{Mat}\;\Conid{Leaf}\;\Conid{Leaf}\;\Varid{a}{}\<[E]%
\\
\>[B]{}\hsindent{3}{}\<[3]%
\>[3]{}\Conid{Row}\;{}\<[9]%
\>[9]{}\!:\!\;{}\<[12]%
\>[12]{}\Conid{Mat}\;\Varid{x}_1\;\Conid{Leaf}\;\Varid{a}\;\to \;\Conid{Mat}\;\Varid{x}_2\;\Conid{Leaf}\;\Varid{a}\;\to \;\Conid{Mat}\;(\Conid{Bin}\;\Varid{x}_1\;\Varid{x}_2)\;\Conid{Leaf}\;\Varid{a}{}\<[E]%
\\
\>[B]{}\hsindent{3}{}\<[3]%
\>[3]{}\Conid{Col}\;{}\<[9]%
\>[9]{}\!:\!\;{}\<[12]%
\>[12]{}\Conid{Mat}\;\Conid{Leaf}\;\Varid{y}_1\;\Varid{a}\;\to \;\Conid{Mat}\;\Conid{Leaf}\;\Varid{y}_2\;\Varid{a}\;\to \;\Conid{Mat}\;\Conid{Leaf}\;(\Conid{Bin}\;\Varid{y}_1\;\Varid{y}_2)\;\Varid{a}{}\<[E]%
\ColumnHook
\end{hscode}\resethooks

\subsection{Boolean Grammars}

\citet{okhotin_parsing_2014} has shown how to extend Valiant's
algorithm to parse Boolean grammars.  Boolean grammars allow to define
the generation of non-terminals not only by union of production rules,
but also intersection and complement. They can characterise non
context-free languages, such as $\{ a^n b^n c^n \mid n ∈ ℕ\}$.  In
this case, the ring-like structure that we have used is not
sufficient: one must apply a Boolean function to all possible
combination of non-terminals before obtaining the parses of a given
substring.

We believe that a straightforward extension of our proof to Okhotin's
variant is possible.

\subsection{Semirings and relatives}

Our formalisation work showed that the specification could be
generalised to work for (matrices over) arbitrary closed semirings
\citep{dolan_fun_2013} and even further.
Comparing to the theory explored in Dolan's paper they mention
already on their page 1:
``If we have an affine map $x → ax + b$ in some closed semiring, then
$x = a{∗~} b$ is a fixpoint''.
It appears that our linear equation \ensuremath{\Varid{x}\;\mathrel{=}\;\Varid{y}\;\Varid{+}\;\Varid{a}\;\Varid{x}\;\Varid{+}\;\Varid{x}\;\Varid{b}} in
\fref{sec:completion} is the natural generalisation of the affine map
fixed point to the case of non-commutative algebra and that (the
corner case \ensuremath{\Conid{V}} of) Valiant's algorithm computes this fixed point for
upper triangular matrices.
Future work includes exploring this relation in more detail and
perhaps generalise out development to arbitrary (non-triangular)
matrices.

\subsection{Sparse matrices}
The efficiency of Valiant's algorithm in the average case relies on
using sparse matrices \citep{bernardy_efficient_2015}.
The above proof does not deal with sparseness.
Yet, it is straightforward to support sparseness as outlined by
\citet{bernardy_efficient_2015}:
one needs to change the \ensuremath{\Conid{U}} type to be a disjunction between the
2-by-2 case and the empty matrix case.


\newcommand{\HREF}[2]{\href{#2}{#1}} 
\bibliographystyle{abbrvnat}

\begin{thebibliography}{21}
\providecommand{\natexlab}[1]{#1}
\providecommand{\url}[1]{\texttt{#1}}
\expandafter\ifx\csname urlstyle\endcsname\relax
  \providecommand{\doi}[1]{doi: #1}\else
  \providecommand{\doi}{doi: \begingroup \urlstyle{rm}\Url}\fi

\bibitem[Bernardy and Claessen(2013)]{bernardy_efficient_2013}
J.-P. Bernardy and K.~Claessen.
\newblock Efficient divide-and-conquer parsing of practical context-free
  languages.
\newblock In \emph{Proc. of ICFP 2013}, pages 111--122, 2013.

\bibitem[Bernardy and Claessen(2015)]{bernardy_efficient_2015}
J.-P. Bernardy and K.~Claessen.
\newblock Efficient parallel and incremental parsing of practical context-free
  languages.
\newblock \emph{J. of Funct. Prog.}, 25, 2015.
\newblock ISSN 1469-7653.
\newblock \doi{10.1017/S0956796815000131}.

\bibitem[Bird and {de Moor}(1997)]{birddemoor96}
R.~Bird and O.~{de Moor}.
\newblock \emph{Algebra of Programming}, volume 100 of \emph{International
  Series in Computer Science}.
\newblock Prentice-Hall International, 1997.

\bibitem[{Bååth Sjöblom}(2013)]{bth_sjblom_agda_2013}
T.~{Bååth Sjöblom}.
\newblock \emph{An {Agda} proof of the correctness of {Valiant}'s algorithm for
  context free parsing}.
\newblock {MSc} thesis, Chalmers University of Tech., 2013.

\bibitem[Chomsky(1957)]{chomsky_syntactic_1957}
N.~Chomsky.
\newblock \emph{Syntactic Structures}.
\newblock Mouton de Gruyter, 1957.

\bibitem[Chomsky(1959)]{chomsky_certain_1959}
N.~Chomsky.
\newblock On certain formal properties of grammars.
\newblock \emph{Information and control}, 2\penalty0 (2):\penalty0 137--167,
  1959.

\bibitem[Coquand and Siles(2011)]{coquand_decision_2011}
T.~Coquand and V.~Siles.
\newblock A decision procedure for regular expression equivalence in type
  theory.
\newblock In \emph{Certified Programs and Proofs}, pages 119--134. Springer,
  2011.

\bibitem[Danielsson(2010)]{danielsson_total_2010}
N.~A. Danielsson.
\newblock Total parser combinators.
\newblock In \emph{Proc. of ICFP 2010}, ICFP '10, pages 285--296. ACM, 2010.

\bibitem[Danielsson and {The Agda Team}(2013)]{danielsson_agda_2013}
N.~A. Danielsson and {The Agda Team}.
\newblock The {Agda} standard library, version 0.7, 2013.

\bibitem[Dolan(2013)]{dolan_fun_2013}
S.~Dolan.
\newblock Fun with semirings: A funct. pearl on the abuse of linear algebra.
\newblock In \emph{Proc. of the 18th {ACM} {SIGPLAN} International Conf. on
  Funct. Prog.}, ICFP '13, pages 101--110. ACM, 2013.

\bibitem[Firsov and Uustalu(2013)]{firsov_certified_2013}
D.~Firsov and T.~Uustalu.
\newblock Certified parsing of regular languages.
\newblock In \emph{Certified Programs and Proofs}, pages 98--113. Springer,
  2013.

\bibitem[Firsov and Uustalu(2014)]{firsov_certified_2014}
D.~Firsov and T.~Uustalu.
\newblock Certified {CYK} parsing of context-free languages.
\newblock \emph{J. Log. Algebr. Meth. Program.}, 83\penalty0 (5-6):\penalty0
  459--468, 2014.

\bibitem[Jourdan et~al.(2012)Jourdan, Pottier, and
  Leroy]{jourdan_validating_2012}
J.~Jourdan, F.~Pottier, and X.~Leroy.
\newblock Validating {LR(1)} parsers.
\newblock In \emph{{ESOP} 2012}, pages 397--416, 2012.

\bibitem[Lange and Lei{\ss}(2009)]{lange_cnf_2009}
M.~Lange and H.~Lei{\ss}.
\newblock To {CNF} or not to {CNF}? an efficient yet presentable version of the
  {CYK} algorithm.
\newblock \emph{Informatica Didactica (8)(2008--2010)}, 2009.

\bibitem[Lee(2002)]{lee_fast_2002}
L.~Lee.
\newblock Fast context-free grammar parsing requires fast boolean matrix
  multiplication.
\newblock \emph{J. of the ACM (JACM)}, 49\penalty0 (1):\penalty0 1--15, 2002.

\bibitem[Mu et~al.(2008)Mu, Ko, and Jansson]{mukojansson08:mpc:dcc}
S.-C. Mu, H.-S. Ko, and P.~Jansson.
\newblock Algebra of programming using dependent types.
\newblock In \emph{Mathematics of Program Construction}, volume 5133/2008 of
  \emph{LNCS}, pages 268--283. Springer, 2008.
\newblock \doi{10.1007/978-3-540-70594-9_15}.

\bibitem[Mu et~al.(2009)Mu, Ko, and Jansson]{MuKoJansson2009AoPA}
S.-C. Mu, H.-S. Ko, and P.~Jansson.
\newblock Algebra of programming in {Agda}: dependent types for relational
  program derivation.
\newblock \emph{J. Funct. Program.}, 19:\penalty0 545--579, 2009.
\newblock \doi{10.1017/S0956796809007345}.

\bibitem[Norell(2007)]{norell_practical_2007}
U.~Norell.
\newblock \emph{Towards a practical programming language based on dependent
  type theory}.
\newblock {PhD} thesis, Chalmers Tekniska Högskola, 2007.

\bibitem[Okhotin(2014)]{okhotin_parsing_2014}
A.~Okhotin.
\newblock Parsing by matrix multiplication generalized to boolean grammars.
\newblock \emph{Theor. Comp. Sci.}, 516\penalty0 (0):\penalty0 101 -- 120,
  2014.

\bibitem[Ridge(2014)]{ridge_simple_2014}
T.~Ridge.
\newblock Simple, efficient, sound and complete combinator parsing for all
  context-free grammars, using an oracle.
\newblock In \emph{Soft. Language Engineering - 7th International Conf., {SLE}
  2014, V{\"{a}}ster{\aa}s, Sweden, September 15-16, 2014. Proc.}, pages
  261--281, 2014.

\bibitem[Valiant(1975)]{valiant_general_1975}
L.~Valiant.
\newblock General context-free recognition in less than cubic time.
\newblock \emph{J. of computer and system sciences}, 10\penalty0 (2):\penalty0
  308--314, 1975.

\end{thebibliography}

\end{document}